\documentclass[manuscript, screen]{jair}

\setcopyright{cc}
\copyrightyear{2026}
\acmDOI{10.1613/jair.1.21111}

\JAIRAE{Ivor Tsang}
\JAIRTrack{Surveys} 
\acmVolume{86}
\acmArticle{17}
\acmMonth{07}
\acmYear{2026}

\RequirePackage[
  datamodel=acmdatamodel,
  style=acmauthoryear,
  backend=biber,
  giveninits=true,
  uniquename=init,
  ]{biblatex}

\usepackage{longtable}

\usepackage[T1]{fontenc}
\usepackage{kantlipsum}
\usepackage[usenames,dvipsnames]{xcolor}
\usepackage[breakable, theorems, skins]{tcolorbox}
\tcbset{enhanced}
\newtcolorbox{specialistbox}[1]{fonttitle=\bfseries,title=#1,colframe=gray!75!white}
\newcommand{\prn}[1]{\left(#1\right)}

\usepackage{paralist}

\newcommand{\eps}{\epsilon}
\newcommand{\del}{\delta}

\newcommand{\Norm}[1]{\left\lVert#1\right\rVert}

\newcommand{\Ncal}{\mathcal{N}}
\newcommand{\cal}{\mathcal{N}}

\newcommand{\noisestate}{\mathbf{\mathcal{S}}}
\newcommand{\optstate}{\mathcal{P}}

\newtheorem{remark}{Remark}
\newtheorem{definition}{Definition}

\usepackage{cleveref}
\usepackage{algorithm}
\usepackage[noend]{algpseudocode}

\def\HiLi{\leavevmode\rlap{\hbox to 0.9\hsize{\color{lightgray}\leaders\hrule height .8\baselineskip depth .5ex\hfill}}}

\def\HiLiLong{\leavevmode\rlap{\hbox to 0.94\hsize{\color{lightgray}\leaders\hrule height .8\baselineskip depth .5ex\hfill}}}

\usepackage{caption}
\usepackage{subcaption}

\usepackage[framemethod=tikz]{mdframed}
\usepackage{multicol}
\usepackage{multirow}
\usepackage{color}
\usepackage{graphicx}
\usepackage{booktabs}
\usepackage{amsmath} 
\usepackage{caption}
\usepackage{rotating}

\usepackage{setspace}
\usepackage{amsthm}
\usepackage{tablefootnote}
\usepackage{dsfont}

\usepackage{pifont}
\usepackage{array}
\usepackage{float}
\restylefloat{table}

\usepackage{cleveref}

\begin{document}

\title[How to DP-fy Your Data]{How to DP-fy Your Data: A Practical Guide to Generating Synthetic Data
With Differential Privacy}

\author{Natalia Ponomareva}
\authornote{Corresponding Author.}
\orcid{0009-0005-6761-1468}
\email{nponomareva@google.com}
\affiliation{%
  \institution{Google Research}
  \country{USA}
}

\author{Zheng Xu}
\orcid{0009-0003-6747-3953}
\email{xuzhustc@gmail.com}
\affiliation{%
  \institution{Google Research}
  \country{USA}
}

\author{H. Brendan McMahan}
\orcid{0009-0003-5892-4193}
\email{mcmahan@google.com}
\affiliation{%
  \institution{Google Research}
  \country{USA}
}

\author{Peter Kairouz}
\orcid{0000-0001-6897-5937}
\email{kairouz@google.com}
\affiliation{%
  \institution{Google Research}
  \country{USA}
}

\author{Lucas Rosenblatt}
\email{lurosenb@google.com}
\affiliation{%
  \institution{NYU}
  \authornote{Work done at Google as part of student researcher engagement} 
  \city{New York}
  \state{NY}
  \country{USA}
}

\author{Vincent Cohen-Addad}
\email{cohenaddad@google.com}
\affiliation{%
  \institution{Google Research}
  \city{New York}
  \state{NY}
  \country{USA}
}

\author{Crist\'obal Guzm\'an}
\authornote{Work done while at Google}
\orcid{0000-0002-1498-2055}
\email{crguzmanp@uc.cl}
\affiliation{
  \institution{Institute for Mathematical and Computational Engineering, Faculty of Mathematics and School of Engineering, Pontificia Universidad Cat\'olica de Chile}
  \city{Santiago}
  \country{Chile}
}

\author{Ryan McKenna}
\orcid{0000-0002-4950-1952}
\email{mckennar@google.com }
\affiliation{%
  \institution{Google Research}
  \country{USA}
}

\author{Galen Andrew}
\orcid{0009-0004-9804-7882}
\email{galenandrew@google.com }
\affiliation{%
  \institution{Google Research}
  \country{USA}
}

\author{Alex Bie}
\email{alexbie@google.com }
\affiliation{%
  \institution{Google Research}
  \country{USA}
}

\author{Da Yu}
\email{dayuwork@google.com }
\affiliation{%
  \institution{Google Research}
  \country{USA}
}

\author{Alex Kurakin}
\orcid{0009-0008-5952-9394}
\email{kurakin@google.com }
\affiliation{%
  \institution{Google DeepMind}
  \city{Mountain View}
  \state{CA}
  \country{USA}
}

\author{Morteza Zadimoghaddam}
\orcid{0000-0003-0717-1120}
\email{zadim@google.com }
\affiliation{%
  \institution{Google Research}
  \country{USA}
}

\author{Sergei Vassilvitskii}
\orcid{0000-0003-0235-1624}
\email{sergeiv@google.com }
\affiliation{%
  \institution{Google Research}
  \country{USA}
}

\author{Andreas Terzis}
\orcid{0000-0002-5681-3399}
\email{aterzis@google.com }
\affiliation{%
  \institution{Google DeepMind}
  \city{Mountain View}
  \state{CA}
  \country{USA}
}

\renewcommand{\shortauthors}{Ponomareva, Xu, McMahan, Kairouz et al.}



\begin{abstract}
High quality data is of vital importance for unlocking the full potential of AI for end users.  Villalobos et al. stated in 2024 that finding new sources of such data is getting harder as most publicly-available human generated data will soon have been used. Additionally, publicly available data often is not representative of users of a particular system --- for example, a research speech dataset of contractors interacting with an AI assistant will likely be more homogeneous, well articulated and self-censored that real world commands that end users will issue. Therefore unlocking high-quality data grounded in real user interactions is of vital interest to both system creators and end users themselves. However, the direct use of user data comes with significant privacy risks, which must be addressed before the data can be used. Differential Privacy (DP) is a well established framework for reasoning about and limiting information leakage, and is a gold standard for protecting user privacy. The focus of this work, \emph{Differentially Private Synthetic data}, refers to synthetic data that preserves the overall trends of source data (often user-generated), while providing strong privacy guarantees to individuals that contributed to the source dataset.  DP synthetic data can unlock the value of datasets that have previously been inaccessible due to privacy concerns. Additionally, DP synthetic data can replace the use of sensitive datasets that previously have only had rudimentary protections like ad-hoc rule-based anonymization. 

In this survey we explore the full suite of techniques surrounding DP synthetic data, the types of privacy protections different generation approaches can offer, and the state-of-the-art for various modalities including image, tabular, text and federated (decentralized) data. We outline all the components needed in a system that generates DP synthetic data, from sensitive data handling and preparation, to tracking the use of synthetic data and empirical privacy testing. 

We hope that work will result in increased adoption of DP synthetic data, spur additional research in still underexplored domains, and additionally increase trust in DP synthetic data approaches.

\end{abstract}

\received{November 20, 2025}
\received[accepted]{25 April 2026}

\maketitle

\section{Introduction}\label{sec-intro}

The modern world thrives on data. Vast quantities of information, encompassing sensitive domains like healthcare records, financial transactions, personal user preferences and others are continuously generated and utilized. Leveraging this data for AI training, fine-tuning, evaluation, or inference can unlock scientific breakthroughs, business insights, and societal advancements. However, these practices carry significant privacy risks that need to be properly addressed. Traditional methods for protecting privacy like data sanitization and de-identification have been proven increasingly inadequate~\cite{brown2022doesmeanlanguagemodel}. Sophisticated attacks, including re-identification through linking anonymized datasets with external information sources (\textit{data linkage attacks}) can successfully uncover individuals' identities and sensitive attributes, undermining privacy protections \cite{4531148}. The inherent difficulty in predicting all possible ways large datasets might be queried or linked makes simple anonymization a fragile defense~\cite{4531148}. Consequently, there is a critical need for more robust, mathematically grounded approaches to enable data analysis and data sharing, while rigorously protecting individual privacy~\cite{royalsoc:synthdata}.

Synthetic data generation has emerged as a compelling approach to address this challenge~\cite{royalsoc:synthdata}. The process involves creating artificial data that mimics the statistical patterns and properties of a real dataset, often using algorithms or generative models trained on the original data. This artificial data ideally contains no real values from the source dataset, offering a potential pathway to share insights, facilitate research, develop and test software, and train machine learning models without directly exposing sensitive individual records~\cite{msr:private_synth}. 

However, a common and dangerous misconception is that synthetic data is inherently private~\cite{royalsoc:synthdata,ricciato2024cautionary}. Generative models, especially powerful modern architectures, can and do inadvertently memorize and replicate specific details from their training data \cite{carlini2021extracting}. Synthetic records therefore might be near-copies of real ones, or the generated dataset might leak aggregate information that still allows inferences about individuals present in the original data. Consequently, simply generating synthetic data based on the original sensitive data does not automatically guarantee privacy; significant care and formal techniques are required to produce synthetic data that is both useful and demonstrably private.

Differential Privacy (DP)  \cite{DworkRothBook:2014} provides the necessary rigorous framework to address the privacy shortcomings of naive synthetic data generation. DP offers a rigorous quantified  guarantee of privacy protection against a wide range of attacks, including those unknown at the time of deployment~\cite{harvard:dp_primer}. The core promise of DP is intuitive yet powerful: the outcome of a differentially private analysis or data release should be roughly the same whether or not any single individual's data was included in the input dataset~\cite{DworkRothBook:2014}. This effectively masks the contribution of any single person, ensuring they remain ``hidden in the crowd'' and limiting what can be learned specifically about them from the output~\cite{harvard:dp_primer}.

This survey provides a holistic overview of the field of DP synthetic data generation.\footnote{Unintended omissions and errors are possible, especially given the breadth and challenging nature of the topic. We welcome feedback on this work from all interested parties.} This is a self-contained guide on how to create DP synthetic data for various modalities including image, tabular and text. It provides analysis and comparison of various methods that exist, highlights current state-of-the-art methods and their ideal usecases as well as shortcomings. We cover the full life cycle of DP synthetic data: from data preparation to safe data handling, best practices for testing the synthetic data and lineage tracking for usage of DP synthetic data. 

Our target audience is practitioners, privacy advocates, and academic researchers alike. Practitioners will benefit from a gentle introduction to DP synthetic data field and clear definitions of what DP synthetic data guarantees and protects, comparisons of competing methods for each modality and an overview of what functionality a DP synthetic data creation system should provide. Additionally we cover important implementation details, that when overlooked could weaken privacy guarantees of the system. For example, the need to understand all the components of a generative model that is being used for creating DP synthetic data and the need to track DP synthetic data lineage. For academic researchers this work can serve as a one-stop in-depth survey of current state-of-the-art and advances in the field of DP synthetic data. We also outline open questions and challenges for each of the modalities, hoping that it will spur additional academic interest for these directions. 

Finally, we note that this survey assumes a background in ML including modern auto-regressive models like large language models (LLMs). Throughout the content we mark some sections as {\color{red}*} to indicate that they contain additional in-depth theoretical details and can be safely skipped if reading/skimming for the broad picture. 
\begin{specialistbox}{Attention boxes}
We also introduce grey boxes like this one to highlight an important statement, conclusion or suggestion.
\end{specialistbox}

\subsection{Preview of the Later Sections}
This survey is organized as follows.

\begin{compactenum}
    \item We begin this Section with an introduction to DP synthetic data and position it w.r.t. other types of synthetic data in Section \ref{sec-types-syn-data}. Section \ref{sec-syn-motivation} clarifies the motivation behind DP synthetic data and outlines its pros and cons. Section \ref{sec-syn-data-challenges} introduces challenges that need to be overcome to obtain useful DP synthetic data for any modality. Finally, Section \ref{sec-fidelity-generic} discusses in broad strokes the evaluation of DP (and non DP) synthetic data for any modality, with later sections (Sections \ref{tabular-metrics}, \ref{sec-image-quality}, \ref{sec-text-metrics} and \ref{sec:fl-eval}) covering modality specific metrics.  

    \item Section \ref{sec-dp-overview} provides background information on Differential Privacy and DP synthetic data. We start with the definition of DP itself (Section \ref{sec-dp-def}) and introduce the concept of the privacy unit, which is paramount for understanding DP guarantees (Section \ref{sec:priv_unit}). We touch upon important properties and mechanisms of DP in Section \ref{sec:dp_property}{\color{red}*}. Section \ref{sec-ml-lifecycle} compares DP synthetic data with alternatives like direct DP training of ML models and outlines the types of methods for DP data synthesis that we will be using for categorizing methods for each modality. We also briefly introduce a technique called \textit{DP-Training} (Section \ref{sec-dpsgd}) which is a workhorse for DP synthetic data generation for all modalities and will be mentioned throughout this work extensively. Next, Sections \ref{sec:tabular} through \ref{sec:text} focus on various data modalities.

    \item In Section \ref{sec:tabular} we tackle the ubiquitous case of \textit{tabular data}. Section \ref{sec-tabular-unit} introduces a discussion of appropriate privacy unit for tabular domain. We introduce foundations of \textit{Query release} (the task of creating useful DP synthetic tabular data for answering statistical queries) and the concept of a query \textit{workload}  in Section \ref{sec-tabular-foundations}. Section \ref{sec-workfload-based} focuses on practical algorithms and outlines the prominent Select-Measure-Estimate paradigm that these algorithms often follow. This class of algorithms still dominates the field of DP tabular data synthesis. For an interested reader, we also dive into less scalable algorithms that are nevertheless are of theoretical interest in Section \ref{sec:theoretical_hist}{\color{red}*} and discuss data preprocessing (like imputation, domain calculation etc.) needed for both empirically and theory oriented algorithms in Section \ref{tabular-feature-handling}. We explore the new end-to-end approaches that employ the power of foundational models like LLMs for tabular data synthesis in Section \ref{tabular-generative} and offer a fair comparison of all the above methods in Section \ref{tabular-comparison}. Section \ref{tabular-metrics} surveys metrics that can be used for evaluating the quality of tabular synthetic data w.r.t the original data, and Section \ref{tabular-open-questions} concludes with open questions and challenges for DP data synthesis methods for this domain.

    \item Section \ref{sec:image} is devoted to the \textit{image} modality. We start with discussing an appropriate privacy unit for visual data in Section \ref{dp-image-privacy-unit} and introduce broad categorization of methods for DP image synthesis in \ref{sec-image-base-methods}. Section \ref{sec-dp-training-image} is a deep dive on \textit{DP-Training} of vision models -- the workhorse method for DP synthetic image synthesis -- where we explore \textit{DP-Training} and \textit{DP-finetuning} of GANs and Diffusion models. We additionally explore training-free method called \textit{Private Evolution} in Section \ref{sec-pe-image} and outline alternative methods (Section \ref{image-alternative-methods}), including new promising line of work that generates images via intermediary representation (Section \ref{sec-image-through-intermediary}). Section \ref{sec-image-comparison} offers a comparison and outlines ideal use cases for each of the explored methods, and Section \ref{sec-image-quality} investigates metrics for evaluating the quality of synthetic images. We conclude with open questions and challenges in Section \ref{sec-image-open-questions}.

    \item We explore the text modality in Section \ref{sec:text}. We outline difficulties in choosing an appropriate privacy unit for text data (Section \ref{dp-text-privacy-unit}), and explore \textit{DP-finetuning} -- the method that produces the best quality synthetic data given given sufficient amount of original data and compute in Section \ref{sec-dp-finetuning}. We outline alternative methods that don't require computationally expensive modifications to LLMs - namely DP inference (\ref{sec-dp-inference}) and Private evolution algorithms (Section \ref{private-evolution-text}). Proxy metrics that can be used to gauge the quality of the synthetic text are presented in Section \ref{sec-text-metrics}. We draw readers' attention to open research problems pertinent to text DP synthetic data generation in Section \ref{sec-text-open-questions}.

    \item While previous sections assumed \textit{Centralized deployment setting}, Section \ref{sec-fl} explores decentralized data in Federated Learning deployment setting. We first introduce \textit{Federated Learning} (Section \ref{sec-fl-primer}) and highlight difficulties of achieving DP in FL in Section \ref{sec:dpfl}. We examine the meaning of DP synthetic data in FL in Section \ref{sec:app_fl} and outline unique challenges one encounters when attempting to generate DP synthetic data from federated datasets. We briefly survey work on FL DP synthetic data generation in Section \ref{sec:method_fl} and compare and discuss such methods in Section \ref{sec:fl-discussion}. Section \ref{sec:fl-eval} introduces evaluation challenges unique to distributed nature of the sensitive data. 

    \item Section \ref{sec-practical} is devoted to practical considerations that designers of a DP synthetic data end-to-end generation system need to take into account. Section \ref{privacy-guarantees-decisions} highlights the most important questions that should be answered, including which privacy unit and privacy guarantees to target (Section \ref{privacy-guarantees-decisions}). We then focus on data preparation, including user-level contribution bounding (Section \ref{user-contrib-bounding}), and safe sensitive data handling practices (Section \ref{sec-safe-data-handling}). We investigate a crucial component that needs to be in place before any DP synthetic data is used in downstream tasks or shared -- namely \textit{Empirical Privacy Auditing} (Section \ref{sec:empirical-privacy-auditing}). We finally bring forward the need of lineage tracking for DP synthetic data in Section \ref{lineage-tracking}. 
 
\end{compactenum}

\subsection{Comparison With Other Surveys on DP Synthetic Data}
There have been many great surveys on DP synthetic data in the recent years. Some surveys focus on a particular method e.g. DP GANs \cite{Fan2020ASO} or data modalities (e.g. \citet{Bowen_2020,10.1007/978-3-031-82349-7_12} cover tabular data exclusively). Two recent comprehensive surveys that are most comparable with our work are by \citet{hu2023sokprivacypreservingdatasynthesis} and \citet{10.1007/978-3-031-82349-7_12}.

\textit{SoK: Privacy-Preserving Data Synthesis} \cite{hu2023sokprivacypreservingdatasynthesis} classifies existing methods for tabular, trajectory and graph data modalities, and surveys various methods including end-to-end methods like GAN-based and auto regressive generation. Their intended audience is academic researchers seeking to identify gaps in the theoretical landscape or the advanced practitioner selecting a model class based on high-level properties.

\textit{Privacy-Preserving Tabular Data Generation: Systematic Literature Review} \cite{10.1007/978-3-031-82349-7_12}  is a comprehensive literature review solely focused on tabular data generation methods. Their analysis provides a snapshot of what researchers are publishing, but offers limited guidance on how to operationalize these models in a production setting. It serves as an excellent historical record of academic trends but lacks the engineering "glue" required for deployment. \citet{yang2024tabulardatasynthesisdifferential} also focuses on tabular data and surveys marginal and deep learning based methods, and touches on decentralized settings (federated learning) for tabular data generation. This work, similarly to ours, provides metrics for evaluation fidelity and utility, and also identifies research gaps and future directions.

In contrast, 
\begin{compactenum}
    \item Our work is organized by data type (Tabular, Image, and Text) reflecting the reality that a practitioner's problem starts with the nature of their data, not a preference for a specific algorithm class. To the best of our knowledge, there is no extensive survey that addresses all the above modalities. For each of the modalities covered, we provide explicit guidance on how to chose the most appropriate method based on various constraints (e.g. amount of private data, yield required, etc.) and reason about fidelity and utility of the synthetic data.
    \item Our work covers both centralized and federated learning deployment settings.
    \item Our tabular chapter reconciles disparate lines of works (theoretical vs practical) in tabular marginal-based synthetic data generation, and compares them with end-to-end methods like recent auto-regressive models-based generation.
    \item Our work serves less as a catalog of algorithms and more as a blueprint for end-to-end system design, introducing critical but previously under-discussed components such as lineage tracking to ensure compliance and reproducibility, empirical privacy auditing to complement theoretical guarantees and provide more clear view of actual practical risks of using synthetic data, safe private data handling and preparation like user contribution bounding etc. Such "Practical Components" mark this work as a potential handbook for system architects rather than just a literature review purely for scientists.
\end{compactenum}
Given the breadth and depth of the topic, we are unable to cover all possible data modalities, with structured data types such as graph and trajectory data being a notable omission. We refer the reader to the survey by \citet{hu2023sokprivacypreservingdatasynthesis} for these topics.

\subsection{Types of Synthetic Data}\label{sec-types-syn-data}

Synthetic data is admittedly an overused term that can mean different things to different people. In this work, we assume that some ``real'' data (drawn from the ``original distribution'') exists or could exist, for example from real-world physical interactions or end-user interactions with a product or service. However, the real data cannot be used directly, perhaps because it is too limited in quantity, expensive to collect, or direct use would compromise privacy. Instead, alternative \emph{synthetic data} is desired. We assume the goal of synthetic data is to match some important properties of this real data. Such synthetic data can be obtained in many ways which can be in general categorized into two broad groups.

\paragraph{Methods that don't directly utilize real data.} For example, simply choosing a readily available proxy dataset (e.g., using the public corpus of Enron emails rather than real user emails) might be sufficient;  alternatively synthetic data can be generating by running a physical (e.g. robotic arm manipulation experiments) or virtual simulations; by prompting a generic LLM or other GenAI model to produce data with specific desired properties (e.g. "write a scientific article about types of synthetic data, include abstract, main body and conclusion"); by asking writers or contractors to produce the data; and many more. Importantly, none of these techniques directly utilize real data. Depending on the original distribution, it may be very hard to use these techniques to generate high-quality synthetic data (data that preserves the desired properties of the real data), or in fact to even evaluate the quality of the synthetic data.

\paragraph{Methods that use real data.} Grounding synthetic data using real data allows for much hider fidelity and utility of the synthetic data. 
One example of techniques in this groups is fitting parameters of a physics simulation based on the real data; classic techniques like SMOTE \cite{DBLP:journals/corr/abs-1106-1813}; prompting a human or an LLM to "Produce more documents like this text, but with different tones and styles, and on slightly different topics" or "Write chat bot interaction with a customer who is having problem with a banking interface"; or fine-tuning a generative model on the real data, and then using that fine-tuned model to sample as many examples as necessary. For any approach that utilizes the real data in producing the synthetic data, one must address the question of privacy: could private information in the real data be revealed in the synthetic data? DP, introduced more formally in Section \ref{sec-dp-def}, provides a rigorous framework for addressing these privacy concerns, and forms the basis for the DP Synthetic Data approaches described in this work which can generate high-quality synthetic data with strong privacy guarantees.

\citet{ricciato2024cautionary} proposes a more nuanced taxonomy, denoting methods that don't directly utilize real data as ``genuinely synthetic'', and distinguishing different methods that use real data depending on whether they conduct low-dimensional model fitting, human-designed logic, or high-dimensional model training.  \citet{ricciato2024cautionary} emphasizes that without additional mitigation, high-dimensional fitting introduces higher privacy risks, motivating approaches that offer differential privacy.

\subsection{Motivation}\label{sec-syn-motivation}
In context of ML, Differential Privacy \cite{DworkRothBook:2014} previously was applied mostly during the training stage by using specialized DP-algorithms like DP-SGD \cite{dpsgd_2016} or DP-FTRL \cite{kairouz21practical} in lieu of standard training techniques like SGD. The state-of-the-art for using differently-private training techniques has substantially advanced over the past decade, from early research that showed such approaches could be practical \cite{dpsgd_2016} to present time, when a variety of advanced approaches are available and DP-trained models have been used in large-scale production deployments \cite{xu23gboard,zhang23flgboard}. Nevertheless, directly training production models on privacy-sensitive data raises numerous practical challenges --- most importantly, only training infrastructure and models that support DP can be used, and because the private sensitive data cannot be inspected, there is no ability to debug with, filter, or human-label sensitive training data, and any analysis of sensitive data (e.g. like calculating counts, variance etc) should be done with DP.

The success of generative AI models has led to an alternative, 2-stage approach: we first create DP synthetic version of the private data that is highly representative of the raw private data. With suitable protections including appropriate DP parameters and audits, this DP synthetic data can be treated as fully anonymous. Thus, it can be used relatively freely in any standard ML workflow: filtering, augmentation, human-labeling or analysis, etc. The (still DP) outputs of any of these processes can then be used freely for tasks including model evaluation, fine-tuning of a production model (possibly one where its size or infrastructure would preclude DP training which is extremely computationally expensive and requires custom implementation of data processing and training), and more. In its ideal form, the formation of DP synthetic data concentrates all the challenges of anonymization in a single step, and provides an output that can be much more useful than a single DP model fine-tuned for a specific purpose.

DP synthetic data offers other advantages, on top of aforementioned ease-of-use, as well. From a privacy point of view, when fine-tuned GenAI models (e.g. LLMs) were used to create this data, the derived DP synthetic data offers a strictly smaller attack surface than the DP GenAI model from which it was derived. For example, numerous papers \cite{nasr2025scalable, carlini2022membership, carlini2021extracting}~have now shown that clever and malicious attacks can extract information from LLMs. However once DP synthetic data is generated, such attacks are structurally not possible --- only DP synthetic data is released and all the artifacts for preparing this data (including GenAI models if they were employed) can be erased or remain locked down. Further, the DP synthetic data itself can be additionally audited for potentially privacy-sensitive information before it is used for any downstream purpose.

The area of DP synthetic data has been extensively explored in context of tabular data, where workload-based algorithms have been perfected over the last 2 decades (Section \ref{sec:tabular}). (Non-DP) Image synthetic data, which requires larger and complex ML models like GANs, saw limited success until the emergence of powerful models like Diffusion, VAE and Autoregressive models (Section \ref{sec:image}). The field of synthetic text benefited immensely from new powerful autoregressive models like LLMs (Section \ref{sec:text}). Nevertheless, the area of DP synthetic data generation for complex modalities like text and images is very much still nascent. 

While there are indications of increased trust in DP synthetic data for navigating the inherent trade-off between data utility and individual privacy protection~\cite{msr:private_synth}, the adoption is not yet wide and DP synthetic data field remains contentious even among the experts. For example, a recent interview study with ``data experts'' (defined as professionals like researchers, data scientists, and practitioners in fields such as economics, medicine and public policy who directly engage with data) highlighted the challenge in practical adoption of differentially private synthetic data \cite{rosenblatt2024data}. This study found that while experts recognize the potential for broad data access through privacy preserving mechanisms, they express many concerns about the potential for DP noise to lead to incorrect conclusions. Recommendations in the face of these concerns included (1) validating DP synthetic data utility against real world use cases and (2) establishing clear standards of use.

We hope that this self-contained work will serve to increase the trust of experts (and non experts alike) in technology that underpins DP synthetic data generation, spur wider adoption of DP synthetic data in industry, and encourage research into methods and questions that are still not resolved. 

\subsection{Challenges in DP Synthetic Data Generation}\label{sec-syn-data-challenges} 
Generating high-quality DP synthetic data, especially for complex modalities like images, long text, multi-modal daata, still faces significant hurdles. 
\begin{itemize}
    \item \textbf{Balancing Utility and Privacy}  remains the paramount challenge~\cite{royalsoc:synthdata}. The noise or constraints introduced by DP inevitably distort the data~\cite{royalsoc:synthdata}. The key difficulty lies in minimizing this distortion to ensure the synthetic data retains sufficient fidelity (e.g. visual quality for images, statistical accuracy, and usefulness for downstream tasks while adhering to a chosen privacy budget $(\eps,\del)$~\cite{pubmed:synth_discoveries}.
    \item \textbf{Preventing Memorization:} Generative models that are commonly used for DP synthetic data creations, particularly large ones like LLMs, can memorize parts of their training data \cite{carlini2021extracting}. A DP mechanism must effectively prevent the model from simply outputting copies or near-copies of sensitive training examples. While DP provides a theoretical guarantee against exact replication influencing the output distribution, empirical validation (Section \ref{sec:empirical-privacy-auditing}) is often needed to assess practical memorization risks, especially under weaker privacy settings (larger $\eps$) which are commonly used in practice.
    \item \textbf{High Dimensionality of original sensitive data:} Both text and images data have high dimensionality representations and capturing the intricate distributions of such data requires complex models. Applying DP noise in high-dimensional spaces (either original space or large models gradient space during training) can easily overwhelm the signal, leading to poor quality synthetic images~\cite{chen:dpgen}. 
    \item \textbf{Computational Cost:} Training large-scale generative models like GANs, diffusion models or LLMs is computationally intensive. DP training (Section \ref{sec-dpsgd}, which is one of the main workhorse methods for DP synthetic data creationn, often adds significant overhead due to per-example gradient computations, clipping, and noise addition, need to scale batch size and tune the hyperparameters 
    increasing training time substantially \cite{Ponomareva_2023}
\end{itemize}

\subsection{Evaluating Synthetic Data Quality}\label{sec-fidelity-generic}

Finally, the evaluation of synthetic data is also a complex problem, as the data must be assessed on a wide range of metrics to ensure its multi-purpose utility.

The utility of DP synthetic data is not absolute but highly dependent on the specific context and intended application. Data generated with a certain DP guarantee might exhibit good visual fidelity (e.g., low FID score for image data \cite{NIPS2017_8a1d6947}) but perform poorly when used to train a downstream classifier, or vice-versa~\cite{lin:pe_images_pdf}. Similarly, data suitable for general trend analysis might yield incorrect results in specific statistical hypothesis tests due to DP-induced distortions~\cite{pubmed:synth_discoveries}. This implies that evaluating DP synthetic data requires a use-inspired approach~\cite{nasr:eval_dp_synth}. Simply stating data is "high quality" based on generic metrics is insufficient; its fitness for the specific purpose must be validated.

When choosing standard evaluation metrics that often work well, \citet{yang2024tabulardatasynthesisdifferential} suggests to think about \textit{fidelity} and \textit{utility} angles.

\paragraph{Fidelity} Fidelity metrics evaluate how closely the synthetic data matches statistical properties of the original data. 
Fidelity metrics are therefore often specific to each modality of the data we consider.  

\paragraph{Utility} For many practical applications the goal is to use synthetic data for some downstream task, for example training/finetuning an ML model. However obtaining such ultimate quality metric can be expensive during modeling and hyperparameter tuning process, which can require a large number of iterations.

A lot of works go the route of directly checking the performance of DP synthetic data on some simplified downstream tasks that server as proxy for the actual application. \citet{borisov2023languagemodelsrealistictabular} referred to the metric that compares performance of various ML models on original and synthetic data as \textit{Machine Learning efficiency (MLE)}. There is a design choice of what actual model one should train here - while academic works often train simple classifiers like Bert or XGBoost, the model should be as close in architecture and behavior to the final downstream model that will use this data.

Additionally to utility and fidelity angles, \citet{nasr:eval_dp_synth} suggests to use the following metrics to gauge usefulness of the synthetic data. 
\paragraph{Privacy leakage metrics}
Accessing potential privacy leakage from the sensitive data to the synthetic data is a very important task. There are a number of metrics that fall into this category (e.g. \textit{Membership Inference} or \textit{reconstruction attacks} success rates), and we we cover the ways to calculate them in depth in Section \ref{sec:empirical-privacy-auditing}

\paragraph{Fairness metrics} In labeled original data (or data with some attributes that can be treated as labels of interest, e.g. cluster assignments) with available demographic information (e.g. race of humans in pictures, sex in tabular data, pronouns used with professions in text data etc.) fairness metrics provide an important angle. \cite{DBLP:journals/corr/abs-2105-02778,nasr:eval_dp_synth}~suggest
metrics that evaluate the sum of absolute value  difference between the whole dataset and each subgroups (e.g. \textit{false positive equality difference (FPED)} and  \textit{false negative equality difference (FNED)}) and \textit{equalized odds} which evaluates whether false positive and true positive rates are the same for all groups.

We will cover modalities specific utility and fidelity metrics in subsequent Sections (\ref{tabular-metrics}, \ref{sec-image-quality}, \ref{sec-text-metrics} and \ref{sec:fl-eval}), however we want to emphasize once again that the most suitable metric for evaluating DP synthetic data quality will be always application specific. If for example downstream use of the data relies on preservation of temporal component between rows in a tabular dataset, standard fidelity and utility metrics will not be sufficient and metrics incorporating temporal correlations should be used for guiding synthetic data quality improvements. 

\section{A Brief Overview of Differential Privacy}\label{sec-dp-overview} 
Differential Privacy is a well established framework for reasoning about information leakage \cite{DworkRothBook:2014}.
In this section we provide a concise overview of differential privacy (DP) concepts pertinent to DP synthetic data. First we introduce a definition of what DP guarantees (Section \ref{sec-dp-def}) and a concept of privacy unit that we will be using throughout this paper (Section \ref{sec:priv_unit}). We will talk about where DP is commonly introduced in a lifecycle of machine learning and explain what DP synthetic data means (Section \ref{sec-ml-lifecycle}). We also briefly explore the concept of DP-Training, which is the workhorse algorithm that will be utilized as one of the main methods for DP synthetic data generation for all modalities in Section \ref{sec-dpsgd}.
For a more comprehensive introduction of DP, we refer the reader to \cite{DworkRothBook:2014,Ponomareva_2023}.

\subsection{Definition of Differential Privacy}\label{sec-dp-def}

Differential privacy (DP) is not a single technique but rather a provable promise. DP provides a formal, mathematical definition of privacy centered on the principle of  \emph{indistinguishability}. The guarantee is that the outcome of a computation will remain almost exactly the same, regardless of whether any single record is included or not. This promise can be captured by the definition of $(\epsilon, \delta)$-DP~\cite{DworkRothBook:2014}, also known as \emph{approximate DP}. 

\begin{definition}[$(\epsilon,\delta)$-Differential Privacy] \label{def:dp}
  A randomized algorithm $\mathcal{M}$ satisfies  ($\epsilon$, $\delta$)-DP if for any two neighboring datasets $\mathbb{D}$, $\mathbb{D}'$ and for all $\mathcal{S}\subset \text{Range}(\mathcal{M})$: 
\[
\textstyle{\Pr[\mathcal{M}(\mathbb{D}) \in \mathcal{S}] \leq e^{\epsilon}\Pr[\mathcal{M}(\mathbb{D}') \in \mathcal{S}] +\delta}
.\]
\end{definition}

In \cref{def:dp}, the randomized algorithm $\mathcal{M}$ is the process imposing differential privacy and whose outcome is released. For machine learning  (ML), $\mathcal{M}$ is often the model training process; for synthetic data generation, $\mathcal{M}$ can be the data generation process and the DP synthetic data is the outcome that is being released. The notion of \emph{neighboring datasets} (datasets that are exact copies of each other and only differ by a single record, so called ``privacy unit’’) is discussed further in \cref{sec:priv_unit}.

The two parameters $(\epsilon,\delta)$, quantify the level of privacy protection. $\epsilon$ is the privacy budget, where a smaller non-negative value implies a stronger privacy guarantee. $\delta$ is a relaxation term, representing a probability of failure of satisfying the strict $\epsilon$-bound. 
The presence of  $\delta$ is why \cref{def:dp} is also called approximate DP. This relaxation is crucial for the practical application of many important DP mechanisms, such as the Gaussian mechanism common in ML. Practically speaking, $\delta$ is typically set to be small, and often smaller than the inverse the dataset size \footnote{Dataset size is measured as a number of privacy units (Section \ref{sec-privacy-unit-general}) in the dataset}, for example $1/|\mathbb{D}|^{1.1}$ (\cite{Ponomareva_2023} Section 5.3.2), 
so that the violation of pure $\epsilon$-DP is rare. For $\epsilon$, values below ten are considered a reasonable choice, while those under one offer a strong privacy guarantee. We provide a discussion on what values of $(\epsilon, \delta)$ are appropriate for DP synthetic data generation in practice in Section \ref{privacy-guarantees-decisions}. 

While many other DP definitions exist (e.g. RDP \cite{Mironov_2017}, zCDP \cite{bun2016concentrateddifferentialprivacysimplifications}, PLD \cite{doroshenko2022connectdotstighterdiscrete}), $(\epsilon,\delta)$-DP remains the popular choice for its flexibility and practical implication and this is the type of guarantee that is most often reported in context of DP synthetic data. We use the $(\epsilon,\delta)$-DP definition throughout this work unless otherwise specified. 

DP provides a \textbf{worst-case guarantee}, protecting an individual's data against an adversary even if an attacker has access to all other data in the dataset, possesses large computational resources, and has access to external datasets to link information. The threat model can be hypothetical and in practice the attacker often only has access to a single released outcome from $\mathcal{M}(\mathbb{D})$ - in context of this survey, this is DP synthetic data.  DP guarantee is also data-independent, as it makes no assumptions about the underlying distribution of the dataset $\mathbb{D}$, or the one record in the adjacency definition. While the strong, future-proof guarantee simplifies the analysis and comparison of different mechanisms, the pessimistic assumptions can make achieving a favorable privacy-utility trade-off challenging. As a result, applying DP in practicee, including for creating DP synthetic data, requires a large amount of computation. 

\subsection{Neighboring Datasets and the Privacy Unit} \label{sec:priv_unit}
The guarantee of differential privacy hinges on the definition of ``neighboring datasets’’ (the datasets are the same and only differ in one record). This relationship is determined by two factors: the privacy unit, which specifies what constitutes one record, and adjacency, which describes how two neighboring datasets can differ.

The privacy unit is the fundamental element of data being protected and defines what constitutes ``a single record’’ within the dataset. The choice of a privacy unit is a critical modeling decision that shapes the privacy guarantee. For example, under \emph{instance/example-level DP}, the neighboring dataset differs on a single data entry. This is a straightforward approach, but it can be insufficient if one person contributes multiple records. In \emph{user-level privacy}, all data points belonging to a single user are grouped and treated as a single privacy unit. This ensures the protection of an individual's entire contribution, offering a stronger and more intuitive guarantee.

The adjacency definition determines how two datasets differ in one record: whether it is an addition or removal of one record (\textit{add-or-remove adjacency}), \textit{replace-one} (e.g. datasets are the same but one record in one dataset is replaced with another record to produce the second dataset) or \textit{zero-out} (one record in one dataset is replaced with a dummy record representing a missing record) \cite{Ponomareva_2023}. The \textit{replace-one} definition is considered a stronger adjacency definition (''twice as strong'' as zero-out or add-or-remove), essentially representing a combination of an addition and a removal of one record.

\begin{specialistbox}{Pay attention to adjacency definition}
  When reporting privacy guarantees for DP artifacts including DP synthetic data, it is important to take note of the adjacency definition used. While \textit{zero-out} and \textit{add-or-remove} adjacency definitions have comparable semantics for a fixed $\epsilon$, the guarantees for \textit{replace-one} are considered approximately twice as strong \cite{Ponomareva_2023}. Nevertheless, care should be taken when attempting to compare performance or guarantees under different adjacency definitions. 
\end{specialistbox} 

There are many important considerations in properly choosing the DP setting, and guarantees reported under different privacy units and adjacency definitions are generally not comparable. While it is possible to convert DP guarantees from example-level to user-level via group privacy theorems \cite{DworkRothBook:2014,charles2024finetuninglargelanguagemodels}, the resulting bound can be loose and the reverse conversion is not obvious. For a detailed discussion on selecting and reporting DP settings, we refer the reader to \cite{Ponomareva_2023}. In the following sections, we will assume \textit{add/remove adjacency}, and discuss privacy unit setting as a key consideration for various synthetic data modalities.

\subsubsection{Privacy Units for DP Synthetic Data}\label{sec-privacy-unit-general} 
In subsequent section for each modality (Image, Tabular, Text, Federated Learning) we will provide a discussion of the most commonly used privacy unit and the ramifications of this choice. Below we will familiarize the reader with types of privacy units that will be encountered throughout this paper.

\begin{specialistbox}{Choice of privacy unit}
One of the most important decisions for any application of DP, including DP synthetic data generation, is the appropriate choice of privacy unit. The privacy unit determines the scope of the privacy guarantee (e.g., protecting examples versus protecting collections of examples owned by a user versus collections of examples that belong to a group of users who are part of an organization etc.). The choice of privacy unit also determines how the original data must be prepared for creating its DP synthetic version, influencing for example how training examples are formed and how data is  sampled in DP fine-tunining. 
\end{specialistbox}

\citet{Ponomareva_2023} outlined several choices for the unit of protection in context of DP. Most of the time, in context of DP synthetic data generation, the choice will be among the following four:

\begin{compactenum} 

    \item \textit{Example-level privacy unit} - in this setting, neighboring datasets from DP definition are defined as two datasets that differ in one record/instance/example only \cite{Ponomareva_2023}. A definition of a record also offers some leeway based on the end use of the data. For example, for tabular data, each record is a row; however for a collection of the text documents each record can be defined as one full text document or a portion of a text document. For example, if the downstream task is to mark a paragraph as a summarizing paragraph vs a paragraph introducing new details, paragraph-level definition of record might be appropriate. 

    \item \textit{User-level privacy unit}. If the original private data was generated by a large number of users who each may have contributed multiple records, this privacy unit might be more appropriate than example-level. Neighboring datasets in this setting will differ in inclusion/exclusion of \textit{all} the data from a particular user. Consider the case of a user interacting with a banking assistant. In this setting, each user might contribute multiple interactions to the private dataset. Example-level DP protection can be insufficient to prevent the model from memorizing information common to multiple examples from the same user, like the bank account number or preferred bank location. Achieving user-level DP becomes more challenging for domains like chats or emails, where an individual example (chat or thread) might be associated with multiple individuals. For such domains, \emph{multi-attribution user-level privacy} \cite{ganesh2025itsdatatooprivate} is appropriate; achieving this requires more complex contribution-bounding strategies, which we discuss in Section \ref{user-contrib-bounding}.  Figure  \ref{emails-privacy-figure} from \citet{ganesh2025itsdatatooprivate} compares various privacy units using an example of emails dataset.

    \begin{figure}
    \includegraphics[width=\textwidth]{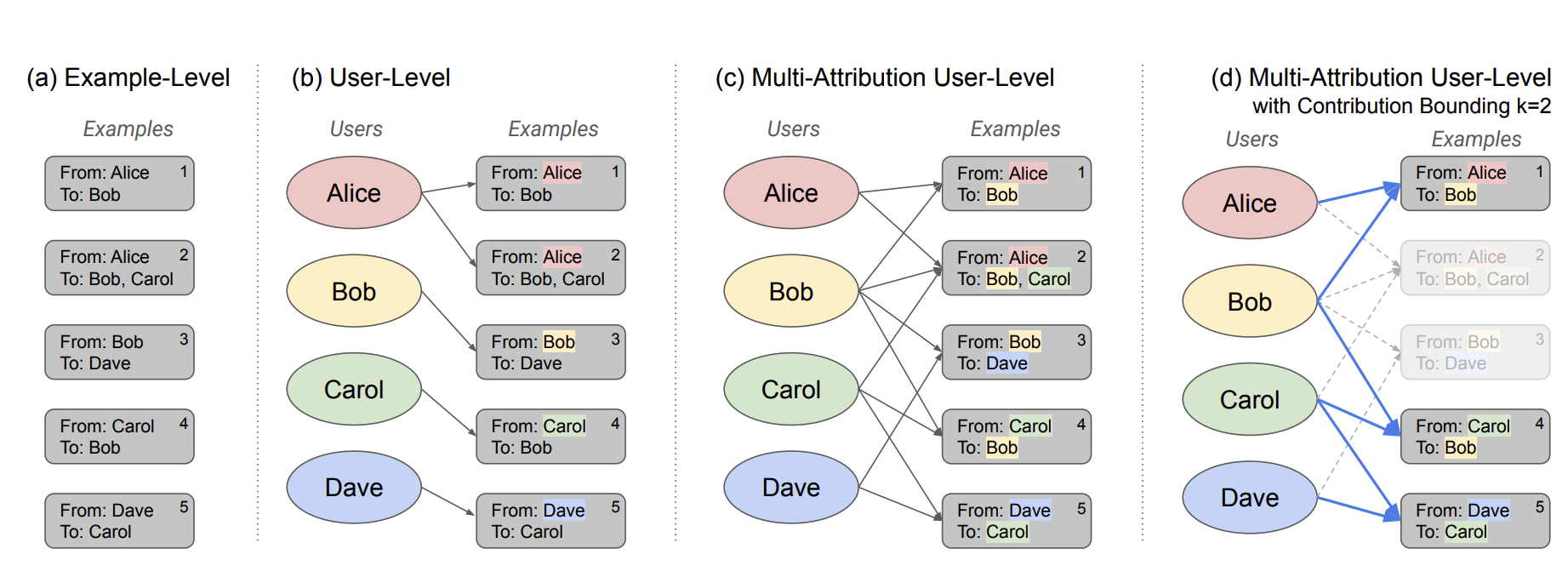}
    \caption{
        Visual representation of choice of privacy unit for a toy private dataset of emails, reproduced with permission from \cite{ganesh2025itsdatatooprivate}. Figure (a) treats the dataset as a flat collection of emails  (\textit{example-level privacy}). In this setting, secrets occurring in multiple emails are not protected. Figure (b) demonstrate the meaning of \textit{user-level privacy unit} where each email is considered to belong only to the sender (\textit{single-attribution user-level} privacy unit). User's secrets that are also present in emails the user received but didn't author might be not protected. Figure (c) attributes emails to all senders and receivers, and (d) bounds the contribution of each user to at most $k=2$ examples from (c) by selecting a subset of emails that will be used for training. 
    \Description{Tradeoffs and meaning of various privacy units.}    
    \label{emails-privacy-figure} }
\end{figure}

    \item \textit{Larger than user-level privacy unit}. Finally, a larger privacy unit can be appropriate when private data is constructed from data from multiple groups, for example various specialized hospitals can contribute their patients data for creation of DP Synthetic data and each hospital wants to be assured that all their patients are protected as a whole group. However, this very-large privacy unit makes achieving DP much more difficult. For example, one typically needs a dataset of at least double digit ($XX$) thousands privacy units in order to utilize state-of-the-art techniques like DP finetuning of LLMs. While a dataset of this number of users in an industry setting might be considered ``small'', a dataset of tens of thousands distinct hospitals or businesses might be very hard to obtain.

    \item \textit{Sub-user-level privacy units}. Units smaller than user level, e.g. user-month and other weaker notions of privacy can be suitable for some applications, particular when complimented by empirical privacy auditing (see Section \ref{sec:empirical-privacy-auditing}) against real-world attacks. However in general we advocate for user-level and larger privacy units for DP synthetic data whenever possible
\end{compactenum} 

\paragraph{Discussion} In order to implement DP algorithms, it must be possible to decompose the private dataset into these privacy units, or more generally, reason about how a change to any one unit might influence the final output. Thus, achieving true person-level privacy would require dataset where each example was perfectly annotated with unique and stable identifiers for each real person associated with that example. This is essentially never the case in practice. For example, user datasets might have metadata like user account ids, email address, but some people might have multiple accounts (weakening the formal privacy guarantee), while others might share an account. Thus, formally we will often only achieve ``user-id-level DP'' but not the true ``person-level DP.'' Fortunately, DP guarantees are well-behaved with respect to small changes in the privacy unit. \textit{Group privacy} \cite{650758} theorems can be utilized to convert guarantees for a “smaller” unit of privacy to the ones for a “larger” unit of privacy.  For example, if one person had two accounts in a system offering account-level DP, their effective DP $\epsilon$ might go from say 1.0 to 2.0 but it would not go to $\infty$. 
A related issue is that even if the user-attribution metadata was perfect, it is possible that some private fact occurs in examples associated with multiple users --- for example, many members of a family might independently discuss one family member's medical condition. Again, as long as such information only occurs in say 5 user's data (if using user-level DP), group privacy still offers protection for such information (though weaker than that offered to information only in one user's examples).

This discussion suggests that one must select the unit-of-privacy, as well as the strength of the protection ($\epsilon$), with pragmatism. In particular, the sensitivity of the source data, the potential harm that would be caused by privacy leakage, and the likelihood and strength of potential attacks should all be considered when evaluating the privacy protections of an end-to-end DP synthetic data deployment. Empirical privacy auditing (Section \ref{sec:empirical-privacy-auditing}) is a valuable tool in calibrating these assessments, as carefully constructed audits can both measure the effectiveness of attacks that are more realistic than the worse case assumptions of DP, and can also be used to measure risks that go beyond the chosen unit-of-privacy (for example, auditing risk when one real person contributed 10 examples, even when contribution bounding and the DP guarantee correspond to one example per username).

It is also worth noting that some methods allow one to simultaneously make DP claims for multiple units of privacy (e.g., a stronger example-level guarantee, and a somewhat weaker user-level guarantee), which can allow different people with different concerns to pick the guarantee that best protects against the threats they care about.

\subsection{Key Properties and Mechanisms of Differential Privacy{\color{red}*}}  \label{sec:dp_property}
DP has several key properties that enable its practical adoption. We highlight a few key properties and discuss their relevance for synthetic data generation. 

\paragraph{Postprocessing.} One powerful property of DP is its invariance to post-processing. Any computation performed on the output of a DP algorithm (without re-accessing the original private data) incurs no additional privacy cost. This property is a natural result of the worst-case consideration of DP definition, and is exceptionally useful for synthetic data generation. Once a synthetic dataset is created using a DP algorithm, it can be analyzed, shared, and used to train any number of models without weakening the privacy guarantee. Moreover, as we will discuss in detail later, many modern synthetic data generation techniques focus on training a generative model with DP. Because of post-processing, one can safely sample from a DP generator to create an infinite number of synthetic data tailored for intended usage, and the resulting DP synthetic dataset will satisfy the same DP guarantee as the generator.

\paragraph{Composition{\color{red}*}} properties provide a framework for analyzing DP guarantees when multiple mechanisms are applied together.  Common examples of composition include the following
\begin{compactenum}
\item \textit{Parallel Composition:} If a dataset is partitioned into disjoint subsets and DP mechanisms are applied independently to each subset, the overall privacy cost is determined by the worst privacy cost of any single computation. Formally, if $(\epsilon_i, \delta_i)$-DP are satisfied on $k$ disjoint data partitions, the combined mechanism satisfies $(\max_{i} \epsilon_i, \max_{i}  \delta_i)$-DP.
\item \textit{Sequential Composition:} When a sequence of mechanisms is applied to the same dataset, their privacy costs accumulate. For the combined mechanism of a sequence of $(\epsilon_i, \delta_i)$-DP mechanisms, \emph{basic composition} provides a loose bound from simple summation of the privacy parameters $(\sum_{i} \epsilon_i, \sum_{i}  \delta_i)$. \emph{Advanced composition} offers a tighter bound, which is essential for iterative algorithms. For example, repeating the $(\epsilon, \delta)$-DP mechanism for $k$ times results in a mechanism satisfies $(\sqrt{2k \ln(1/\delta’)} \epsilon + k\epsilon(e^\epsilon-1), k\delta + \delta’)$-DP for all $\delta’ \geq 0$ using Theorem 3.20 of \cite{DworkRothBook:2014}. 
\end{compactenum}

These composition principles are critical for machine learning and synthetic data generation, which often use complex algorithms that access the data multiple times and in different partitions. They are often used with \textbf{privacy amplification by sampling} \cite{KasiviswanathanLNRS08}, another powerful tool to make DP analyses more efficient: a key technique where applying a mechanism to a random data sample, rather than the whole dataset, significantly reduces the privacy cost. Together, these tools allow for a principled analysis (called \textit{accounting}) of the rigorous DP guarantee for the complete learning process across all steps, which is fundamental to algorithmic design \cite{dpsgd_2016}. Accounting is often done using alternative definitions of DP (e.g. RDP \cite{Mironov_2017}, zCDP \cite{bun2016concentrateddifferentialprivacysimplifications}, PLD \cite{doroshenko2022connectdotstighterdiscrete}, and $(\epsilon, \delta)$-DP guarantees then are derived from those definitions.

\paragraph{Gaussian Mechanism} \label{sec:gauss_mech}
The Gaussian mechanism is a foundational building block for many differentially private algorithms in machine learning. We discuss it here to explain core DP principles. We first focus on applying Gaussian mechanism once, and then discuss the usage in iterative algorithms like DP-SGD \cite{dpsgd_2016} and DP-FTRL \cite{kairouz21practical}. For a broader discussion of various DP mechanisms, we refer the reader to \cite{Ponomareva_2023,DworkRothBook:2014}.

The Gaussian mechanism achieves DP by adding noise from a Gaussian distribution to the output of a real-valued function $f$. The amount of noise added depends on the function's $\ell_2$-sensitivity, which bounds the maximum possible impact of any single privacy unit on the function's output.
\begin{definition}[$\ell_2$-sensitivity] \label{def:sens}
For a function $f:\mathcal{D} \rightarrow \mathbb{R}^d$, the $L_2$-sensitivity is:
$$\Delta_2(f) = \max_{\mathbb{D}, \mathbb{D}'} \|f(\mathbb{D}) - f(\mathbb{D}')\|_2$$
where the maximum is taken over all neighboring datasets $\mathbb{D}$ and $\mathbb{D}’$.
\end{definition}
A single application of Gaussian mechanism is defined as:
$$\mathcal{M}(\mathbb{D}) = f(\mathbb{D}) + \mathcal{N}(0, \sigma^2 \cdot \Delta_2(f)^2 \cdot \mathbf{I})$$
Here, noise is drawn from a zero-mean Gaussian distribution with a standard deviation of $\sigma \Delta_2(f)$, where $\sigma$ is the \emph{noise multiplier}. A larger $\sigma$ provides a stronger privacy guarantee of smaller $\epsilon$, and we can use $\sigma =  \sqrt{2 \ln(1.25/\delta)} /\epsilon $ to achieve $(\epsilon, \delta)$-DP when $\epsilon < 1$ , refer to Theorem A.1 of \cite{DworkRothBook:2014}.

\subsection{DP Synthetic Data and Methods for Creating It}\label{sec-ml-lifecycle}

DP synthetic data, from the point of view of the consumer of the data, can be viewed as a privatized version of the original dataset (this corresponds to introducing DP at the input level/dp-fying the data \cite{Ponomareva_2023} Section 3.2). Once the data is dp-fied, anything done with this data is also DP due to postprocessing (Section \ref{sec:dp_property}) - any downstream ML model trained on DP synthetic data is DP w.r.t. the original training data, and any prediction of such model is also DP. 

An alternative DP synthetic data is training each downstream ML model with DP via algorithms like DP-SGD (\textit{DP-Training}) which we discuss next. This was previously prevalent in academic literature on DP ML. In this setting, by postprocessing property, once an ML model is dp-fied, any output of such ML model is DP as well (so both the model and model's predictions are also DP w.r.t the original training data). Another alternative to DP synthetic data is introducing DP at the level of prediction of each ML model (\textit{DP prediction or DP inference}) of an ML is the ``weakest'' form of DP protection - in this setting only predictions of an ML model are DP and can be shared. However in theory this type of DP entry point might require much less noise to realize, which can potentially mean much better utility of the predictions \cite{Ponomareva_2023}.

\textbf{DP synthetic data offers the strongest protection among the alternatives we discussed}. The flip side is that creating a useful DP synthetic dataset is a task that in general is harder to achieve than for example DP-Training a particular ML model on private data with DP, as a synthetic dataset must serve multiple, even unforeseen, purposes. The future-proof, worst-case nature of the DP guarantee makes the privacy-utility-computation trade-off difficult to navigate.  However, a major advantage of DP synthetic data is its utility as a safe and shareable asset, a direct result of DP's worst-case guarantee and invariance to post-processing, as discussed in Section \ref{sec:dp_property}. Synthetic data without explicit privacy protection are vulnerable to attacks; for example, LLMs can memorize and reconstruct training data 
\cite{carlini2021extracting,nasr2025scalable}.

While DP can be applied to directly train downstream models to be deployed, or during inference time for a specific task, DP synthetic dataset is more versatile. Once a DP synthetic dataset is generated, it can be distributed, analyzed, and used for unlimited analysis and modeling tasks without incurring additional privacy costs. DP synthetic data additionally can allow one to reduce computation cost by creating DP synthetic dataset once and reusing it (without DP) for training various ML models, instead of training each model with DP (a computationally expensive process).

Broadly speaking, there are two groups of methods for creating DP synthetic data: \textbf{those that involve modification of the training process of models that will be used to generate DP synthetic data}, to ensure differential privacy (DP-Training, Section \ref{sec-dpsgd}) and \textbf{those 
that avoid training ML models on private data} and instead rely on inference-only access to trained ML models. This categorization will be present when we describe methods for modalities like Images (Section 
\ref{sec:image}) and Text (Section \ref{sec:text}). 
Tabular data, which is unique modality with well established history of theory backed research into dp synthesis method, is the only modality that enjoys a more diverse suite of methods that we explore in Section \ref{sec:tabular}.

\begin{figure}
   \includegraphics[width=\textwidth]{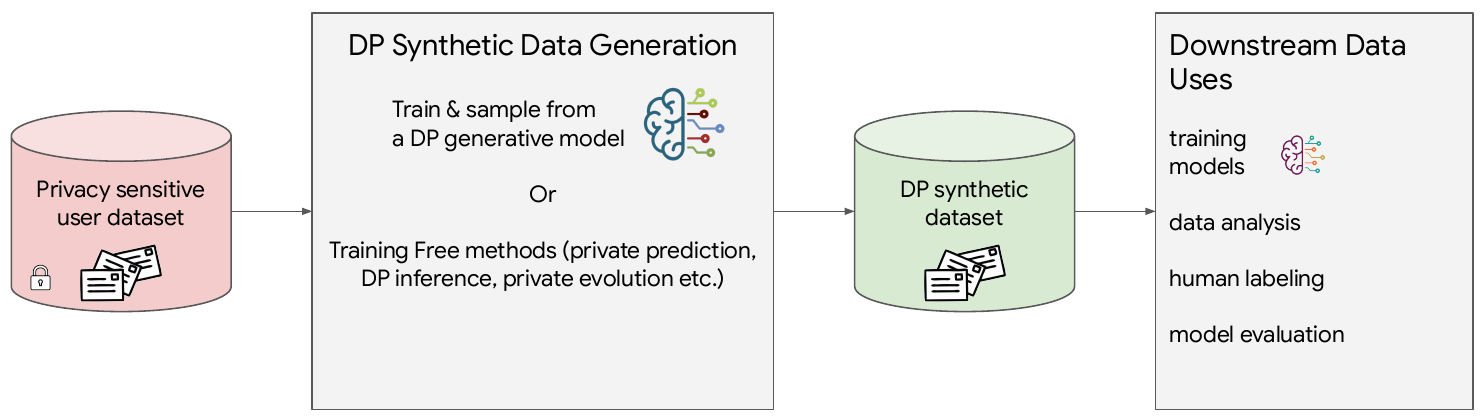}
   \caption{DP synthetic data can be used as a drop-in replacement for the original private data and can be safely used for training downstream models, data analysis, human labeling and other applications. DP synthetic data is created either with the help of DP-Training of ML models or Training free methods. }
   \Description{Privacy sensitive user dataset can be used for DP synthetic data generation. Such dp-fied dataset can then serve as a drop-in replacement for the original data, including for training downstream models, data analysis, human labelling, model evaluation etc.}
\end{figure}

\subsection{DP-Training} \label{sec-dpsgd} 
\textit{DP-Training} is a workhorse algorithm for obtaining differentially private ML models that, as we will see in subsequent sections, can be successfully employed to create DP synthetic data.

\textit{DP-Training\footnote{Throughout this work, we will use terms like \textit{DP-Training} and \textit{DP-finetuning}. Both terms imply an application of DP algorithm like DP-SGD. DP-training is most commonly used to describe training a model from scratch with DP, whereas DP-finetuning often is used to describe finetuning a publicly pretrained model like an LLM with DP. Some works use these terms interchangeably.} } commonly refers to a DP variant of SGD-like algorithms (DP-SGD \cite{dpsgd_2016} and DP-FTRL~\cite{kairouz21practical}) that iteratively apply Gaussian mechanism to achieve DP protection of ML model gradients. DP-fied gradients are then used to do gradient step update, resulting in all the gradient sequences, and consequently, all the checkpoints of a trained ML model to be DP w.r.t. the original training data.

Algorithm \ref{algo:dpsgd} demonstrates modifications that are introduced to SGD algorithm to ensure DP. The changes include per example gradient clipping to limit $\ell_2$ sensitivity and an addition of the noise via Gaussian mechanism to aggregated gradients. Similar modifications can be used to obtain DP versions of algorithms like Adafactor, Adam etc. Clipping the gradients is computationally expensive step and is the slowest part of DP-SGD, since most modern auto-differentiation based frameworks do not provide easy access to per example gradients. Additionally, in order to obtain good utility, DP-training requires significantly larger batch sizes in order to tolerate introduced noise, and potentially more iterations to converge \cite{Ponomareva_2023}. Finally, other hyperparameters like learning rate and clipping norm require extensive hyperparameter tuning to achieve good utility. All of the above make DP-Training much more computationally expensive than training an ML model without DP. 

In order to provide DP guarantees of DP-SGD algorithm, tracking the cumulative privacy cost across multiple iterations of DP-SGD needs to be performed (this task is often referred to as \emph{privacy accounting}). Combining Gaussian mechanism $\epsilon = \sqrt{2 \ln(1.25/\delta)} /\sigma $ and advanced composition $\epsilon’ = \sqrt{2T \ln(1/\delta’)} \epsilon + T\epsilon(e^\epsilon-1)$ from Section \ref{sec:dp_property} gives us a rough understanding of the relationship between DP guarantee and the noise multiplier for full-batch DP-SGD. The number of iterations $T$ also affects the DP guarantee, and data processing pattern is another important design choice when applying advanced techniques of privacy amplification by sampling \cite{dpsgd_2016,balle2018privacy,Chua2024} or correlated noise mechanisms \cite{kairouz21practical,mcmahan2024hassle}.

While basic and advanced compositions (Section \ref{sec:dp_property}) provide a theoretical understanding, they often yield loose bounds in practice. Modern privacy accounting is typically handled by specialized libraries \cite{pldlib} that implement more sophisticated methods to achieve tighter bounds. These methods evolved from moments accountant and Rényi Differential Privacy \cite{dpsgd_2016}, to DP variants (Gaussian DP~\cite{dong2022gaussian} and zero-concentrated DP \cite{bun2016concentrated}) more suitable to Gaussian mechanism and can be cleanly converted to $(\epsilon, \delta)$-DP.  The recent development of Privacy Loss Distribution (PLD) accounting tracks a more detailed representation of the privacy loss at each step, often yielding the tightest known $(\epsilon, \delta)$ bounds for a given process \cite{pldlib}. These advanced accounting methods are essential for designing high-utility DP mechanisms, including training synthetic data generators, within a reasonable privacy budget. 

\begin{algorithm}[htbp]
\caption{The DP-SGD algorithm, adapted from \cite{Ponomareva_2023}, based on \cite{dpsgd_2016}.}\label{algo:dpsgd}
\begin{algorithmic}
\Require Training data, consisting of features  $X := \{x_1, x_2, ..., x_N\}$ and labels $Y := \{y_1, y_2, ..., y_N\}$. \\
$h(x; \theta)$ is the model applied to an input $x$ and parameterized by $\theta$. \\
$L(y, y')$ is the loss function for label $y$ and prediction $y'$.\\
SGD hyperparameters: $\eta$ learning rate, $T$ number of iterations, $B$ batch size. \\
DP hyperparameters: $C$ clipping threshold, $\sigma$ noise multiplier, $\delta$ (for privacy accounting).
\Ensure $\theta_T$ final model parameters
\State $\theta_0 \gets $ initialization
\For{$t \gets 1$ to $T$}                     
     \State Randomly sample a batch $B_t$ with sampling probability $B/N$ for each data point. 
     \State Data are sampled with replacement for each batch.
     
     \For{$i \in B_t$}  
        \State$g_t(x_i) \gets \nabla_{\theta_t}{L(y_i, h(x_i; \theta_t))}$
        \Comment{Compute per-example gradient wrt the weights }
        
         \State  \HiLi $g_t(x_i) \gets g_t(x_i)/\max(1, \frac{||g_t(x_i)||_2}{C})$
        \Comment Clip the per-example gradient
        
     \EndFor
     
     \State  \HiLiLong $\bar g_t  \gets \frac{1}{B}(\sum_i{g_t(x_i)}+\mathcal{N}(0,\,\sigma^{2} C^2 \mathbf{I} ) )$
     \Comment Add noise

    \State $\theta_{t+1} \gets \theta_t - \eta \bar g_t $
    \Comment Gradient descent step
     
\EndFor
\end{algorithmic}
\end{algorithm}

\section{DP Synthetic Tabular Data}\label{sec:tabular} 
\textit{Tabular} data (data that can be represented as tables of values organized into rows and columns), has long been the most prominent way to organize data in various domains like the sciences, healthcare, business, finance, retail and others. Unsurprisingly, initial attempts at generating DP synthetic data focused on tabular data, which has developed into one of the richest lines of work in the DP synthesis literature.  

The shared goal among the various works in this space is to generate a synthetic data that shares the same ``statistical characteristics'' as the real data, such that it can be a drop-in replacement for the real-data in downstream pipelines with minimal utility degradation.  Exactly what form these ``statistical characteristics'' take varies across approaches, typical evaluations focus on distributional similarity of low-order marginals or train-on-synthetic, test-on-real classification tasks.  The space of mechanisms can be partitionined into two groups: \emph{workload-adaptive} methods, where the mechanism tailors the synthetic data towards a specific (finite) set of downstream tasks, and \emph{workload-agnostic} methods, where the mechanism aims for general distributional similarity, without targeting accuracy w.r.t. any specific set of tasks.

For example, some algorithms focus on ensuring the synthetic data distribution is ``close'' to the real one in a principled statistical distance, like Wasserstein distance. One could argue that this \textit{implicitly} defines the workload as a broad (potentially infinite) class of smooth, Lipschitz queries. If the two distributions are close in Wasserstein distance, then we also get a bound on the error of \textit{any} of these smooth queries, thus giving general assurances against unanticipated future analyses of this kind. However, this does not \textit{adapt} to the workload in that this broad guarantee surely covers a greater class of queries than the user could reasonably pre-specify. It is also worth mentioning that this same rationale applies more broadly to \textit{\textbf{any}} \textit{integral probabilility metric}, e.g. Kolmorogov distance and total variation distance, for a suitable class of (infinitely-many) linear queries  \citet{Muller:1997}.

\begin{specialistbox}{Marginal-based algorithms}
Tabular mechanisms typically operate by taking noisy measurements and creating synthetic data that conforms to them.  These measurements may or may not be related to a user's potential workload. When an algorithm measures marginal queries, they are often referred to as \textbf{marginal-based algorithms}. 
   
\end{specialistbox}

Mechanisms can also be classified as \textit{theoretically oriented algorithms}, which aim to provide rigorous bounds on the utility of the algorithm across all possible datasets, and \textit{empirically oriented algorithms}, which aim to provide good utility across a representative set of benchmark datasets and measurements. Empirically-oriented algorithms often outperform theoretically-oriented algorithms on real world datasets, both in terms of utility and scalability, at the cost of limited (if any) formal guarantees on the quality of the generated synthetic data.  Indeed, it is not even possible to run many theoretical algorithms on real-world benchmark datasets due to challenges of scale. While the algorithmic ideas underlying both classes of approaches are largely similar, small implementation details can make \textit{big} differences in practice.

\begin{remark}[Workload vs. Measurements]
   In the context of DP tabular data synthesis, it is crucial to distinguish between two concepts. The \textbf{workload} refers to the set of statistical queries a synthetic dataset is intended to be accurate on for downstream analysis. The \textbf{measurements} are the set of queries an algorithm actually executes on the private data to guide the generation process. All algorithms (apart from some of the end-to-end models from Section \ref{tabular-generative}) take measurements; for instance, even methods that provide distributional distance guarantees can be seen as implicitly targeting certain workloads (e.g. the Chebyshev moments of a distribution \cite{musco2024sharper}).
\end{remark}

\begin{remark}[Workload-Adaptive vs. Workload-Agnostic]
The philosophies behind workload-adaptive and workload-agnostic approaches are fundamentally different. The workload-adaptive approach tailors the synthetic data to a known set of analytical needs. By leveraging this explicit guidance, these mechanisms can strategically allocate the privacy budget, significantly enhancing accuracy for the tasks that matter to the downstream analyst. In contrast, the workload-agnostic philosophy aims for a more general, ``one-size-fits-all'' solution by attempting to learn the most salient features of the joint probability distribution automatically. However, this ambitious goal can lead to higher-than-expected error on a specific downstream workload that was not implicitly prioritized by the algorithm's \textit{measurement strategy}. While the workload-adaptive approach risks poor performance on ``off-workload'' queries, this can be mitigated by providing a broad workload (like \textit{all} 3-way marginals).

The core design choice in existing algorithms for DP tabular synthetic data is not  \textit{whether} to take measurements, but  \textit{how the measurement strategy is selected in relation to a potential workload}.
\end{remark}

\paragraph{Roadmap} As a roadmap for this chapter, we will explore methods that fit into both of the aforementioned categorizations. We must first, however, discuss the fundamental notion of the \textbf{privacy unit} for tabular data in Section \ref{sec-tabular-unit}, as this underpins all subsequent algorithms. Additionally, in Section \ref{sec-tabular-foundations}, we describe the problem of \textit{query release} that lays the foundation of many methods for DP tabular data synthesis, as well as the \textit{histogram} representation of tabular data that many methods use either implicitly or explicitly. 

In Section~\ref{sec-workfload-based}, we will delve into the family of \textbf{select-measure-estimate} algorithms, which represent the most prominent and successful class of practical, empirically-oriented  methods. We subsequently cover theoretically-oriented algorithms in Section~\ref{sec:theoretical_hist}, detailing the foundational algorithms that come with strong theoretical guarantees: those targeting additive error and those targeting distributional distance.
We defer details on data preparation and discretizations, which are often necessary for practical implementations, to Section~\ref{tabular-feature-handling}.

Finally, in Section~\ref{tabular-generative}, we return to empirically motivated algorithms and introduce a newer strain of work based on \textbf{end-to-end generative models} that take advantage of modern advances in deep learning, another important class of potentially practical algorithms that generally offer a different structure than traditional algorithms from the select-measure-estimate paradigm. We give a brief comparison of these different families of methods in Section~\ref{tabular-comparison}, and finally, in Section~\ref{tabular-metrics}, discuss methods for evaluating the quality of synthetic tabular data, which is itself an active research area.

\subsection{Privacy Unit for Tabular Data}\label{sec-tabular-unit}
For tabular DP synthetic data, the privacy unit is most commonly assumed to be on an \textit{example-level (row-level in context of tables)} i.e. where each user contributes at most 1 row to the dataset. As such, DP synthetic tabular data guarantees that the resulting synthetic dataset would not be significantly different whether a particular user contributed their row to the original tabular dataset or not. Neighboring datasets are then two tabular datasets that differ in at most $1$ row. It is worth noting that supporting user-level privacy with each user contributing at most $x$ rows could be handled by creating one aggregate row for each user using $x$ user's datapoints, or by using a group privacy property and running the algorithms we describe below with $\epsilon/x$ and $\delta/x$ in order to achieve $x$ group level privacy of ($\epsilon, \delta$) \cite{DworkRothBook:2014}. Ensuring that each user has at most $x$ rows can be done with appropriate user contribution bounding (Section \ref{user-contrib-bounding}). Developing better user-level algorithms that directly handle varying number of rows per each user remains an open research direction.

We note here an alternative, distinct privacy unit setup, which to the best of our knowledge remains relatively unexplored in the literature, where each user is assumed sole authorship of an entire table i.e. the privacy unit is \textit{itself} a full tabular dataset. In this setup, each user can contribute one or more tabular dataset(s), over potentially varying domains (e.g. different columns headers). Thus, the neighboring definition is two tabular collections differing in at most 1 (or $x$) tabular datasets. The goal is then still to create synthetic data that satisfies a DP guarantee, but in this case the output is a \textit{collection} of tabular datasets (as opposed to a \textit{single} tabular dataset). While such a use case might be boiled down to the previous setup (e.g., if each tabular dataset has the same columns and their domains), in general this problem differs fundamentally and is significantly harder than the single table input/output model. All methods discussed in the following sections assume the former \textit{row} level privacy unit definition, where we have a single input table on a fixed domain and output a synthetic version of it over the same domain with a DP guarantee, though we believe exploring the \textit{table} level privacy unit is a promising avenue for future work.

\subsection{Foundations of DP Tabular Synthetic Data}\label{sec-tabular-foundations}

DP synthetic tabular data was originally explored in context of \textit{query release} -- the task of creating useful synthetic data that gives accurate answers to a number of predefined statistical queries (\textbf{workload)} performed on this data, e.g. sums and counts. While one can answer such statistical queries by adding appropriate noise to each query result, DP synthetic tabular data allows one to both answer the predefined queries and perhaps additional queries that will be defined later (there are generally no guarantees about the utility of answers to such queries). 

\subsubsection{Foundation of Workload-Adaptive Algorithms: The Query Release Problem{\color{red}*}}\label{sec:query_release_problem}
More formally, the synthetic tabular data \textit{query release} problem can be formulated as follows (refer to Table \ref{tabular-notations} for the full list of notations throughout this chapter). Let $\mathcal{X}$ be the data universe from which rows of the tabular dataset are drawn (for now we assume it to be a finite set); for example, if our data contains $d$ binary attributes, then ${\cal{X}} = \{0, 1\}^d$. Let $\mathcal{F}$ be a finite set of functions that we will refer to as the \textit{counting} {\em queries} or a {\em (query) workload}, where for each $f \in \mathcal{F}$, $f : \mathcal{X} \mapsto \{0, 1\}$ and $f$ is a per-row predicate, and we desire an estimate of the sum $f(D)=\sum_{i=1}^{n}f(x_i)$ (the count of input rows that satisfy the predicate $f$). Note that the choice of (binary) counting queries can be easily extended to continuously-valued queries $f:{\cal{X}} \mapsto [0, 1]$, with the same error rates. This follows from the fact that these queries enjoy the same sensitivity bounds as their binary counterparts.

Recall that, as we discussed at the beginning of this section, the relationship between a DP synthetic data algorithm's internal \textit{measurement strategy} and a user-defined evaluation \textit{workload} is the core of our taxonomy, a distinction that has its roots in the query release problem. 
As we noted previously, a \textit{workload-adaptive} algorithm is one whose measurement strategy can be guided by the user's workload to optimize accuracy (e.g., an algorithm where we could explicitly align workload queries with measurement queries). A \textit{workload-agnostic} algorithm employs a measurement strategy designed to preserve general distributional properties, irrespective of any specific downstream workload (e.g., an algorithm that does not incorporate the workload queries into noisy measurements explicitly). 

The goal of the workload-adaptive strategies for \textit{query release} problem is to design an algorithm ${\bf{A}}: \mathcal{X}^n \mapsto \mathcal{X}^n$ (here, $n$ represents the number of rows in the original dataset)\footnote{We opt here for algorithms whose input and output have the same number of rows. This is made for simplicity and can be generalized. Furthermore, it may be the case (e.g., for add/remove neighboring relations) that the number of rows is itself private; in this case, one can privatize the number of rows and produce synthetic data with that number of rows. Finally,
for datasets with varying cardinality one needs to approximately preserve averages, rather than counts, as done in Equation~\ref{eq:accuracy_synthetic_data}.} that is $(\epsilon,\delta)$-DP (w.r.t. the privacy unit) and that is $\alpha$-accurate with respect to ${\cal F}$ for some $\alpha \geq 0$ and some norm function $|| \cdot ||$. Namely, for any input tabular dataset $D \in \mathcal{X}^n$, and its DP synthetic counterpart $\widehat{D}={\bf{A}}(D)$, it must hold that its error over the finite set of queries is bounded in \textbf{expectation} by $\alpha$. With some abuse of notation, letting  ${\cal F}(D)=(f(D))_{f\in {\cal F}}$ be the vector of workload query answers, this utility measure can be expressed as:
\begin{equation} \label{eq:accuracy_synthetic_data}
\textit{\em Err}_{p}({\bf{A}}, D) := \mathbb{E}_{\widehat{D} \sim {\bf{A}}(D)}\Big[ \Big| \Big| {\cal F}(\widehat D)- {\cal F}(D) \Big|\Big|_p\Big] \leq \alpha~, 
\end{equation}
Taking $p = \infty$ is a common choice in the theory literature, while $p=1$ or $p=2$ is common in the empirical literature. All of these options have their merits and the right choice ultimately depends on the downstream use case. We note that this objective is an \textit{absolute} error rather than a relative one -- it is typically easier to analyze and optimize mechanisms for this setting. In the rest of the section, we will drop $p$ from $|| \cdot ||$ with the understanding that it can in principle be anything, unless it is important to explicitly call out.

\renewcommand{\arraystretch}{1.2}
\begin{longtable}[htb]{p{0.1\linewidth} p{0.8\linewidth} }
\caption{Table of notations for tabular data discussion}\label{tabular-notations} \\
 \toprule
 \textbf{Symbol} & \textbf{Description}  \\
\midrule

$\mathcal{X}$ &  The data universe from which tabular dataset rows are drawn; for example, if our data contains $d$ binary features, then ${\cal X}=\{0,1\}^d$. \\ 

$n$ & Number of rows in the original tabular dataset and number of rows in the synthetic dataset. \\ 
$D$ & Original tabular dataset, $D \in \mathcal{X}^n$ \\ 
$\widehat D$ & DP synthetic tabular dataset,  $\widehat D \in \mathcal{X}^n$ \\ 

$f$ & Row-level predicate  $f:\mathcal{X}\mapsto\{0,1\}$  \\ 
$f(D)$ & $f(D)=\sum_{i=1}^{n}f(x_i)$ - number of rows in the dataset satisfying predicate $f$ \\ 

$\mathcal{F}$ & Finite set of queries $f$ that synthetic tabular data must perform well on \\ 
${\bf A}$ &  $(\epsilon,\delta)$-DP mechanism ${\bf A} :\mathcal{X}^n\mapsto \mathcal{X}^n$  for generating DP synthetic tabular data from the original tabular dataset. Alternatively, when talking about privatizing histograms, ${\bf A}: \mathcal{X}^n\mapsto  \mathbb{N}_+^{|{\cal X}|}$
\\ 
$\alpha$ & Expected error (expected distance between synthetic and private datasets, in some norm) of the DP synthetic dataset \\ 
$h(D)$ & Histogram representation of a dataset $D$, $h(D)\in \mathbb{N}^{|\mathcal{X}|}$, where $h(D)_x:=|\{i\in[n]: x_i=x\}|$. This histogram essentially represents, for each combination of column values, number of rows that have this combination. For example  for a dataset with columns A and B (and possible values $a_1, a_2$ and $b_1, b_2$ respectively), the histogram would have bins $(a_1, b_1)$, $(a_1, b_2)$, $(a_2, b_1)$ and $(a_2, b_2)$ with counts of rows that fall into such combination of attribute values. \\ 
$\widehat h$ & DP synthetic data histogram \\
\bottomrule
\end{longtable}
\renewcommand{\arraystretch}{1.0}

\subsubsection{Histogram Representation: From Data and Back to Data} \label{sec:histogram_representation}
Most DP synthetic tabular data generation approaches leverage a linear-algebraic reformulation of the problem by representing tabular datasets as \textit{histograms}. Some methods directly instantiate a histogram while others do not; in either case it is important to understand how a dataset could be represented as a histogram and how one can ``go back'' from this histogram representation to form a corresponding tabular synthetic dataset (with a DP guarantee).

Let $D \in \mathcal{X}^n$ be a dataset with $n$ records, and let $h(D) \in \mathbb{N}^{|\mathcal{X}|}$, where the histogram value for a bin $x$ is $h(D)_x := |\{i \in[n]: x_i=x\}|$. This histogram representation counts the frequencies of rows for all possible combinations of column values, and it is useful as it linearizes the search space. Namely, using this histogram representation, query results can be expressed as inner products: $f(D)=\sum_{x \in {\cal{X}}} h_x f(x) = \langle h(D), f\rangle$.

There is a one-to-one correspondence between datasets $D \in {\cal{X}}^n$ and \textit{integer-valued} histograms $h \in \mathbb{N}^{|\cal{X}|}$. However, many approaches require relaxing integrality constraints, i.e., working with real-valued histograms $\hat{h} \in \mathbb{R}_+^{|\cal{X}|}$, which prevents a direct construction of a dataset given a histogram. We highlight some techniques in the following bullet points to bridge this gap.
\begin{compactitem}
    \item In the context of linear programming approaches, there exist a host of {\em rounding techniques}, which approximate a continuously-valued vector $\hat{h} \in \mathbb{R}_+^{|\cal{X}|}$ by an integer-valued vector $\hat{h}' \in \mathbb{N}_+^{|\cal{X}|}$ \cite{Williamson:2011}. The latter can be directly used to construct a dataset. Examples of this technique for DP synthetic tabular data can be found in e.g.~\cite{Barak:2007,Dwork:2009}, as well as in the Top Down Algorithm used in 2020 US Census Disclosure Avoidance System \cite{TopDown}.
    \item A more general approach is, given a continuously-valued histogram $\hat{h} \in \mathbb{R}_+^{|\cal{X}|}$, its normalization $\hat{h}/\|\hat{h}\|_1$ induces a probability distribution over ${\cal X}$. Sampling $n$ datapoints, which we can call a sample dataset $\tilde{D} = (\tilde{x}_1, \ldots, \tilde{x}_n)$ i.i.d.~from this distribution satisfies the following approximation bounds with high probability (this is known as Maurey's Empirical Method or the Approximate Carath\'eodory Theorem \cite{Pisier:1980})
    \[ \Big| \Big| \big(f(\tilde{D}) - \langle \widehat{h},f\rangle \big)_{f\in {\cal F}} \Big|\Big|_{\infty} = O\Big( \sqrt{n\log |{\cal F}|}\Big).\]
    This approach is advised in settings where the number of rows is relatively small compared to the number of bins or queries.
\end{compactitem}
These techniques can be applied in a post-hoc fashion, and their results will be DP as long as the histogram $\hat{h}$ was computed privately, due to the standard DP post-processing argument.

\begin{remark}
Histogram representations of data provide significant mathematical convenience for designing and reasoning about mechanisms, although their linear size in $| \cal X |$ (which results in an exponential size in the number of attributes) can be prohibitive in high-dimensional settings.
\end{remark}

\subsection{Practical Workload-Adaptive Algorithms}\label{sec-workfload-based}
The following methods exemplify a practical, empirically-driven class of \textbf{workload-adaptive} algorithms that prioritize utility for a user-specified workload. Many mechanisms in this class are members of what we refer to as the \textit{select-measure-estimate} paradigm. \footnote{This paradigm has been referred to under different names in the literature, including select-measure-generate \cite{mckenna2021winning}, select-measure-reconstruct \cite{mckenna2018optimizing}, and Adaptive Measurements \cite{liu2021iterative}.}

\subsubsection{Practical Select-Measure-Estimate Algorithms} \label{select-measure-estimate}

\begin{figure}
    \includegraphics[width=\textwidth]{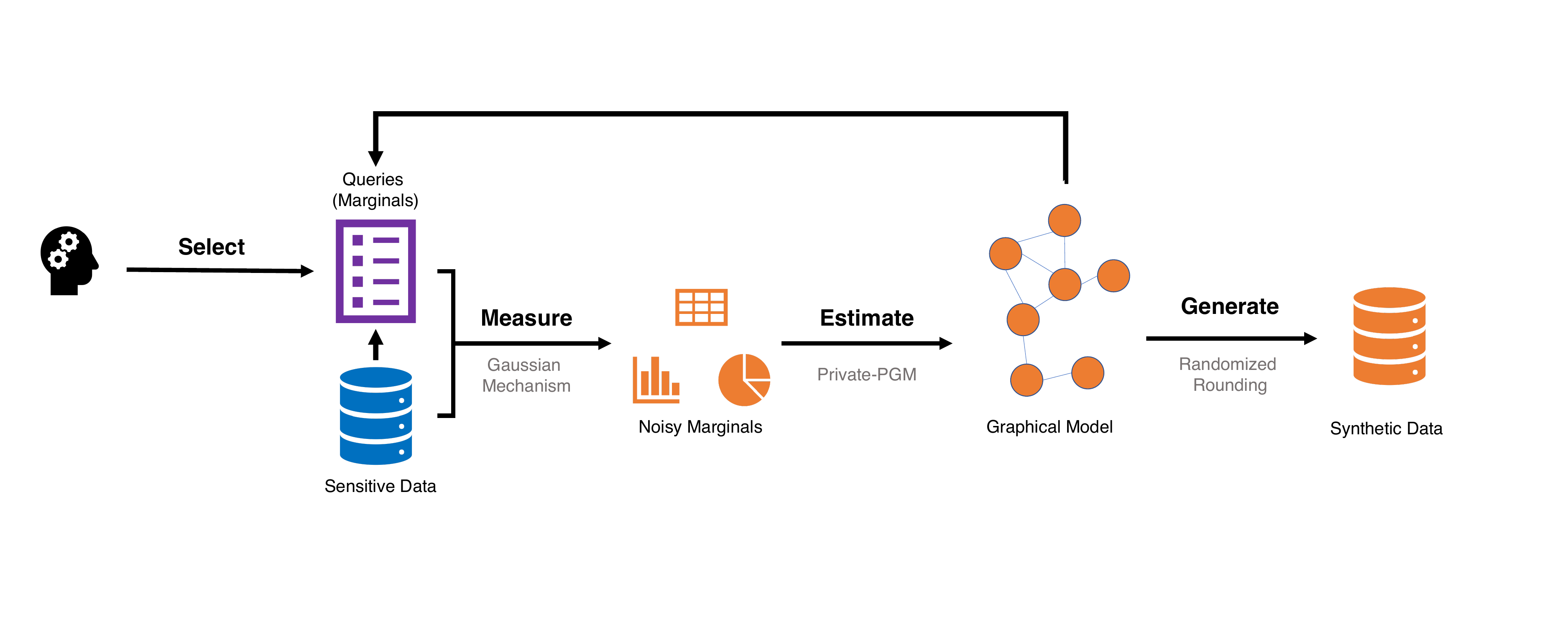}
    \caption{\label{fig:select-measure-reconstruct} The select-measure-estimate paradigm. (1) A mechanism or domain expert \emph{selects} a set of queries, possibly based on the workload and/or data.  (2) The selected queries are measured by a standard noise addition mechanism like the Gaussian mechanism. (3) The data distribution is \emph{estimated} from the noisy measurements. (4) The process optionally repeats from step (1) until the privacy budget is spent. (5) Synthetic data generated from the data distribution via a sampling or randomized rounding procedure.} \label{select-measure-fig}
    \Description{This figure demonstrates the select-measure-estimate paradigm.}
\end{figure}

The general framework is given schematically in Figure~\ref{select-measure-fig}. This framework encompasses a majority of practical algorithms for tabular data generation. Some additional algorithms that can't be easily represented by this paradigm will also be discussed in \Cref{sec:theoretical_hist}.

This framework provides a modular approach to the privacy-preserving synthetic tabular data generation problem. It decomposes the challenge into a sequence of more tractable subproblems, most notably decoupling the task of \textit{query selection} (i.e., deciding what statistical properties of the data to preserve) from \textit{data estimation} (i.e., reconstructing a full data distribution from noisy answers to those queries). This separation of concerns has proven to be a powerful research catalyst, allowing for independent innovation on selection strategies \cite{DBLP:journals/corr/abs-2201-12677,cai2021data,cai2023privlava,fuentes2024joint} and estimation engines \cite{mckenna2019graphical,mckenna2021relaxed,Aydore:2021,liu2021leveraging,liu2021iterative,mullins2024efficient}, which can then be combined in a ``plug-and-play'' manner. This modularity has enabled the development of general-purpose tools that free the mechanism designer to focus on the critical question of what statistics to measure to maximize the utility of the final synthetic data for downstream tasks.

The paradigm is generally understood through a sequence of five foundational steps, which together transform a sensitive dataset into a shareable, synthetic one. The process begins with \textbf{Select}, where a collection of queries is chosen to be measured on the private data. In the context of tabular data, these queries are typically low-dimensional marginals, which are contingency tables that capture the counts of various combinations of attribute values. The choice of these marginals fundamentally determines which correlations and statistical patterns will be preserved in the final output. The second step is \textbf{Measure}, where the selected queries are executed on the sensitive dataset, and carefully calibrated noise is added to the true answers to satisfy $(\epsilon, \delta)$-differential privacy. The Laplace and Gaussian mechanisms are frequently employed for this purpose, with the Gaussian mechanism often being preferable when a larger number of marginals need to be measured \cite{DBLP:journals/corr/abs-2201-12677}.

The third step is \textbf{Estimate}, which is a post-processing phase that takes the noisy, and often inconsistent, measurements from the previous step and seeks to find a coherent probability distribution over the entire data domain that best explains them. This is typically formulated as a maximum likelihood estimation (MLE) problem, where the goal is to find a distribution that minimizes the distance (e.g., $\ell_2$ distance for Gaussian noise) to the noisy measurements. For some advanced strategies, this process is not a one-shot operation but is instead iterative, leading to the optional fourth step, \textbf{Repeat}. In adaptive mechanisms like PMW or MWEM \cite{hardt2012simple} and its more sophisticated and scalable variants \cite{DBLP:journals/corr/abs-2201-12677,cai2021data,Aydore:2021,liu2021iterative}, information from the current estimated distribution is used to guide the selection of the next marginal query to measure, often targeting the one for which the current model exhibits the highest error. This creates a cycle of progressive refinement, focusing the privacy budget where it is most needed. The final step is \textbf{Generate}, where the fully estimated probability distribution can be converted to either a sampling process or randomized rounding process as described in \Cref{sec:histogram_representation}.

\newcommand{\pgm}{\text{PGM}} 
\newcommand{\norm}[1]{\left\|#1\right\|} 
\begin{algorithm}[tb]
    \caption{\label{alg:aim_simplified} AIM (Simplified)} 
    \begin{algorithmic}[1]
        \Require Dataset $D$, workload $W$, privacy parameter $\rho$
        \Ensure Synthetic Dataset $\hat{D}$
        \Statex \textbf{Hyper-Parameters:} rounds $T$, $\alpha=0.9$, MAX-SIZE (in Megabytes)
        \State Initialize $\hat{p}_0$ using an independent set procedure (see \cite{DBLP:journals/corr/abs-2201-12677} for exact algorithm)
        \State $\rho_{\text{used}} = 0$, $t = 0$
        \State Initialize schedulers for $\epsilon_t$ and $\sigma_t$
        \While{$\rho_{\text{used}} < \rho$}
            \State $t \leftarrow t + 1$
            \State $\rho_{\text{used}} \leftarrow \rho_{\text{used}} + \frac{1}{8}\epsilon_t^2 + \frac{1}{2\sigma_t^2}$
            \State $C_t = \{r \in W \mid \text{size}(r_1, \dots, r_t) \le \frac{\rho_{\text{used}}}{\rho} \cdot \text{MAX-SIZE}\}$
            \State \textbf{select} $r_t \in C_t$ using exponential mechanism with $\epsilon_t$ budget:
            \[ q_r(D) = w_r \left( \norm{M_r(D) - M_r(\hat{p}_{t-1})}_1 - \sqrt{2/\pi} \cdot \sigma_t \cdot n_r \right) \]
            \State \textbf{measure} marginal on $r_t$:
            \[\tilde{y}_t = M_{r_t}(D) + \mathcal{N}(0, \sigma_t^2 \mathbb{I})\]
            \State \textbf{estimate} data distribution using \pgm:
            \[ \hat{p}_t = \operatorname*{argmin}_{p \in S} \sum_{i=1}^t \frac{1}{\sigma_i} \norm{ M_{r_i}(p) - \tilde{y}_i}_2^2 \]
            \State \textbf{anneal} parameters $\epsilon_{t+1}, \sigma_{t+1}$ using a scheduling procedure (see \cite{DBLP:journals/corr/abs-2201-12677} for exact algorithm) 
        \EndWhile
        \State \textbf{generate} synthetic data $\hat{D}$ from $\hat{p}_t$ using \pgm
        \State \Return $\hat{D}$
    \end{algorithmic}
\end{algorithm}

A cornerstone of this entire paradigm is the post-processing property of differential privacy. The ``Estimate'' and ``Generate'' steps operate exclusively on the already-privatized noisy measurements and do not access the original sensitive data. Consequently, any computational effort expended during these steps -- no matter how complex the optimization or how sophisticated the generative model -- incurs no additional privacy cost. This crucial property has motivated research focused on building powerful, general-purpose reconstruction algorithms. By creating robust and efficient estimation engines, the paradigm allows mechanism designers to abstract away the complexities of data generation and concentrate on the distinct and equally challenging problem of query selection, which is more important for synthetic data utility \cite{rosenblatt2024epistemic}.

\subsubsection{Selection Methods}

The \textbf{Select} step is the first and arguable most critical component of the select-measure-estimate paradigm \cite{DBLP:journals/corr/abs-2201-12677,rosenblatt2024data}. The principle is simple: we cannot guarantee that information that is not \textit{directly} measured will be \textit{preserved}. Even a perfect, error-free estimation algorithm is powerless to reconstruct statistical patterns if they were not captured by the noisy measurements taken. This reality underscores the importance of developing intelligent strategies for selecting which queries to measure, a challenge that has motivated a significant and evolving line of research. The quality of a selection strategy can be evaluated along multiple dimensions; the work on the \textbf{AIM} mechanism (\Cref{alg:aim_simplified}) proposes a useful taxonomy of four criteria for an ideal strategy: \textit{workload-awareness}, \textit{data-awareness}, \textit{budget-awareness}, and \textit{computation-awareness} \cite{DBLP:journals/corr/abs-2201-12677}.

\textbf{Workload-awareness} refers to a mechanism's ability to prioritize measurements that are most important for a known, downstream analytical task, (user-specified ``workload''). Mechanisms like \textbf{AIM} are highly workload-aware; they employ a quality score that explicitly weights candidate marginals based on their relevance to the user's specified workload queries. Similarly, \textbf{MWEM+PGM} and \textbf{RAP} are designed to optimize for a given set of queries \cite{Aydore:2021}. In contrast, methods like \textbf{MST} and \textbf{PrivBayes} are workload-agnostic \cite{PrivBayes:2017,mckenna2021winning}. They select marginals to measure based on general statistical properties of the data, such as attribute correlations, without considering any specific end use. This makes them suitable for general-purpose data exploration but potentially suboptimal and wasteful of the privacy budget when the analytical goal is known in advance. The trade-off is between specialized, high-utility data for a specific task and more broadly applicable, but potentially less accurate, data for general use.

\textbf{Data-awareness} describes the extent to which a selection strategy adapts to the underlying statistical properties of the data itself. Early strategies were often data-oblivious; a major step forward came with data-driven methods like \textbf{PrivBayes} and \textbf{MST}, which first use a portion of the privacy budget to privately learn a dependency structure from the data -- a Bayesian network or a maximum spanning tree, respectively -- and then use this structure to guide the selection of marginals to measure. A more advanced form of data-awareness is found in iterative and adaptive mechanisms like \textbf{AIM} and \textbf{PMW}. These algorithms greedily select the next marginal to measure based on a quality score that evaluates the error of the \textit{current} synthetic data model. This allows them to focus the privacy budget on the aspects of the data distribution that are currently least well-approximated, leading to a more efficient learning process.

\textbf{Budget-awareness} concerns the intelligent allocation of the privacy budget, $\epsilon$. Naive strategies might simply divide the total budget evenly among a predetermined number of measurements. More sophisticated mechanisms, however, recognize that the available budget dictates the achievable signal-to-noise ratio and should influence the selection strategy. For instance, with a larger budget (and thus lower noise), it becomes feasible to accurately measure larger, higher-dimensional marginals. Mechanisms like \textbf{PrivBayes} and \textbf{PrivSyn} are considered budget-aware because they adjust the number and size of selected marginals based on the total budget \cite{zhang2021privsyn}. \textbf{AIM} introduces a particularly advanced form of budget-awareness with its ``annealing'' procedure. It adaptively adjusts the per-round privacy budget during its iterative selection process. If the model's accuracy is not improving, it increases the budget for subsequent rounds, effectively ``unlocking'' new candidate marginals that were previously too noisy to be useful. This allows \textbf{AIM} to perform robustly across a wide range of $\epsilon$ values without manual hyperparameter tuning. A related concept, ``frugal budgeting,'' appears in \textbf{JAM-PGM}, which can use public data for some measurements \cite{fuentes2024joint}. When a noise-free public measurement is chosen, the privacy budget allocated for that round is saved and rolled over, concentrating the budget for more difficult private measurements later on.

\textbf{Computation-awareness} is a crucial practical consideration, especially for mechanisms that rely on \textbf{Private-PGM} for estimation. The selection of marginals directly determines the structure of the graphical model used in the estimation step. If the selected marginals induce a graph with high treewidth, the estimation phase can become computationally intractable, failing due to excessive time or memory requirements. Computation-aware mechanisms are designed to prevent this. Methods like \textbf{MST}, \textbf{PrivMRF}, and \textbf{AIM} have this awareness built into their selection process, greedily adding marginals only if they do not violate a constraint on the complexity of the resulting graphical model \cite{cai2021data}. \textbf{AIM} makes this safeguard explicit by maintaining a hard filter on its candidate set, discarding any marginal that would cause the estimated junction tree size to exceed a predefined memory limit (line 7 in \cref{alg:aim_simplified}). This guarantees a predictable runtime and memory footprint, a feature lacking in some other adaptive methods like \textbf{MWEM+PGM}, which can fail on certain workloads.

The design of the \textbf{AIM} \cite{DBLP:journals/corr/abs-2201-12677} algorithm represents a culmination of these ideas, demonstrating that the four ``awareness'' criteria are not merely an independent checklist of desirable features but are deeply interconnected aspects of a single, holistic optimization problem. The core of \textbf{AIM}'s selection process is its quality score function, where here, we let $M_r(D)$ denote the vector of counts for the marginal query $r$ on the dataset $D$. The quality score is then given by,
\begin{equation}
q_r(D) = w_r \times (\|M_r(D) - M_r(p_{t-1})\|_1 - \sqrt{2/\pi} \times \sigma_t \times n_r)~,
\end{equation}
which co-optimizes for these competing pressures in each greedy decision. The $w_r$ term directly encodes workload-awareness by weighting candidates based on their relevance to the user's task. The error term, $\|M_r(D) - M_r(p_{t-1})\|_1$, provides data-awareness by measuring the deficiency of the current data-dependent model. The noise penalty term, $\sqrt{2/\pi} \times \sigma_t \times n_r$, instills budget-awareness, as the noise scale $\sigma_t$ is a direct function of the available privacy budget, naturally penalizing large, noisy marginals when the budget is tight. Finally, this entire selection process is constrained by an explicit check for computation-awareness, ensuring tractability at subsequent estimation steps.

\subsubsection{Estimation Methods: From Structured Optimization to Generative Models}\label{sec:estimation_methods}
Once a set of queries has been chosen in the \textbf{Select} step and measured privately, the next challenge is to reconstruct a coherent data distribution from those noisy answers. This is the goal of the \textbf{Estimate} step, which can be formally cast as a constrained optimization problem. Given a vector of noisy measurements $y$, produced by applying a DP mechanism $\mathbf{A}$ to the true data's histogram $h(D)$, the objective is to find an estimated probability distribution $p$ that best explains the observations. This is a classic inverse problem, often framed as finding $p$ that maximizes the likelihood of observing $y$, e.g. $\max_p \text{Likelihood}(p | y)$. In the common case where the mechanism ${\bf{A}}$ adds zero-mean Gaussian noise to a set of answers to linear queries (marginals) represented by a matrix $Q$, this simplifies to a least squares minimization problem over the probability simplex, or,
\begin{equation}
 \min_p \| Q(p) - y \|_2^2~.
\end{equation}
While the objective function is convex, the primary challenge is the dimensionality of the search space for $p$, which is the size of the data universe (i.e., $|\mathcal{X}|$) and thus exponential in the number of attributes. Many works that crucially rely on a histogram representation of the dataset have been proposed to solve variants of this problem \cite{hay2010boosting,nikolov2013geometry,lee2015maximum,li2015matrix,abowd2019census}; again, we defer a discussion of these methods to \Cref{sec:theoretical_hist}.

An alternative and principled approach to this problem is \textbf{Private-PGM} \cite{mckenna2019graphical}. Instead of optimizing over the full probability vector, Private-PGM leverages probabilistic graphical models (PGMs). A key insight is that when the measurements are a set of low-dimensional marginals, an optimal solution to the $\ell_2$ minimization problem is guaranteed to be a distribution that can be represented by a PGM whose structure (i.e., its factors) is determined by the measured marginals. By optimizing over the parameters of this compact graphical model, Private-PGM can achieve exponential savings in computation. Critically, it solves a convex optimization problem over the \textit{marginal polytope}, which ensures the resulting estimated distribution is \textit{globally consistent}. Private PGM was validated by its use in winning entries of the 2018 and 2020 NIST DP competitions \cite{mckenna2021winning}. The main trade-off is scalability: the computational cost is tied to the graph's treewidth, making it intractable if the measured marginals induce a dense dependency graph.

To address these scalability limitations, a family of approximation methods has emerged. \textbf{Approx-Private-PGM (APPGM)} \cite{mckenna2021relaxed} relaxes the global consistency requirement, instead enforcing only a specified set of \textit{local consistency} constraints over the simpler \textit{local polytope}. Other methods, like the \textbf{Gradually Update Method (GUM)} \cite{zhang2021privsyn}, are more heuristic, iteratively ``massaging’’ a dataset to conform to the noisy marginals. 

A distinct line of research bypasses the explicit structure of PGMs by using alternative, highly expressive representations. The \textbf{Relaxed Adaptive Projection (RAP)} framework \cite{Aydore:2021} uses a relaxed tabular representation amenable to gradient-based optimization. Similarly, \textbf{GEM (Generative networks with the Exponential Mechanism)} \cite{liu2021iterative} parameterizes the data distribution with a generative neural network. Both offer great flexibility but operate in a non-convex landscape, lacking the formal guarantees of Private-PGM. Other methods, such as \textbf{PrivBayes} \cite{PrivBayes:2017}, use a Bayesian Network, though its heuristic estimation step can be improved by replacing it with a principled engine like Private-PGM \cite{DBLP:journals/corr/abs-2201-12677}. Finally, some approaches leverage a \textbf{public dataset} to improve utility, e.g., by restricting the synthesizer's domain to values seen in the public data \cite{liu2021leveraging}, though this carries the risk of degrading performance if the public and private distributions are dissimilar.

More recent innovations have introduced entirely new reconstruction primitives. \textbf{GREM (Gaussian Residuals-to-Marginals)} \cite{mullins2024efficient} proposes reconstructing marginals not from noisy measurements of other marginals, but from noisy measurements of \textit{residuals}. Residuals are a different basis of linear queries that are mutually orthogonal and possess a Kronecker product structure, properties that allow for extremely efficient and scalable reconstruction using pseudo-inverse operations. \textbf{GReM-LNN} further refines the output to enforce local non-negativity and consistency, which can significantly reduce error. It is important to note, however, that GREM is a mechanism for \textit{marginal reconstruction} (i.e., query answering) and does not inherently produce a generative model. While it can produce highly accurate answers for a workload of marginals, it does not directly yield a synthetic dataset, limiting its use for analysts who need a dataset for downstream model training.

The evolution of these estimation methods highlights a fundamental design tension between rigor and scalability. Principled methods like \textbf{Private-PGM} perform an ``inner approximation,’’ guaranteeing a valid probabilistic model at the cost of high computational complexity. Their search space is restricted to the set of valid, globally consistent probability distributions, and the resulting solutions are guaranteed to be a realizable probabilistic model from which one can directly sample. The cost of this rigor is computational complexity, as the search space (the marginal polytope) is difficult to navigate. To achieve scalability, methods like \textbf{APPGM}, \textbf{GUM}, and \textbf{GReM} perform an outer approximation. They relax the problem by optimizing over a larger space of ``pseudo-marginals,'' which are only required to be locally consistent. This makes the optimization problem much easier to solve but comes with a crucial caveat: the solution may not correspond to any single, valid data distribution. This introduces a new challenge, as a final (often heuristic) step is required to project the inconsistent model back into a valid dataset, a process that can introduce its own unquantified approximation errors. Therefore, the choice of an estimation method involves a deep architectural decision between a slower, principled approach that guarantees a valid model and a faster, more scalable approach that may fit the noisy data better but whose resulting model may not be mathematically sound. This tension has motivated the exploration of entirely different paradigms, such as end-to-end deep generative models, which offer an alternative path to capturing complex data distributions, as we will discuss further in \Cref{tabular-generative}.

\subsection{Algorithms With Strong Worst-Case Guarantees{\color{red}{*}}}
\label{sec:theoretical_hist}

Having detailed the paradigm that underpins many empirically-oriented algorithms, we now turn to their theoretical counterparts. Unlike methods that are optimized for performance on benchmark datasets, the algorithms in this section are designed to provide strong, provable error guarantees and give rise to precise error theorems which allow a theoretical comparison between workload-adaptive and workload-agnostic methods.

Specifically, many theoretically oriented algorithms, like their empirical counterparts discussed in Section \ref{sec-workfload-based}, operate in a setting where the goal is to provide high accuracy for a given workload of queries $\mathcal{F}$, aiming to minimize the additive error as defined in Equation \ref{eq:accuracy_synthetic_data}. Some of these foundational methods achieve their guarantees by operating on a full histogram representation of the data to produce a valid and globally consistent probability distribution. While this theoretical rigor is their primary strength, it often comes at the cost of scalability, making many of these methods impractical for large-scale problems without significant modification.

We begin by discussing a class of foundational algorithms that are data-independent. In this context, ``data-independent’’ means that the choice of which statistics to measure is fixed in advance and does not depend on the input data $D$.

These approaches are often simpler to analyze and produce unbiased estimates of the workload-query answers (before post-processing, such as enforcing non-negativity constraints). One could view them as basic instances of the select-measure-estimate paradigm, but with a trivial ``select'' or ``estimate'' step. As the nuances are important and they are foundational, we believe it is appropriate to separate them out and describe them directly.

\paragraph{Noisy Histograms.} A quintessential \textit{workload-agnostic} approach, this method's measurement strategy is fixed and exhaustive: it measures the count of every possible record in the universe ${\cal X}$. One directly adds discrete Laplace noise to each histogram bin: for each bin $x \in {\cal X}$, compute $\widehat{h}_x = h(D)_x + \mbox{Lap}(1/\epsilon)$, and post-process it by making the noisy counts nonnegative. This approach is effective when the number of bins is much smaller than the number of queries, as its error rates scale polynomially with the former but only poly-logarithmically with the latter (e.g., Theorem 2.9 in \cite{vadhan17complexity}).

\paragraph{Noisy Query-Answering.} In direct contrast, this is a canonical \textit{workload-adaptive} algorithm. Here, the algorithm's measurement strategy is one and the same as the user's evaluation workload. It adds noise (e.g., Laplace or Gaussian) directly to the answers of the queries in ${\cal F}$: $\tilde{h} = (\langle h(D),f\rangle+\mbox{Lap}(\Delta {\cal F}/\epsilon))_{f\in {\cal F}}$. The entire privacy budget is spent on preserving information known to be critical for the downstream task. To recover a histogram from the noisy query answers, one can solve an (integer) optimization problem over the histogram variable $\widehat{h}$ that aims to minimize the difference between its query answers and the noisy ones, as proposed by \cite{Barak:2007}:
\begin{equation} \label{eq:estimation}
\min_{\widehat{h} \in \mathbb{R}_+^{\cal X}:\,\|\widehat{h}\|_1=n}
 \Big|\Big|\Big\langle \widehat{h}-\tilde{h},f\Big\rangle_{f \in \cal F}  \Big| \Big|.
\end{equation}
While this approach is suboptimal in general, it has proved effective in settings such as synthetic graphs that are accurate for cut queries \cite{GuptaRU12}. More involved randomization strategies \cite{steinke16between,GaneshFORC:2021,DK22,GKM21}, including the generalized Gaussian mechanism and the sparse vector technique, can yield optimal rates. Error rates can be made independent of the universe size (number of bins) and the number of rows, making this approach advisable for a relatively small number of queries, such as low-dimensional marginals as we discussed in \Cref{select-measure-estimate}.

\paragraph{The Matrix Mechanism.} The Matrix mechanism \cite{li2015matrix} interpolates between these two baselines and includes them as special cases. Its key idea is to compile the workload into a different set of queries known as the \textit{strategy}, add noise to the strategy query answers, and then post-process them to estimate the workload answers and the underlying histogram. Setting the strategy to be the identity matrix recovers the noisy histogram approach, while setting it to the workload recovers the noisy query-answering approach. The mechanism also proposes a \textit{strategy optimization} problem to Pareto dominate both baselines. However, it requires representing the workload and strategy in matrix form, which can be prohibitive for large domain sizes. The high-dimensional matrix mechanism (HDMM) \cite{mckenna2018optimizing} addresses this by utilizing implicit matrix representations, enabling its use in higher-dimensional settings when the workload consists of conjunctive queries.

\paragraph{Data-dependent Approaches with Guarantees}
If we relax the requirement of unbiasedness, we can trade a small amount of bias for a large reduction in variance. This is often a worthwhile trade-off in practice, especially for smaller datasets and privacy budgets.

{
\small
\begin{longtable}[htb]{p{0.3\linewidth} p{0.3\linewidth}  p{0.2\linewidth} }
\caption{Best-known upper bounds on the error for tabular synthetic data. Algorithms from the first two rows work for answering counting queries, but they can be turned into synthetic data generators by solving an integer linear program over $n\times \log|{\cal X}|$ boolean variables. The rest of the algorithms directly provide synthetic data. For approximate-DP, the optimal error rates are determined by the minimum of the corresponding rates in the table. } \\
\toprule
\textbf{Reference} & \textbf{$ \ell_{\infty} $ Upper Error bound} & \textbf{DP} \\
\midrule 

\citet{steinke16between} & $O\Big({\frac{|{\cal F}|}{\epsilon}}\Big)$ & $\epsilon$-DP \\ \hline

{\citet{DK22}, \citet{GKM21}}  & $O\Big({\frac{\sqrt{|{\cal F}| \log \frac 1\delta}}{\epsilon}}\Big)$ & $(\epsilon, \delta)$-DP  \\ \hline

{\citet{vadhan17complexity}} &   $O\prn{\frac{\sqrt{|{\cal X}| \log |{\cal F}|}}{\epsilon}}$ & $\epsilon$-DP  \\ \hline

\citet{BlumLR13,HardtR10,HardtLM12} &  $O\prn{n^{\frac23} {\frac{\log |{\cal F}| \log|{\cal X}|}{\epsilon}}^{\frac13}}$ & $\epsilon$-DP \\ \hline

 \citet{HardtR10,HardtLM12} & $O\left({n^{\frac12}\prn{\frac{\sqrt{\log |{\cal X}|} \log |{\cal F}| \log \frac1\delta}{\epsilon}}^{\frac12}}\right)$ & $(\epsilon, \delta)$-DP \\ 

 \bottomrule
    
\label{tab:rates_synthetic_data}
\end{longtable}
}

One such data-dependent approach is known as \textbf{Private Multiplicative Weights (PMW)} \cite{HardtR10}. \textbf{PMW} works with a histogram variable initialized with uniform weights and subsequently updates the histogram to progressively make the counts more accurate (see Algorithm \ref{alg:pmw}). The updates are based on the most inaccurate queries for the current histogram, which must be identified privately using methods like Report Noisy Max or the Exponential Mechanism \cite{McSherry:2007}. \textbf{PMW} attains error rates that are poly-logarithmic with respect to $|{\cal X}|,|{\cal F}|$, but polynomial in $n$, making it advisable for a relatively small number of rows. These running times are generally unavoidable \cite{Dwork:2009,Ullman:2011}. Considering that the data domain or the number of queries can be very large, this limits its practicality. 

Following work on \textbf{PMW}, \citet{HardtLM12}~consider the case where the domain can be decomposed in product form, ${\cal X} = \mathcal{X}_1 \times \cdots \times \mathcal{X}_t$, such that each query depends only on a single ${\cal X}_i$, and propose a Multiplicative Weights-Exponential Mechanism (\textbf{MWEM}) approach. \textbf{MWEM} allows for parallelization and a much smaller product-distribution representation of the histogram, though it rules out queries across attributes from different groups. 

Another alternative is to solve the dual problem to \eqref{eq:zero_sum_synthetic_data} \cite{Gaboardi:2014}. This approach, based on the Minimax Theorem \cite{vonNeumann:1928}, performs PMW updates on a distribution over \textit{queries}. Privacy is achieved by sampling from the resulting query distribution (an application of the exponential mechanism). With privacy guaranteed by sampling, the inner minimization subproblems can be solved with effective but \textit{nonprivate} heuristics for NP-hard problems (e.g., integer linear programs). This is advantageous when the number of queries is much smaller than the  number of bins. 
\begin{equation} \label{eq:zero_sum_synthetic_data} 
\min_{\widehat{h} \in \mathbb{R}_+^{\cal X}:\,\|\widehat{h}\|_1=n} 
\Big|\Big|\Big(\big\langle \widehat{h}, f \big\rangle - \big\langle h(D),f\big\rangle\Big)_{f\in {\cal F}}\Big| \Big|.
\end{equation} 

\begin{longtable}[htb]{p{0.4\linewidth} p{0.5\linewidth} }
\caption{Advised choice of approach, based on the error rates from Table \ref{tab:rates_synthetic_data}.} \\
 \toprule
 \textbf{Preferred Approach} & \textbf{Regime}  \\
\midrule
Noisy histograms &  $|{\cal X}|\ll \max\{ n,|{\cal F}|\}$  \\ 
Noisy-query answering & $|{\cal F}|\ll \max\{n,|{\cal X}|\}$ \\ 
Private multiplicative weights & $n\ll \max\{|{\cal F}|,|{\cal X}|\}$ \\ 
\bottomrule
\label{tab:comparison_regimes_synthetic data}

\end{longtable}

We emphasize that as far as the theoretical guarantees are concerned -- there is no one approach (noisy histograms, noisy-query answering or PMW) that outperforms uniformly across different problem settings \cite{PrivBayes:2017,Rosenblatt2020differentially}. This conclusion is consistent with the fact that methods' performance differ on different regimes of data size, data domain, and number of queries\footnote{It may seem odd to refer here to `number of queries'. For clarifications on what queries may be of interest here, see \Cref{sec:query_release_problem}.}. \Cref{tab:comparison_regimes_synthetic data}
provides rough recommendations on which approach is preferred in which regime.

\begin{algorithm} 
\small
\caption{Private Multiplicative Weights Update Method}\label{alg:pmw}
\begin{algorithmic}[1]
\Require Private dataset: $D\in\mathcal{X}^n$, query workload: finite set ${\cal F}\subseteq \{0,1\}^{\cal X}$, stepsize $\eta>0$.
\Ensure Privatized histogram: $\widehat{h}\in \mathbb{R}_+^{\cal X}$, $\|\widehat{h}\|_1=n$, representing DP synthetic tabular data
\State $h^1\gets \frac{n}{|{\cal X}|}\mathbf{1}$
\For{$t = 1$ to $T$}
    \State Select worst performing query $f_t$ that approximately solves $\max_{f\in{\cal F}}|\langle h^t-h(D),f\rangle|$ (in DP manner), together with $v_t$ private estimate of query error $\langle h^t-h(D), f_t\rangle$
  \If{$|v_t|\leq \alpha$}
    \Return $\widehat{h}=h^t$
  \Else
    \State $h^{t+1}= n \dfrac{\big(h_x^t\cdot\exp(-\eta\cdot\mbox{\small sgn}(v_t)\cdot f_t(x))\big)_{x\in{\cal X}}}{\sum_{y\in{\cal X}} h_y^t\cdot\exp(-\eta\cdot\mbox{\small sgn}(v_t)\cdot f_t(y)) }$
  \EndIf
\EndFor
\State \Return $\widehat{h}=\frac1T\sum_{t=1}^T h^t$
\end{algorithmic}
\end{algorithm}

\subsubsection{Extensions and Variants{\color{red}*}}\label{sec-extensions-tabular}

\paragraph{Relative error guarantees}    
    Instead of focusing on achieving worst-query utility in additive terms (e.g. under the $\ell_{\infty}$-norm in \eqref{eq:accuracy_synthetic_data}), 
    recent work has focused on 
    assessing utility in terms of {\em relative error}.
    Here, the goal is to construct synthetic histograms with a mixed multiplicative/additive approximation guarantee: for error parameters $\zeta, \alpha>0$, the goal is computing a synthetic histogram $\widehat{h}$ such that for all $f\in{\cal F}$
    \[ (1-\zeta)\langle h(D),f\rangle-\alpha \leq \langle \widehat{h},f\rangle \leq (1+\zeta)\langle h(D),f\rangle+\alpha. \]
    These guarantees are particularly relevant for analytics settings, where the goal is to distinguish `large' from `small' counts, rather than accurately predicting each count  \cite{Cormode:2012,Qardaji:2013,Qardaji:2013,EpastoMMMVZ23,Ghazi:2023}. Recently, it has been shown that in this setting there exist DP algorithms with constant multiplicative approximation (i.e., $\zeta$ constant) whose additive error (i.e. $\alpha$) scales only poly-logarithmically with $n,|{\cal X}|,|{\cal F}|$ \cite{GhaziPREM:2025}.

\paragraph{Algorithms targeting distributional guarantees.} 
Algorithms in this group can be considered as having an infinite workload: this analogy is precise for the broad class of {\em integral probability metrics} \cite{Muller:1997}. As mentioned earlier, a notable and widely studied particular case is synthetic data which are accurate in terms of the Wasserstein/Kantorovich distance, characterized by approximation of expectations over arbitrary 1-Lipschitz functions \cite{Villani:2008}. This example is relevant, as Lipschitz queries commonly arise as the result of stable data analyses, e.g.~models trained on regularized convex losses. In this context, early hierarchical histogram algorithms already approached a $\tilde{O}(1/n)$ Wasserstein-1 error bound with heirarchical histogram methods in one dimension, although they did not state the analysis as such \cite{hay2010boosting}. More recently, a variety of results have matched or improved these early results such that they have achieved nearly optimal Wasserstein error rates with algorithms that run in polynomial time \cite{Boedihardjo:2024,He:2023,musco2024sharper,feldman2024instance}; these algorithms also leverage effective ways to hierarchically discretize the continuous space, together with noise addition approaches to privately count the number of data points falling into each bucket for the discretizations. Some take the approach of constructing a random walk that diverges logarithmically with the number of steps (as opposed to with the square root of the step count) \cite{Boedihardjo:2024}, while others look at matching a class of distributional moments closely tied to Wasserstein distance \cite{musco2024sharper}, and still others look at alternative heirarchical histogram constructions \cite{feldman2024instance}.

Unfortunately, the error rates in this setting suffer from the curse of dimensionality, and these rates are essentially unimprovable in general (there is a known $\Theta(1/n^{1/d})$ lower bound \cite{Boedihardjo:2024}). Improvements have been obtained under additional assumptions, such as (intrinsic) low dimensionality of the data \cite{DP_heatmaps:2022,Donhauser:2023,He2023lowdim}, as well as extensions to online settings \cite{HeOnlineDP:2024}. Characterizations of the sample complexity of DP synthetic data generation that are accurate for infinitely-many counting queries have also been obtained in the literature \cite{Bousquet:2020}.

\subsection{Numerical Feature Handling} \label{tabular-feature-handling}

Almost all of the methods we've described so far assume the data comes from a categorical domain with a fixed, known set of possible values. Many real-world tabular datasets have a mix of both categorical and numerical features. One simple way to bridge this gap is to discretize numerical attributes to intervals. This numerical-to-categorical transformation yields a fully categorical dataset and enables using the mechanisms described earlier. There are multiple strategies one might use to discretize numerical attributes, the simplest being a uniform discretization which maps values to fixed-width intervals. The benefit of this strategy is that it is data-independent, and only requires knowledge of the minimum and maximum possible value for a numerical attribute. This number of bins is an important hyper-parameter to consider. Larger values yield finer-grained discretizations that may better reflect the original distribution, but results in bins with smaller counts which are not preserved as well in the presence of noise, and larger categorical domains, which can increase runtime of downstream mechanisms that operate on the categorical data. When the distribution is heavily concentrated on a single value, uniform discretizations sometimes have undesirable properties. For example, the capital gain attribute of the adult dataset spans from $0$ to $100,000$, but $91\%$ of the values are $0$. Under a uniform discretization with $100$ bins, these $0$-values will be mapped to the interval $[0, 1000]$, and a lot of information is lost. Converting this interval back to a numerical value can be done in a number of ways, choosing the midpoint deterministically ($500$) or sampling uniformly at random from the interval are two natural options. In both cases, the numerical data is heavily skewed away from the distribution of the real data.

Adaptive discretization techniques \cite{zhang2016privtree,gillenwater2021differentially} can overcome this problem, by identifying a partition of the domain into roughly equal-density intervals. Under this approach, smaller intervals will be chosen where data is more concentrated, while larger intervals will be chosen where it is more spread out. In the example above, the interval $[0,0]$ would typically be chosen because it contains so much probability mass, and hence this class of approaches do not suffer from the same distribution-skew as uniform discretization.

Mapping numerical attributes to categorical ones is just one way they can be handled. However, it's important to note that this discretization step can introduce inaccuracies and affect the utility of the resulting synthetic data, as highlighted by \cite{ganev2025importance}. The number of bins and the choice of discretization strategy can significantly impact the quality of the generated data. As shown by \cite{ganev2025importance}, optimizing both the discretizer (uniform, quantile, k-means, or PrivTree) and the number of bins can lead to substantial improvements in utility. Technically speaking, one must apply DP to the discretization process itself, to avoid introducing new privacy vulnerabilities.

Alternatively, some mechanisms, like the \textbf{RAP++} algorithm \cite{Vietri:2022} are designed to natively handle numerical attributes \textit{without} requiring discretization. This is a key advantage, as it allows for preserving more information about the original data distribution. \textbf{RAP++} achieves this by directly optimizing for ``class conditional linear threshold queries'' and ``mixed marginal queries,'' which can capture the relationships between numerical features and class labels without relying on a binning strategy. These types of queries are defined directly on the numerical values, instead of on categories. This allows for a more accurate representation of the underlying data, and can lead to faster runtimes compared to methods that require discretization.

\subsection{Constraint Handling}

An often-overlooked aspect of synthetic data generation is handling of structural constraints, which are very common in real-world settings.. That is, the setting of one or more attributes might limit the set of possible values another attribute might take. It is important to respect these constraints for the purposes of data integrity.  When constraints are violated, downstream analyses of the data can be confusing and misleading. A naive approach to handling constraints is by down-sampling: i.e., discarding impossible synthetic records. This simple approach is compatible with any synthetic data generator, but might increase the bias of the mechanism in ways that are hard to quantify and limit, and if a large percentage of records violate the constraints, can be expensive as well.  Another related approach, used by Kamino \cite{ge2021kamino} (to be discussed in the next section), is to incorporate constraints in the inference / sampling step, after estimating the model with DP ignoring the constraints.

Another approach is available for methods that use Private-PGM \cite{mckenna2019graphical,mckenna2022aim}, where constraints can be directly incorporated into the model estimation step.  This approach bakes the constraints directly into the optimization problem to find the best-fitting distribution, yielding better utility than solving an unconstrained problem then projecting it onto the constraint set or enforcing the constraints at inference time. This approach ensures that no invalid records will be generated in the first place, bypassing the issues of down-sampling. This method is most effective when constraints involve a small number of attributes, as it requires materializing a `masking` tensor of that dimensionality.
    
\subsection{Modern End-To-End Methods for Tabular Data Generation}\label{tabular-generative}
In contrast to the \textit{workload-adaptive} algorithms that take explicit guidance on which statistics to preserve (Section \ref{sec-workfload-based}), another significant class of practical algorithms is fundamentally \textit{workload-agnostic}. These methods use modern, end-to-end deep generative models. Their ``measurement strategy'' is implicit in the model's architecture and the training objective (e.g., minimizing an adversarial loss or a reconstruction loss), which aims to learn a holistic, flexible representation of the joint data distribution. This strategy is designed for general-purpose fidelity and is not structured to adapt to a user-provided workload of specific statistical queries (unless the method allows the user to specify measurement queries of interest, which is sometimes the case.) The hope is that a model that accurately captures the overall distribution will prove 
useful for a wide range of future, unanticipated analyses. 

These approaches often rely on DP-finetuning or DP-training (from scratch) techniques, just like the winning methods in image (Section \ref{sec-dp-training-image}) and text (Section \ref{sec-dp-finetuning}) data synthesis. With privacy unit being a row in table, these models conceptually learn to generate one row of synthetic data, rather than 1 full image or a piece of text. 

We survey evolution of these approaches next.

\textbf{Early attempts}: In non-DP settings, both GANs (Generative Adversarial Networks) \cite{goodfellow2014generativeadversarialnetworks} and Variational Autoencoders (VAEs) \cite{kingma2022autoencodingvariationalbayes} have been successfully applied to generative tasks, especially in image domains. GANs are adversarial networks that approximate the joint data distribution through an adversarial training process that pits a generator, which attempts to produce a synthetic sample good enough to fool a discriminator, against a discriminator that learns to distinguish real vs. synthetic samples. VAEs, on the other hand, attempt to model the joint distribution of data by mapping a sample to a distribution in the latent space (encoding step), then sampling from this distribution (decoding step).

DP-GANs either employ DP-SGD like training: DP-GAN \cite{dpgans}, RDP-CGAN \cite{Torfi_2022}, CTAB-GAN$+$ \cite{beb4dd08069842f485cfcd92e2adbd3f}, or attempt private inference via ensembling (PATE-GAN \cite{yoon2018pategan}). However, an empirical study that looked at the performance of downstream ML models trained on DP synthetic data (\textit{Machine Learning Efficiency} metric) by \citet{DBLP:journals/corr/abs-2112-09238} demonstrated that previously discussed workload-based methods outperform GAN-based models, with the latter not being able to accurately preserve even one-dimensional statistics of the tabular data. GAN-based methods may be hard to train with DP-SGD due to GANs' adversarial nature and their inherent instability (even in a non-DP setting) \cite{tran2024differentiallyprivatetabulardata}. VAE-based approaches include DPGM \cite{acs2018differentiallyprivatemixturegenerative}, DP-SYN \cite{dpsyn}, and P3GM \cite{DBLP:journals/corr/abs-2006-12101}; DP combinations of both GANs and VAEs also exist \cite{yang2024tabulardatasynthesisdifferential}. Another approach, Kamino \cite{ge2021kamino}, fits the distribution of one column (or small group of columns) given all previously generated columns using any discriminative model with DP-SGD, and splitting the privacy budget between the trained models.

\textbf{Modern models}: \textit{Diffusion models} have shown great promise in image domains recently \cite{nakkiran2024stepbystepdiffusionelementarytutorial}, and these approaches have also been applied in the tabular domain as well. Diffusion models learn to progressively distort the input by introducing noise over a number of steps and then learning to reverse this process for generation. For example, the TableDiffusion model \cite{truda2023generatingtabulardatasetsdifferential} outperformed DP-GAN-based models in a limited study involving two datasets \cite{truda2023generatingtabulardatasetsdifferential}.
The following two groups of generative models (Autoregressive transformer-based and Pretrained LLM-based) employ LLM-like architectures and model the tabular data via textual encoding, then essentially employing the chain rule on the tabular data on a token-by-token basis, modeling next token prediction based on previous token values.

\textbf{Autoregressive models}:
\citet{castellon2023dptbarttransformerbasedautoregressivemodel} introduced a probabilistic model DP-TBART, an autoregressive transformer-based model. The difference between this model and a standard LLM is the exact architecture, which is a 3-layer decoder closely resembling DistilGPT-2, and a custom tokenizer that assigns different tokens to each column's distinct values (after discretization) so no two column encodings share the same tokens. Each tabular row is then encoded as values of columns (assuming some ordering of features), with feature names not being used in this encoding. DP-SGD is used to train the model. During sampling, to force the output to be valid, post-processing is employed, removing all disallowed tokens for each column. The authors provide a comparison with a number of methods including AIM, with AIM's workload-based method outperforming the proposed methods (and all other considered baselines). As was noted in~\Cref{select-measure-estimate}, higher-order marginals suffer from exponential scaling, and in practice, this leads to the use of lower-order marginals, which can fail to capture complex interactions of the features. To demonstrate this phenomenon, \citet{castellon2023dptbarttransformerbasedautoregressivemodel} propose two new datasets with complex joint feature interactions -- Dyck-k, an artificially constructed dataset representing a worst-case dataset for workload-based methods, and WikiText-103 (a real-world dataset of Wikipedia articles). DP-TBART outperforms AIM by a significant margin on both these challenge sets.

Recently, \citet{tiwald2025tabularargnflexibleefficientautoregressive} introduced a Tabular AutoRegressive Generative Network (TabularARGN), departing from the trend of using complex models like Transformers to model joint distribution. They argued that treating tabular data as text and using LLMs is counter-intuitive. Their architecture is essentially a multi-tower model with each tower being dedicated to each feature. At the top of the model, a dedicated embedding encodes each feature; the embeddings are then combined (with some permutation of order) and forwarded to the dedicated towers. Each tower sees only the feature values on which it conditions: for the first feature, no additional embeddings are passed; the second feature is conditioned on the first and receives its embedding; the third is conditioned on the first two, etc. These towers are standard feedforward layers with ReLU and dropout. Each tower has a softmax head, and the cross-entropy loss of all heads is summed. During sampling, the column values are inferred one by one and passed as the input to the next column. To achieve DP, DP-SGD can be used. While the simplicity and performance of the model in a non-DP setting are appealing, no competing DP models were used for comparison in the DP setting, making it hard to judge the effectiveness of this method for DP synthetic tabular data generation.

\textbf{Pretrained LLM based}:
With the rise of LLMs, research focus shifted towards attempting to harness their capabilities for tabular synthetic data generation.
Using LLMs can allow researchers to forego extensive preprocessing needed for query release methods (categorization of continuous numeric attributes, data imputation and normalization, outlier removal, and data smoothing). Additionally, purely numerical data representations like histograms erase subtle connections that even a rudimentary language model can infer. For example, \citet{borisov2023languagemodelsrealistictabular} highlighted that for the Adult dataset \cite{adult_2}, features like age, marital status, and education exhibit clear hidden relationships that a pretrained LLM can uncover and utilize easily: e.g., higher degrees cannot be obtained at an early age, and in many countries, it is illegal for teens to be married.
While this knowledge can be discovered given sufficient data by histogram-based methods, LLMs can readily take feature names into account.

Early attempts at harnessing the emerging powers of using standard LLMs included pioneering work by \citet{borisov2023languagemodelsrealistictabular} in the \textit{non-DP setting}. Their method GReaT works by first creating a purely textual encoding of all tabular data attributes, including the label attribute (e.g., ``Bachelors Education, Adult male, income <50K''), while permuting the order of the attributes in the encoding. Then, a pretrained LLM is finetuned on such textual encodings. Permutation of feature names is introduced to avoid the model's reliance on a pseudo-ordering of attributes that the model would learn otherwise. Additionally, it unblocks conditional sampling by providing either any feature name or any combination of feature names and initial values. Sampled text data is then converted back to tabular, with samples that do not conform to the expected format (e.g., out of range, wrong names of attributes) discarded.

While this approach can be easily converted to a DP setup by swapping an optimizer with its DP version during LLM finetuning, \citet{tran2024differentiallyprivatetabulardata} argue that it will not work out of the box and the DP-finetuned model will not conform to the expected format due to the noise introduced during DP training. \citet{tran2024differentiallyprivatetabulardata} instead separate the DP finetuning into two stages: learning format compliance and actual private data modeling. For the first stage, \citet{tran2024differentiallyprivatetabulardata} treat feature names and categorical values, as well as the range of values for numerical attributes, as non-sensitive data and create random data represented in a similar format. This data does not exhibit realistic feature distributions and dependencies and serves to learn the proper output format of the data. An LLM is then finetuned non-privately on this data with LoRA. The second stage is DP LoRA finetuning on the actual private data textual encodings. These encodings are similar to \cite{borisov2023languagemodelsrealistictabular} (<attribute> is <value>). For this stage, the authors additionally replace the standard next token Cross Entropy loss with a combination of weighted cross-entropy (WCEL) and numerical understanding loss (NUL). WCEL up-weights the actual private tokens and down-weights formatting tokens (feature names, connection 'is', commas, etc.). NUL, on the other hand, attempts to solve the problem that CE-based learning does not distinguish the magnitude of error when it comes to numerical data: an error in only one token can have a different magnitude of error when the prediction is converted to a numeric format. For example, assuming each digit is a token, for a 10.0 ground truth, predictions 19.0 and 10.1 both have one token error, while the magnitude of the numerical error is widely different. NUL loss thus penalizes the squared errors between predicted numerical tokens and actual ground truth numerical values. The authors compare their method DP-LLMTGen with GAN-based methods, PATE-based methods, and some workload-based methods (RAP \cite{Aydore:2021}, RAP++ \cite{Vietri:2022}, and GSD \cite{liu2023generatingprivatesyntheticdata}), with DP-LLMTGen exhibiting impressive results, outperforming competitive methods in terms of statistical fidelity in most cases, often by a large margin, with RAP++ being the only method that comes close. However, when considering the ML efficiency metric (performance of a gradient boosted tree model trained on synthetic data), DP-LLMTGen outperforms workload-based methods in only 3 out of 10 cases. While the comparison is in no way exhaustive and some modern methods like AIM \cite{DBLP:journals/corr/abs-2201-12677} are missing, this study hints at the potential of LLMs for DP tabular data synthesis.

Concurrently, \citet{afonja2025dp2stageadaptinglanguagemodels} reached similar conclusions and proposed a two-stage method similar to \citet{tran2024differentiallyprivatetabulardata}. For the first stage, they considered using either public data encoding (that does not reference correct feature names, ranges, or categorical values but conforms to the same encoding scheme) or random data generated using the aforementioned knowledge. Their findings indicate that a public dataset works sufficiently well. They also argued that column shuffling makes learning in DP harder and advocate against such shuffling (doing this, however, will fix the order of feature names during sampling). For DP finetuning, they use only a weighted loss that similarly down-weights formatting tokens. For sampling, they suggest a modification of the sampling mechanism where, instead of fully discarding the sample that does not conform to the expected formatting, only the offending tokens are discarded, and sampling is re-attempted by prompting with valid tokens (imputation). Their ablations on weighted loss suggest that it is beneficial for situations when a public dataset was used in the first Stage, and less needed for when a uniformly generated dataset was used. They compare their method to modern methods like MST and AIM, and while their method demonstrates good performance across most metrics, underperforming on certain fidelity metrics, especially for large dimensional datasets, AIM  \cite{DBLP:journals/corr/abs-2201-12677} still significantly outperforms it.

\begingroup
\setlength{\tabcolsep}{3.5pt} 
\renewcommand{\arraystretch}{1.1} 
\footnotesize

\begin{longtable}{p{5.5cm}|c|c||c|c|c|c}
\caption{Taxonomy over algorithms for DP tabular data synthesis. A checkmark (\checkmark) indicates the algorithm's measurement strategy is designed to adapt to a user's downstream workload.}
\label{tab:taxonomy_updated} \\

\hline 
\textbf{Method} & \textbf{Year} & \textbf{Paradigm} & \begin{tabular}[c]{@{}c@{}}\textbf{Workload}\\\textbf{Adaptive}\end{tabular} & \begin{tabular}[c]{@{}c@{}}\textbf{Data}\\\textbf{Aware}\end{tabular} & \begin{tabular}[c]{@{}c@{}}\textbf{Budget}\\\textbf{Aware}\end{tabular} & \begin{tabular}[c]{@{}c@{}}\textbf{Computation}\\\textbf{Aware}\end{tabular} \\ \hline
\endfirsthead

\hline 
\textbf{Method} & \textbf{Year} & \textbf{Paradigm} & \begin{tabular}[c]{@{}c@{}}\textbf{Workload}\\\textbf{Adaptive}\end{tabular} & \begin{tabular}[c]{@{}c@{}}\textbf{Data}\\\textbf{Aware}\end{tabular} & \begin{tabular}[c]{@{}c@{}}\textbf{Budget}\\\textbf{Aware}\end{tabular} & \begin{tabular}[c]{@{}c@{}}\textbf{Computation}\\\textbf{Aware}\end{tabular} \\ \hline 
\endhead

\multicolumn{7}{c}{\textit{Marginal-Based Mechanisms}} \\ \hline
PMW / MWEM \cite{HardtR10}, \cite{HardtLM12} & 2010/12 & Marginal & \checkmark & \checkmark & & \\
Noisy Query-Answering \cite{Barak:2007} & 2007 & Marginal & \checkmark & & & \\
Matrix Mechanism \cite{li2015matrix} / HDMM \cite{mckenna2018optimizing} & 2015/18 & Marginal & \checkmark & & & \\
RAP \cite{Aydore:2021} & 2021 & Marginal & \checkmark & \checkmark & & \checkmark \\
GEM \cite{liu2021iterative} & 2021 & Marginal & \checkmark & \checkmark & & \checkmark \\
AIM \cite{DBLP:journals/corr/abs-2201-12677} & 2022 & Marginal & \checkmark & \checkmark & \checkmark & \checkmark \\
JAM-PGM \cite{fuentes2024joint} & 2024 & Marginal & \checkmark & \checkmark & \checkmark & \checkmark \\ \hline
PrivBayes \cite{PrivBayes:2017} & 2017 & Marginal & & \checkmark & \checkmark & \checkmark \\
MST \cite{mckenna2021winning} & 2021 & Marginal & & \checkmark & & \checkmark \\
PrivSyn \cite{zhang2021privsyn} (w/ GUM) & 2021 & Marginal & & \checkmark & \checkmark & \checkmark \\
PrivMRF \cite{cai2021data} & 2021 & Marginal & & \checkmark & \checkmark & \checkmark \\ \hline
\multicolumn{7}{c}{\textit{End-to-End Generative Mechanisms (Workload-Agnostic by design)}} \\ \hline
GAN-based \cite{dpgans}, \cite{yoon2018pategan} & 2018/2019 & Generative & & \checkmark & & \checkmark \\
VAE-based \cite{acs2018differentiallyprivatemixturegenerative},\cite{dpsyn} & 2018 & Generative & & \checkmark & & \checkmark \\
Kamino \cite{ge2021kamino} & 2021 & Generative & & \checkmark && \checkmark \\
Diffusion-based \cite{truda2023generatingtabulardatasetsdifferential} & 2023 & Generative & & \checkmark & & \checkmark \\
LLM-based \cite{tran2024differentiallyprivatetabulardata}, \cite{afonja2025dp2stageadaptinglanguagemodels} & 2024/25 & Generative & & \checkmark & & \checkmark \\ \hline
\end{longtable}
\endgroup

\subsection{Comparison of the Methods}\label{tabular-comparison}
When comparing these diverse approaches, it is helpful to return to our initial categorization. On one hand, practical workload-based methods have solid foundations, impressive empirical performance and, in some cases, come with strong performance guarantees. On the other hand, generative models, particularly those based on LLMs, lack this theoretical backing but show great promise and may implicitly leverage knowledge that is difficult to capture with explicit marginal measurements.

Non-DP tabular data generation is a rapidly developing area, and many models that are introduced and successful in the non-DP setting were either not yet attempted and adapted for DP (e.g., TabPFNGen \cite{ma2024tabpfgentabulardata}) or may require modifications to make them powerful under the noise introduced by DP. This was already demonstrated for GReaT-inspired models \citet{borisov2023languagemodelsrealistictabular}, where DP variants had to be trained in a two-stage setup. Additionally, many evaluations are not exhaustive, often offering comparisons only against other generative models of the same type and excluding well-established emprically oriented workload-based methods like AIM \cite{DBLP:journals/corr/abs-2201-12677}. For example, a recent benchmark study by \citet{chen2025benchmarkingdifferentiallyprivatetabular} that showed AIM's superiority specifically excluded LLM-based methods from comparison, arguing that such methods use knowledge embedded in pretrained LLMs that prevents fair comparison.

There is, however, evidence that LLM-based modeling can have an edge over workload based methods. Firstly, one interesting observation by \citet{borisov2023languagemodelsrealistictabular} was that for tabular data generation, training generative models using meaningful feature names like Age, Education, etc., results in better synthetic data compared to the same data that had random feature names assigned. This demonstrates that LLMs utilize pretraining knowledge, which is different from GANs, VAEs, and workload-based methods. Further, pretrained (but not SFT or RLHF LLM checkpoints) work marginally but not significantly better, suggesting that pretraining is important. \citet{castellon2023dptbarttransformerbasedautoregressivemodel} pointed out the exponential scaling of workload-based methods to high-order marginals and demonstrated two use cases where an LLM-based method outperforms them significantly. 

While finetuning-based approaches clearly exhibit potential, early attempts at using an LLM without finetuning have encountered difficulties in obtaining good quality tabular synthetic data, e.g., a private evolution-based approach \cite{swanberg2025apiaccessllmsuseful} and in-context learning based \cite{huincontext}. This highlights that tabular data is often out-of-distribution for pretrained LLMs, and thus finetuning is necessary. 

To summarize, practical workload-based methods (Section \ref{sec-workfload-based}) have demonstrated impressive performance, solid  foundations and can come with performance guarantees. While LLM-based methods currently lack wide adoption and rich body of research, and while practical workload-based methods like AIM \cite{DBLP:journals/corr/abs-2201-12677} outperform generative-based methods currently on simpler and/or more realistic datasets, generative models, in particular LLM-based ones, show great promise.

\subsection{Evaluating Synthetic Tabular Data Quality} \label{tabular-metrics} 
On top of the generic metrics applicable to all data modalities mentioned in Section~\ref{sec-fidelity-generic}, below we outlined metrics specific to tabular synthetic data. 

\textbf{Fidelity}: 
For workload adaptive algorithms which create synthetic data to answer a set of predefined queries accurately, the fidelity of achieved synthetic data can be evaluated and iterated upon using the Equation \ref{eq:accuracy_synthetic_data} (namely, the performance of synthetic data on workload queries). This metric also gives an idea of performance on unseen queries that were not in the query set for the algorithm, albeit no guarantees are provided.

Alternatively, for all types of tabular data, the following metrics can be employed.
\citet{Aydore:2021} introduced \textit{Statistical Fidelity} which is an average of total variation distances of joint distributions (1-5 way marginals) between synthetic and set aside real data. Similarly, 
\textit{Pairwire Attribute Distribution Similarity} measures the similarity of all two way marginals  by averaging histogram intersections with numerical attributes discretized into bins \citet{afonja2025dp2stageadaptinglanguagemodels} also suggests \textit{pairwise correlation similarty}, which estimates how well the synthetic data preserves pairwise column correlations. 

\citet{castellon2023dptbarttransformerbasedautoregressivemodel} suggested \textit{Kolmogrov-Smirnov test} for numerical attributes and Chi-square test for cateogrical columns. 
Distributional distance metrics like  \textit{Maximum mean discrepancy (MMD)} and \textit{$\alpha-Precision$} can also be used for this purpose and beta-recall  \cite{alaa2022faithfulsyntheticdatasamplelevel} 

Many additional metrics for comparing tabular datasets surfaced in non-dp setting, for example the $\ell_1$-distances of synthetic examples to the closest real datapoint \cite{borisov2023languagemodelsrealistictabular}. However they are hard to properly dp-fy. 

\textbf{Utility}: 
For many practical applications the goal is to use synthetic data for some downstream task, for example training/finetuning an ML model. None of the workload based algorithms we discussed previously directly optimize for this measure, however all the algorithms for query release can be used to create tabular synthetic data in this case, and Equation \ref{eq:zero_sum_synthetic_data} can still be used as a proxy metric that is assumed to be correlated with downstream ML model performance. \textit{This is what commonly happens in practice.}

\footnote{Alternatively, the case of creating synthetic data to maximize downstream ML model utility can be understood as a situation with queries that preserve the excess risk of the solution of empirical risk minimization problems with respect to a class of loss functions. It turns out that the private multiplicative weights update method can be applied in greater generality to address convex minimization queries \cite{UllmanPMW:2015}. This algorithm however suffers the same limitations mentioned for private multiplicative weights for linear queries and is not considered practical. }

\textit{Systematic}: Some papers propose utility evaluations that are systematic, in that they consider fully specified downstream use cases of differentially private synthetic data as the metric itself. For example, \citet{hod2024differentially} perform a national scale data release of sensitive medical data using DP synthesizers; as part of their real world criteria, they considered \textit{faithfulness}. This is a finer granularity metric, requiring that some proportion of the released dataset records ``look like’’ a proportion of the original records (as opposed to just matching moments or aggregate statistics). In particular, they operationalize their metric as ($\alpha, \beta$)-faithfulness. Given a cost function between two records $c : X \times X \rightarrow \mathbb{R}_{\geq 0}$, a dataset $S \in X^n$ is $(\alpha, \beta)$-faithful with respect to a dataset $R \in X^n$ if there exists a bijection $\pi : S \rightarrow R$ such that
\begin{align}
\frac{1}{n} \sum_{i=1}^{n} \mathbb{I}[c(s_i, r_{\pi(i)}) \leq \alpha] \geq \beta,
\end{align}
where $\mathbb{I}$ is the indicator function. In words, at least a $\beta$ proportion of records in $S$ are matched to records in $R$ with a cost no greater than $\alpha$. Additionally, they define maximal-$\beta$-faithfulness of a dataset $S \in X^n$ with respect to a dataset $R \in X^n$ and a cost function $c$ as,
\begin{align}
\beta_{\max}(R, S) = \max_{\pi \text{ matching}} \frac{1}{n} \sum_{i=1}^{n} \mathbb{I}[c(s_i, r_{\pi(i)}) \leq 1],
\end{align}
which is the highest possible proportion of records in $S$ that can be matched to records in $R$ with a cost no greater than 1. 

\citet{rosenblatt2024epistemic} take a broader approach than a single real data release. They consider an \textit{epistemic parity} metric, which evaluates the ability of a differentially private (DP) synthetic dataset to reproduce scientific findings by assessing whether a functionalized version of a statistical test yields the same conclusion when applied to both the original and the DP synthetic data (within some context-specific tolerance $\alpha$ accounting for sampling error and noise introduced by DP). The epistemic parity, or level of agreement, is the proportion of statistical tests that result in the same conclusion when using the real and DP synthetic datasets, and is a measure of how well a DP synthetic data approach works for reproducing the types of statistical analyses present in the paper in question.

Finally, we want to highlight that standard metrics outlined Section \ref{sec-fidelity-generic} remain the mostly commonly used metrics for utility evaluation of DP synthetic tabular data.

\subsection{Open Questions and Challenges}\label{tabular-open-questions}

Despite the significant amount of work in this area, various basic and central questions in private synthetic tabular data generation remain open. 
From a theoretical perspective, the main open problem is the optimal accuracy rates for pure DP algorithms in terms of worst-case error as in \eqref{eq:accuracy_synthetic_data} where $p=\infty$. Currently there is a polynomial in $n$ gap between best upper and lower bounds \cite{HardtPhDThesis:2011,DPorg-open-problem-optimal-query-release}.

Another underexplored direction relates to the computational complexity of producing such data for specific query families. While the hardness result of  \citet{Ullman:2011} indicate that queries corresponding to hard constraint satisfaction problems (CSP) are necessarily hard for synthetic data generation (in fact, this argument implies that producing synthetic data that approximates 2-way marginal queries is hard), it is not known whether easy CSPs are such that producing synthetic data is easy. On the other hand, hardness results on synthetic data are based on cryptographic assumptions (more precisely, the existence of one-way functions), rather than more traditional (computational) hardness of approximation.

Many mechanisms assume that the data lives on a closed domain, where each attribute is categorical and can assume a fixed number of (reasonably small) possible values. While numerical attributes can be nominally supported via discretization, it remains relatively unexplored to handle other data types, like open set categorical attributes, set-valued attributes, and free-form text attributes.

While the techniques described in this section focus on the single-table setting, generating DP synthetic databases according to a factored multi-table schema is a much less explored area. Early work includes PrivLava and PrivPetal \cite{cai2023privlava,cai2025privpetal}.

From a practical perspective, closing the gap in performance between generative models and workload based methods is an important question. Furthermore, assessments of the data quality generated by generative models are still done on an ad-hoc basis, and more systematic methods are needed.

Additionally, we want to emphasize that the problem of generating a collection of tabular synthetic datasets (a setting where privacy unit is a full tabular dataset) is very much unexplored.

\section{DP Synthetic Image Data} \label{sec:image}
Image data is ubiquitous in the modern age, resulting in the very large volume of available training data. However creating privacy-preserving (DP) realistic looking synthetic images remains a challenge even after years of research and several families of models (GANs, Diffusion). Generating high-quality DP synthetic images data faces several significant hurdles:
\begin{compactenum}
    \item \textbf{High Dimensionality of Images:} The high dimensionality of image data poses a major obstacle~\cite{chen:dpgen}. Capturing the intricate distributions of natural images requires complex models. Applying DP noise in such high-dimensional spaces (either pixel space or gradient space during training) can easily overwhelm the signal, leading to poor quality synthetic images~\cite{chen:dpgen}. Techniques to counter this often involve working in lower-dimensional latent spaces \cite{ghosh:dpldm} or using specialized DP mechanisms~\cite{chen:dpgen}.
    \item \textbf{Utility Evaluation Difficulties:} Assessing the quality of DP synthetic data is multifaceted and challenging.
        Standard metrics like FID for image quality might not correlate well with performance on specific downstream tasks~\cite{nasr:eval_dp_synth}. Furthermore, the noise introduced by DP can distort statistical properties, potentially leading to misleading results or false discoveries (e.g., inflated Type I errors) when analyzing the synthetic data~\cite{pubmed:synth_discoveries}.
    \item \textbf{Computational Cost:} Training large-scale generative models like GANs and diffusion models is computationally intensive. DP training, particularly using DP-SGD, often adds significant overhead due to the need for larger batch sizes and per-example gradient clipping, potentially increasing training time substantially~\cite{ghosh:dpldm}. Additionally generative models like GANs and Diffusion are hard to dp-fy properly due to their training complexity, potential use of non-dp-able components (Section \ref{sec-gotchas}) and training process instability.
\end{compactenum}

Nevertheless, the progress in synthetic data synthesis has been short of amazing. Not surprisingly, DP synthetic data synthetic field also demonstrated impressive results recently, which we will explore in this Section.

\paragraph{Roadmap}
First we discuss what a meaningful privacy unit for image data is (Section \ref{dp-image-privacy-unit}). Then we proceed to discuss general approaches for generating DP synthetic data (Section \ref{sec-image-base-methods}). We then focus on 2 prominent methodologies : DP finetuning (Section \ref{sec-dp-training-image}) of GANs (Section \ref{sec-image-gans}) and more recently Diffusion Models (Section \ref{sec-image-diffusion}) and training-free technique called Private Evolution (PE, Section \ref{sec-pe-image}). We also cover methods that don't fit neatly into the above categories in Section \ref{image-alternative-methods}, including PATE-style GANs (Section \ref{sec-dp-pate-image}) and DP-fying images via an intermediary re(Section \ref{sec-image-through-intermediary}).
We then offer a comparative analysis of these approaches (Section \ref{sec-image-comparison}), go over potential metrics that can be used to evaluate the quality of synthetic data (Section \ref{sec-image-quality}) and conclude with a discussion of ongoing challenges and potential future research directions (Section \ref{sec-image-open-questions}).

\subsection{The Privacy Unit for Image Data}\label{dp-image-privacy-unit}
Determining the right privacy unit for image data is an important task. Each image often has a clear and complete semantic meaning making \textit{sub-example-level privacy unit} not appropriate or extremely hard. For example, protecting only people's faces in images might not be sufficient ---  a photo taken in someone’s kitchen may still reveal private information about their home even if faces are blurred. 

Most of the literature treats each individual image as a privacy unit (\textit{example-level privacy unit}). This might be appropriate if each user contributed at most one image to the overall private dataset.

In cases when each user can be associated with more than 1 image, \textit{user-level} privacy unit is a better choice. The notion of a ``user'' however must be chosen carefully. If a user is defined as the creator of an image, each user can be reliably associated with a set of images they contributed to the training dataset. Appropriate single-owner user contribution bounding (Section \ref{user-contrib-bounding}) can then be used to ensure user-level privacy. We might also want to treat someone who directly or indirectly appears in an image (e.g. a person's face, or personal dwelling) as a user associated with the image. In this case, obtaining metadata to assign a set of users to each image may be technically challenging, raise other potential privacy concerns (e.g., the use of facial recognition to cluster images), or even be practically impossible. 
But when such attribution is possible, multi-owner user-contribution bounding techniques will be needed (Section \ref{user-contrib-bounding})

\subsection{Methods for DP Synthetic Image Generation}\label{sec-image-base-methods}
Several paradigms exist for generating synthetic image data that satisfies DP:
\begin{itemize}
    \item \textbf{DP finetuning or training.} DP Training from scratch generative models (e.g. GANs and VAE) using private data or DP finetuning of pretrained models like LLMs or Diffusion using private data are the two most common and powerful approaches for complex data like images. This is achieved via previously discussed DP-SGD methods and its variants. Such training process ensures that the final model parameters are differentially private with respect to the training data. Samples drawn from this trained DP model constitute the synthetic dataset. We cover these methods in Section \ref{sec-dp-training-image}.
     
    \item \textbf{Methods that utilize API-only access of generative models.}
    This is another group of methods that foregos expensive DP-Training and instead leverage pre-existing large foundation models accessible via APIs. DP is introduced during the stage of choosing the best examples generated by public models to best match private data distribution, 
    Private evolution (PE) is one prominent method of this class. This algorithm queries ``generation'' and ``variation'' APIs iteratively, using a DP selection mechanism to guide the process towards the private data distribution~\cite{lin:pe_images_pdf}. We cover this algorithm in Section \ref{sec-pe-image}.
    
    \item \textbf{Private prediction.}
    This group of methods uses ideas of Private Prediction to train GANs models without doing an end-to-end DP-Training with DP-SGD algorithms. We explore PATE-style algorithms for image generation in Section \ref{sec-dp-pate-image}

     \item \textbf{Creating DP synthetic image data via an intermediary text representation}. These types of methods change the modality of the data from image to text via captions and then proceed to create DP synthetic captions using methods for DP synthetic text data. Synthetic images are then created using dp-fied captions. We cover this new but promising class of methods in Section \ref{sec-image-through-intermediary}
    
    \item \textbf{DP data release mechanisms.} This class of approaches  have not been particularly successful.These approaches, instead of training or finetuning a complex generative model privately with DP-Training, focus on releasing certain statistics or intermediate representations of the private data in a DP manner. Synthetic data is then generated based on these sanitized statistics. Examples include:
    \begin{compactenum}
        \item DPGEN \cite{chen:dpgen} avoids DP-SGD entirely and instead privatizes training images themselves by deriving DP-randomized recovery directions. This data is used to train (without DP) an energy network \citet{dockhorn2023differentiallyprivatediffusionmodels} demonstrated that the final trained model was not actually DP and outlined multiple sources of privacy leakage.
        \item DP-DRE \cite{mao:dpdre} which utilizes a publicly pretrained encoder and ICGAN generator. Private data is embedded via an encoder and the distribution is DP-fied via DP density estimator. Samples are then drawn from this distribution and decoded via ICGAN \cite{liu2024icganimplicitconditioningmethod}
        \item DP-MERF models \cite{DBLP:journals/corr/abs-2002-11603} use random feature representations of kernel mean embeddings and learns synthetic data represenation with Maximum Mean discrepancy objective, that pulls the embeddings of synthetic and private data together. Authors "release" mean private data embeddings' with DP and use this representation in closed-form calculation for MMD.

    \end{compactenum}

\end{itemize}

We will next focus on the first four groups of approaches in the upcoming sections.

\subsection{DP Training/Finetuning} \label{sec-dp-training-image}
DP training (from scratch) or finetuning (existing pre-trained models) are currently the most common and prominent approaches for synthetic image data generation. A generative model (e.g., GAN, VAE, Diffusion Model, LLM) is trained or finetuned on the private dataset using an optimization algorithm that incorporates DP, most notably Differentially Private Stochastic Gradient Descent (DP-SGD) and its variants (Section \ref{sec-dpsgd}). As we have seen before, such training process ensures that the final model parameters are differentially private with respect to the training data. Samples drawn from this trained DP model constitute the synthetic dataset.

\subsubsection{GAN-Based Models}\label{sec-image-gans}
Early attempts at generating DP synthetic images used Generative Adversarial Networks (GANs). GAN training involves a generator $G$ trying to fool a discriminator $D$, which tries to distinguish real images from generated ones. The most common method for creating DP-GANs involves integrating Differentially Private Stochastic Gradient Descent (DP-SGD) into the standard GAN training loop~\cite{dpgans} (DP-Training).
Often DP is primarily applied to the discriminator $D$ as it directly processes the real private data, although variants that also apply DP to generator exist (e.g. \cite{chen:gswgan}).

DP-Training GANs however presents significant challenges:
Firstly, standard GAN training is notoriously sensitive to hyperparameters and prone to instability, such as mode collapse \cite{DBLP:journals/corr/abs-1807-04015}. The introduction of noise and gradient clipping via DP-SGD often exacerbates these issues, making it even harder to achieve stable convergence and produce high-quality results~\cite{chen:dpgen}: the already delicate balance between the generator and discriminator is easily disrupted by the DP modifications.

The inherent difficulty of stabilizing GAN training under the noisy and biased gradient conditions imposed by DP-SGD has been a major driver for research in this area. Many subsequent approaches have sought to either refine the application of DP-SGD within the GAN framework or to bypass the standard GAN/DP-SGD combination entirely, seeking more compatible pairings of generative models and DP mechanisms.

Several strategies have been explored to improve DP-GAN  \cite{dpgans} performance and address the aforementioned challenges. \citet{torkzadehmahani2019dp} introduce a framework for training Conditional Generative Adversarial Networks (CGAN) while providing DP guarantees. The authors' primary contribution is a gradient clipping and perturbation strategy that treats the gradients from real and fake data separately within the discriminator, which they demonstrate improves the quality and utility of the generated synthetic data and labels compared to previous DP approaches.

\citet{chen:gswgan} point out that since the generator is the only ``released'' object, DP should be applied only to gradients of the generator, reducing the total noise injected during training. Additionally, their method GS-WGAN \cite{chen:gswgan}  modifies how gradients are processed before noise addition. Instead of relying on clipping arbitrary large gradients to limit their sensitivity (which is needed to scale the noise added to the gradients), an alternative loss (Wasserstein-1 loss) that generates bounded gradients with norms near 1 is used. This allowed authors to forego expensive hyperparameter search for the appropriate clipping norm. Additionally, since gradients were closer to the clipping norm, gradient distortion from DP was much less significant.

\citet{bie2023private} reexamines the DPGAN formulation of \citet{dpgans} where only the discriminator is trained with DP-SGD, and find that taking many discriminator steps between consecutive generator steps (e.g. >50 in some cases) successfully addresses the imbalance between the generator and discriminator, leading to large improvements in quality. With this modification, \citet{bie2023private} report that a well-tuned application of DP-SGD to the GAN discriminator outperforms all alternative GAN privatization schemes.

Recognizing the difficulty of learning complex image distributions from scratch under DP-Training, some methods incorporate public data (DP-finetuning of publicly pre-trained models). 
\textit{DP-GAN-MI} \cite{DBLP:journals/corr/abs-2201-03139} uses a pre-trained feature extractor but trains a standard DP-GAN directly in the feature space to model the private feature distribution. 
DP-DRE \cite{mao:dpdre} first trains an autoencoder (specifically, an Invertible Conditional GAN or IC-GAN) non-privately on public data. Then, both private and public images are mapped to the IC-GAN's feature space. A DP mechanism (specifically, training an auxiliary GAN as a density ratio estimator with DP-SGD) is used to learn the difference between the private and public feature distributions. To generate, features are sampled according to the learned DP ratio from the public feature distribution, and then decoded back to images using the IC-GAN's generator. This also avoids direct DP-GAN training in the high-dimensional pixel space and leverages the structure learned from public data~\cite{mao:dpdre}.
DP-DRE was reported to perform better than DP-GAN-MI~\cite{mao:dpdre}.

Several works avoid adversarial objective of standard GANs altogether, that is hard for DP-Training and finetuning and explore alternative losses than Wasserstein distance. 
\textit{DP-Sinkhorn \cite{gao:dpsinkhorn}:} This method use an optimal transport (OT) based loss, specifically the Sinkhorn divergence, measuring the distance between the real and generated data distributions. The generator is trained to minimize this divergence. The approach relies on a new semi-debiased Sinkhorn loss to better control the bias-variance trade-off inherent in estimating gradients under DP constraints. By avoiding the adversarial objective, DP-Sinkhorn aims for more stable training~\cite{gao:dpsinkhorn}.
        
DP-GANs generally exhibit a pronounced privacy-utility trade-off. While progress has been made, achieving high visual fidelity, diversity, and resolution comparable to non-private state-of-the-art GANs, especially under strong privacy guarantees (low $\eps$), remains a significant challenge. Methods like DP-Sinkhorn \cite{gao:dpsinkhorn} reported improvements over earlier DP-SGD-based GANs. Techniques that effectively utilize public data, like DP-DRE \cite{mao:dpdre} can also boost performance by reducing the learning burden on the private data under DP constraints.

Refer to Table \ref{dp-gan-comparison} for a brief comparison between GAN-based and alternative methods.

{\footnotesize
\begin{longtable}[htbp]{p{2cm} p{1cm} p{3cm} p{4cm} p{3cm} p{0.5cm}}
\caption{Comparison of Selected DP-GAN and GAN alternatives for Image Synthesis \label{dp-gan-comparison}. All methods ensure approximate $(\eps,\del)$ DP.
} \\
\toprule
\textbf{Method} & \textbf{Core Model} & \textbf{DP Mechanism} & \textbf{Advantages} & \textbf{ Challenges}  \\
\midrule
DP-GAN \cite{dpgans} & GAN & DP-SGD on Discriminator (or both) & Foundational approach & Instability, Low Resolution, Poor Utility   \\
\hline
GS-WGAN \cite{chen:gswgan} & GAN (WGAN) & DP-SGD on ensemble image gradients & Noise on lower-dim image gradient; ensemble stability & Still relies on DP-SGD dynamics    \\ 
\hline
DP-DRE \cite{mao:dpdre} & Autoenc. (ICGAN) + Aux GAN & DP-SGD on auxiliary GAN in feature space & Leverages public data; avoids DP-GAN in pixel space; learns density ratio & Requires good autoencoder \& public data  \\
\hline
DP-Sinkhorn \cite{gao:dpsinkhorn}& Generator (OT-based) & DP noise on Sinkhorn loss gradients w.r.t. samples & Replaces adversarial loss with OT; improved stability; semi-debiased loss for DP gradients & OT computation complexity    \\
\bottomrule

\end{longtable} }

\subsubsection{Diffusion Based Models}\label{sec-image-diffusion}
Diffusion probabilistic models have recently emerged as the state-of-the-art for high-fidelity image generation~\cite{ghosh:dpldm}. Naturally, significant effort has been directed towards applying differential privacy to these models to enable the generation of private synthetic images. Diffusion models work by gradually adding noise to data in a "forward process" and then learning a model to reverse this process, starting from pure noise and iteratively denoising it to generate a sample~\cite{ghosh:dpldm}.

Similar to DP-GANs, the predominant technique for achieving DP in diffusion models is DP-Training/Finetuning via DP-SGD~\cite{ghosh:dpldm}.
DP-SGD is applied during the training phase where the model (typically a UNet-like architecture \cite{ghosh:dpldm}) learns to predict the noise added at each diffusion time step $t$, or equivalently, predict the less noisy image $x_{t-1}$ or the original image $x_0$. 
Commonly used loss for diffusion model training is mean squared error between the predicted noise and the actual noise added.

To potentially improve utility under DP-SGD, techniques like "noise multiplicity" (averaging gradients over predictions made with different noise samples added to the same image at time $t$) \cite{ghosh:dpldm} or "augmentation multiplicity" (averaging the loss over multiple random augmentations of an image before computing the gradient) \cite{park:cvpr2024} have been proposed.

\paragraph{DP-training vs DP-finetuning}
 The initial attempts at DP Diffusion models involved training (from scratch) a diffusion model on the private dataset from random initialization using DP-SGD throughout the process (e.g., DPDM \cite{ghosh:dpldm},  DPLDM\cite{ghosh:dpldm}). While feasible, this can be computationally very expensive and often results in lower utility compared to methods using pre-training, as the model has to learn complex image features entirely under the constraints of DP noise~\cite{ghosh:dpldm}. Therefore DP-finetuning of a pretrained Diffusion models has become a dominant and highly effective paradigm~\cite{ghosh:dpldm}. \textit{DP-Diffusion} is a diffusion model that is first pre-trained on a large, publicly available dataset (like ImageNet) without privacy constraints. This allows the model to learn general visual features and structures. Subsequently, this pre-trained model is fine-tuned on the target private dataset using DP-SGD for a smaller number of steps~\cite{ghosh:dpldm}. Since the model starts from a strong initialization, the fine-tuning process requires less learning from the private data, thus accumulating less privacy loss and allowing for better utility for a given budget $\eps$~\cite{ghosh:dpldm}.
 \textit{Privimage \cite{zhu:privimage}} refines this idea by adding a preliminary step: it uses DP queries (based on semantic similarity using CLIP embeddings) to select a small, relevant subset from the large public dataset that semantically aligns with the private data. Pre-training is then performed only on this selected subset before DP fine-tuning. This aims to make the pre-training more targeted and efficient, potentially leading to better results with fewer parameters and less computation compared to pre-training on the entire public dataset~\cite{zhu:privimage}. \textit{DP-RandP \cite{lee:dprandp}} investigates pre-training on synthetically generated random data (e.g., images from random processes) before DP fine-tuning on the private data, offering an alternative when large relevant public datasets are unavailable.
 When only a small amount of in-distribution public data is available (e.g., a small public subset of a larger private dataset), work by \citet{park:cvpr2024} proposes to use a diffusion model trained on this initial small public dataset to synthesize more public-like data. This amplified public dataset is then used for more effective warm-up training before applying DP-SGD fine-tuning on the private data. This addresses the challenge of leveraging public data when large, diverse external datasets are not suitable or available~\cite{park:cvpr2024}.

    \textbf{Efficiency Improvements via Latent Diffusion (DP-LDM):} To tackle the computational cost and high dimensionality of pixel-space diffusion models, DP-LDM \cite{ghosh:dpldm} operates on Latent Diffusion Models (LDMs). LDMs use a pre-trained autoencoder to map images to a lower-dimensional latent space, where the diffusion process takes place~\cite{ghosh:dpldm}. DP-LDM applies DP-SGD during the fine-tuning stage only within this latent space. Furthermore, it drastically reduces the number of parameters updated with DP-SGD (by $\sim$90\%) by fine-tuning only specific components critical for adaptation and conditional generation, namely the attention modules within the UNet backbone and the conditioning embedders (if applicable), while keeping the rest of the UNet frozen~\cite{ghosh:dpldm}. This targeted approach significantly reduces computational cost and has been shown to achieve a better privacy-utility trade-off~\cite{ghosh:dpldm}.
     \paragraph{Leveraging Inherent Diffusion Noise (dp-promise):} This method \cite{wang:dppromise} explicitly leverages the fact that the forward diffusion process itself injects Gaussian noise, which contributes to privacy (satisfying Gaussian DP). The intution behind the method is that during forward diffusion pass, later iterations of images are very close to the noise already, whereas earlier iterations are much more sensitive and closer to the private data. Accordingly, the paper proposes to modify the training that happens during reverse diffusion propose. Authors introduce a two-phase training strategy. Phase I trains the model non-privately on the later, noisier diffusion steps (from time $S$ to $T$), relying on the inherent DM noise for a partial DP guarantee. The gradients from this stage are already noised enough from the noise introduced during the forward pass, and sensitivity bound is derived based on image dimensions. 
     Phase II trains the model using DP-SGD (clipping and injection of the noise) only on the earlier, less noisy steps (from time 1 to $S-1$), where additional privacy protection is needed. By avoiding redundant DP-SGD noise injection in the later stages, dp-promise aims to improve the overall privacy-utility balance~\cite{wang:dppromise}.
     
  \paragraph{Discussion}
The success of many state-of-the-art DP diffusion methods underscores the critical importance of leveraging external information, typically through large public datasets used for pre-training~\cite{ghosh:dpldm}. Pre-training allows the model to learn powerful, general-purpose visual representations without consuming any private budget~\cite{ghosh:dpldm}. The subsequent DP fine-tuning phase then specializes the model to the private data distribution, requiring fewer updates and thus less noise injection compared to training from scratch~\cite{park:cvpr2024}. The relevance and quality of the public data can significantly impact the effectiveness of transfer learning~\cite{zhu:privimage}. This heavy reliance suggests that the vast knowledge encoded in models pre-trained on public data is currently almost essential to counteract the information loss imposed by DP constraints when learning from limited private data, especially for complex tasks like high-fidelity image generation. Achieving high utility purely from private data under strong DP remains largely impractical with current diffusion techniques.

DP diffusion models employing pre-training and fine-tuning strategies like DP-Diffusion and DP-LDM have demonstrated significant progress, achieving state-of-the-art results in DP image synthesis~\cite{ghosh:dpldm}. They often outperform earlier DP-GAN methods in terms of image fidelity (FID scores) and the utility of generated data for downstream tasks (e.g., classification accuracy)~\cite{ghosh:dpldm}. Techniques like DP-LDM also offer substantial improvements in computational efficiency by operating in latent space and fine-tuning only a fraction of the parameters~\cite{ghosh:dpldm}.

Innovations in DP diffusion are increasingly moving beyond simply applying DP-finetuning with DP-SGD as a black box. Instead, they focus on how and where to integrate DP most effectively within the diffusion model framework. This includes operating in latent spaces (DP-LDM \cite{ghosh:dpldm}), selectively fine-tuning parameters (DP-LDM \cite{ghosh:dpldm}), adapting the DP mechanism based on the diffusion time step (dp-promise \cite{wang:dppromise}), and optimizing the use of auxiliary data (Privimage \cite{zhu:privimage}, Park et al. \cite{park:cvpr2024}). This trend signifies a shift towards more model-aware and efficient DP techniques tailored to the specific characteristics of diffusion models.

It is worth pointing out recent success of autoregressive models for image generation in non-DP setting \cite{sun2024autoregressivemodelbeatsdiffusion}. Visual autoregressive models (VARs) apply next-token-prediction paradigm that dominates text domain to vision domain. These models work by training special image tokenizers that generate sequence of image tokens that are then used for training transformer-based models. During generation, sequence of image tokens is generated token-by-token and subsequently decoded into the image by a trained decoder. Early attempts of adopting this paradigm for DP image synthesis \cite{shaikh2025implementingadaptationsvisionautoregressive} however significantly underperform DP-diffusion based models that have been extensively researched and iterated upon in DP setting.

{\footnotesize
\begin{longtable}[htbp]{p{2cm} p{2cm} p{3cm} p{4cm} p{3cm} } 
\caption{Comparison of Selected DP-Diffusion Approaches for Image Synthesis \label{tab:dp_diffusion_comparison}. All methods ensure approximate $(\eps,\del)$ DP.
} \\
\toprule
\textbf{Method Name} & \textbf{Core Technique} & \textbf{DP Mechanism} & \textbf{Advantages} & \textbf{Challenges}  \\
\midrule
DPDM \cite{ghosh:dpldm}& Train from Scratch & DP-SGD & Foundational DP-Diffusion; Noise Multiplicity & High Compute Cost, Lower Utility  \\ 
\hline
DP-Diffusion (Generic) \cite{dockhorn2023differentiallyprivatediffusionmodels}  & Pre-train (Public) / Fine-tune (Private) & DP-SGD during Fine-tuning & Leverages public pre-training for better utility/efficiency & Requires large public dataset; still costly \\ 
\hline
Privimage \cite{zhu:privimage}& Select Public Data / Pre-train / Fine-tune & DP Query + DP-SGD Fine-tuning & Targeted public data selection for efficient pre-training; reduced model size & DP query overhead; relies on embedding   \\ 
\hline
DP-Latent Diffusion Model (LDM) \cite{ghosh:dpldm} & Pre-train LDM (Public) / Fine-tune Attention (Private) & DP-SGD on Attention in Latent Space & Operates in latent space; tunes few parameters ($\sim$10\%); high efficiency \& utility & Requires LDM architecture  \\ 
\hline
dp-promise \cite{wang:dppromise} & Two-Phase Training (DM Noise + DP-SGD) & Partial DP-SGD + Inherent DM Noise & Leverages DM noise for DP guarantee; reduces redundant DP-SGD noise injection & Complex training schedule; choice of S  \\ 
\hline
Park et al. \cite{park:cvpr2024}   & Synthesize Public Data / Warm-up / Fine-tune & DP-SGD during Fine-tuning & Amplifies limited public data for better warm-up; addresses in-distribution public data & Extra step of public data synthesis \\ 
\bottomrule
\end{longtable}
}

\subsection{Methods That Avoid DP-Training: Private Evolution}\label{sec-pe-image}
As we have seen in the previous section, DP training or finetuning models like GANs or Diffusion models is extremely computationally expensive - DP-SGD algorithms require much larger batches, expensive per example clipping and potentially many more iterations to converge \cite{Ponomareva_2023}. Avoiding DP finetuning is thus a promising avenue that has been explored in a method called \textit{Private evolution (PE)} introduced by \citet{lin:pe_images_pdf}, which foregoes tuning of any models completely and relies on existing foundational models' APIs. DP is introduced during the selection stage when best synthetic examples generated by public models are chosen to match private data distribution.

The central idea of PE is to leverage the generative capabilities of powerful, potentially proprietary or computationally expensive-to-train foundation models (like Stable Diffusion) as black boxes~\cite{lin:pe_images_pdf}. Instead of training these models with DP, PE uses an evolutionary algorithm guided by DP principles to iteratively refine a population of synthetic samples steering them towards the distribution of the private data using only API calls for generation and variation~\cite{lin:pe_images_pdf}. The fact that the approach is inference-only simplifies deployment and allows capitalizing on state-of-the-art models that might only be accessible via APIs~\cite{lin:pe_images_pdf}. Additionally, private data is never directly input into the generative model - this means that the API provider can be potentially untrusted \cite{lin:pe_images_pdf}.
This new paradigm makes advanced DP synthesis potentially more accessible, bypassing the need for direct model access, gradient computation, or extensive private training infrastructure, provided suitable APIs are available.

PE relies on two APIs, namely \textit{random api} and \textit{variation api}, that are often directly available for popular models (e.g. DALL-E \cite{openai_dalle}, Stable Diffusion \cite{rombach2021highresolution}) or can be implemented with appropriate prompt engineering (e.g. GPT \cite{chen2020generative}). The \textit{random api} generates random samples (which can be conditioned with text prompts for some models), while \textit{variation api} creates variants of a given sample by generating images similar to the provided image.

The PE algorithm, presented in Algorithm \ref{alg:image_pe}, operates iteratively as follows \cite{lin:pe_images_pdf}:

\begin{algorithm}[tb]
    \caption{\label{alg:image_pe} Private Evolution for Image (PE,  \citet{lin:pe_images_pdf})} 
    \begin{algorithmic}[1] 
        \Require Private data $D$, image embedding model  $\Phi$ (e.g. CLIP, Inception), target number of synthetic samples $N$, evolution rounds $T$, population size multiplier $L$ (number of variations for each synthetic image).
        \Ensure Synthetic dataset $\hat{D}$.
        \State Initialize $\hat{D}_{0}$ of size $L\cdot N$ with Random API.
        \For{$t = 0, \dots, T-1$}
            \State $E_{t}=\Phi(\hat{D}_{t})$ \textsl{\scriptsize //Embedding calculation for  synthetic samples.}
            \State Let each $\Phi(z), z\in D$ vote for the nearest embedding in $E_{t}$.
            \State Privatize the voting results with $(\epsilon, \delta)$-DP to get a DP histogram $H_{t}$.
            \State $\hat{H}_{t}$ = $H_{t}/sum(H_{t})$ \textsl{\scriptsize //Histogram normalization.}  
            \State Get $\hat{D}_{t}^{'}$: sample $N$ images with replacement from $\hat{D}_{t}$ proprotionally to $\hat{H}_{t}$.
            \If{$t<T-1$}
                \State Get $\hat{D}_{t+1}$ of size $L\times N$:  call Variation API to get $L$ variants for each $z\in \hat{D}_{t}^{'}$.
            \Else
                \State \Return $\hat{D}_{t}^{'}$.
            \EndIf
        \EndFor
    \end{algorithmic}
\end{algorithm}

\paragraph{Initialization}
A starting population of synthetic samples, $\hat{D}_{0}$, is created, potentially using a generic generation API of the chosen foundation model. Alternatively, if available, a public dataset that is similar to the sensitive data distribution can be employed as the intial pool.

\paragraph{Iterative Refinement} 
Once the initial pool of image data is created, the algorithm refines it through a repeated series of voting and variation steps.
    \begin{compactenum}
        
        \item \textit{Embedding:} Both the private dataset $D=\{x_1,\dots,x_n\}$ and the current synthetic population $\hat{D_{t}}=\{z_1,\dots,z_m\}$ are mapped into a common embedding space using a pre-trained embedding function $\Phi$ (e.g., Inception network, CLIP embeddings). This allows for semantic comparison~\cite{lin:pe_images_pdf}.
        \item \textit{Selection (Parent Selection via DP Nearest Neighbors Histogram):}
        \begin{compactenum}
            \item \textbf{Nearest Neighbor Identification:} For each private sample $x_i \in X$, its nearest neighbors $z_{\text{NN}(i)}$ is identified within the current synthetic population $\hat{D_{t}}$ based on the $\ell_2$ distance in the embedding space: $d(x_i, z_j) = \Norm{\Phi(x_i) - \Phi(z_j)}_2$ 
            \item \textbf{Histogram Construction:} A histogram is built over the synthetic samples in $\hat{D_{t}}$. The count for each synthetic sample $z_j$ is the number of private samples $x_i$ for which $z_j$ is the nearest neighbor: $\text{Count}(z_j) = |\{i | z_j = z_{\text{NN}(i)}\}|$ 
            
            To ensure differential privacy, noise is added to these counts. 
            Specifically, i.i.d. Gaussian noise $\Ncal(0, \sigma^2)$ is added to each $\text{Count}(z_j)$, where $\sigma$ is a noise multiplier related to the desired privacy budget and $H$ is a threshold parameter. The sensitivity is 1 since each private example gets only 1 vote.
            The noisy counts are optionally thresholded at $H$. Let the resulting noisy, thresholded count for $z_j$ be $\widetilde{\text{Count}}(z_j)$. This DP histogram mechanism provides an $(\eps,\del)$-DP guarantee for the release of the histogram information in each iteration
            \item \textbf{Parent Sampling:} A new set $\hat{D}_{t}^{'}$ of $N$ ``parent'' samples is selected from $\hat{D_{t}}$ by sampling with replacement, where the probability of selecting $z_j$ is proportional to its noisy, thresholded count $\widetilde{\text{Count}}(z_j)$. Samples deemed more similar to the private data (according to the DP histogram) are more likely to be chosen as parents.
        \end{compactenum}
        \item \textit{Variation (Offspring Generation):}
        
        For each selected parent sample $z_p$ from $\hat{D}_{t}^{'}$, a "variation API" of the foundation model is called to generate $L$ "offspring" samples. This API is designed to produce samples that are similar to the input $z_p$ but introduce some diversity or modification (e.g., using image-to-image generation with slight changes, varying guidance scale in diffusion models)~\cite{lin:pe_images_pdf}.
            \item The collection of all generated offspring forms the new synthetic population $D_{t+1}$ for the next iteration.
    \end{compactenum}
    \paragraph{Final Output:} After $T$ iterations, the final population $D_{T-1}$ (or a subset thereof) constitutes the DP synthetic dataset. The overall privacy cost $(\eps_{\text{total}}, \del_{\text{total}})$ is calculated by composing the privacy costs of the $T$ DP histogram releases using advanced composition theorems~\cite{lin:pe_images_pdf}.

The choice of the initial set, the distance metric and the variation API are crucial. Using $\ell_2$ distance in a semantic embedding space (like CLIP~\cite{CLIP}) allows comparison based on meaning rather than superficial pixel differences~\cite{lin:pe_images_pdf}. The paper also proposes a "lookahead" distance metric modification, where the distance for selecting $z_j$ is computed between $\Phi(x_i)$ and the mean embedding of $k$ potential variations of synthetic sample $z_j$ generated by the variation API ~\cite{lin:pe_images_pdf}:
\begin{equation}
d(x_i, z_j) = \Norm{\Phi(x_i) - \frac{1}{k} \sum_{l=1}^k \Phi(\text{VARIATION\_API}(z_j)_l)}_2    
\end{equation}
This heuristic attempts to select parents that are likely to produce good offspring in the next generation, potentially accelerating convergence by anticipating the evolutionary dynamics. The effectiveness of PE inherently depends on the quality of these components: the embedding must capture relevant features, and the variation API must allow effective exploration around promising regions of the sample space~\cite{xie2024differentially}.

\paragraph{User-level privacy unit adjustments}
The original algorithm can realize the user-level privacy unit by normalizing the private votes from samples from the same user to sum to 1 (in $\ell_2$ norm). This ensures the sensitivity w.r.t one privacy unit (user) remains one and that the noise is scaled appropriately.

\paragraph{Discussion}
The initial PE paper reported compelling results for image generation \cite{lin:pe_images_pdf}. On benchmarks like CIFAR10 (using ImageNet as the implicit public data via the foundation model), PE reportedly achieved FID scores comparable to or better than state-of-the-art training-based DP methods (like DP-Diffusion, DP-MEPF) but with significantly smaller privacy budgets (lower $\eps$)~\cite{lin:pe_images_pdf}. For example, an FID score below 7.9 was achieved with $\eps=0.67$, whereas prior work required $\eps$ values around 32 for similar FID levels 
Further, PE was successful at generating high-resolution (512x512) images using Stable Diffusion for inference, even when the private dataset was small (e.g., 100 cat images)~\cite{lin:pe_images_pdf}. This is a challenging setting where training large models from scratch or fine-tuning with DP might be difficult due to data scarcity and computational cost. 
Additionally, PE showed limited adaptability to distribution shift between the foundation model's implicit training data (e.g., ImageNet) and the private target data (e.g., Camelyon17 medical images). While downstream classification accuracy on Camelyon17 (79.56\% at $\eps \approx 7.6$) , as expected, lagged behind a specialized DP fine-tuning approach (91.1\% at $\eps=10$) in this data-rich scenario, the result was still non-trivial, indicating the framework's potential adaptability~\cite{lin:pe_images_pdf}.
Finally, compared to DP fine-tuning methods which can require hundreds or thousands of GPU hours and access to checkpoint weights \cite{ghosh:dpldm}, PE's training-free nature can make it more computationally efficient, depending on the cost and latency of API calls~\cite{ghosh:dpldm}.

The downsides of PE is that its performance depends heavily on the quality of the foundation model APIs (generation, variation, embedding)~\cite{xie2024differentially}. Since those apis can be parameterized by prompts in text-to-image models, extensive prompt engineering must be needed. Additionally, while PE can handle some distributional difference between pretraining data for the foundational models and private data to be mimicked, large distributional shifts will result in poor quality synthetic image data. \citet{lin2025differentially} demonstrated that instead of using foundational models (and relying on private data to be of similar distribution to their pretraining data), 
PE can use \textit{synthesizers that do not rely on neural networks} like graphics-based image generation or physics-based robotics simulators. Such synthesizers can be further combined with foundational models by using simulators during early algorithm iterations to create diverse pool of samples and then using foundation model APIs to enhance quality of the final synthetic images. For situations involving large distributional shifts between pretraining and private data in domains where good image simulators exist, Sim-PE offers a significant boost (improving the accuracy x3 and reducing FID by 80\% at $\epsilon=10$) over the original PE algorithm.

\subsection{Alternative Methods}\label{image-alternative-methods}
We want to briefly mention two additional lines of work that does not fit neatly into DP-Training or Training-Free categories we analyzed so far. 

\subsubsection{PATE-style Models}\label{sec-dp-pate-image}
PATE (Private Aggregation of Teacher Ensembles) 
\cite{papernot2017semi,papernot2018scalable}~is an example of introducing DP at the prediction stage, in contrast to DP-Training that alters training process of the models. 
The gist of PATE is to split the private data into a number of disjoint datasets, train a number of non private models and introduce the noise to their aggregated prediction proportional to the level of agreement of the submodels. 

PATE-style GAN models still train a GAN model, but avoid DP-SGD and instead privatize predictions or gradients of the discriminator using ideas from PATE. \footnote{PATE-style GANs models sit in an awkward place, where one can consider them doing DP-Training with an alternative way/place of introducing the DP.} 

A PATE-GAN model \cite{yoon2018pategan} was initially introduced in context of tabular data generation but subsequently was adopted and used as baselines for a number of DP image synthetic data generation papers. PATE part (DP-inference) is applied to obtaining a DP version of a discriminator: private data is partitioned into a number of disjoint subsets of data, a number of discrimintor teacher models are trained on those subsets, each trained with a goal of improving their loss with respect to the generator $G$. To classify each new sample as either real or synthetic, all teacher models vote on the sample and their output is noisily aggregated. However, since discriminator and generator in GANs settings need to be trained together, a discriminator should be differentiable, which PATE style discriminator isn't. Instead, this collection of submodels is distilled into a student model $S$, which can be differentiated over. Student is trained by taking some public unlabelled data and using prediction from PATE ensemble as pseudo labels. Alternatively when public data is not available, student is trained using the data generated by the Generator and pseudo-labels from teachers. Generator $G$ and student $S$ are co-trained, where generator is trying to fool the student, student is trying to improve its loss w.r.t teachers and teachers are training to improve their loss wrt generator. 

G-PATE \cite{DBLP:journals/corr/abs-1906-09338} argues that making a discriminator differentially private is unnessary - as long the gradients of the discriminator used to update the generator are DP, the whole process is DP. 
Additionally, discriminator can be trained on the private data directly without relying on public data or using samples from the generator itself, as in PATE-GAN. 
Consequently, G-PATE does not dp-fy aggregated predictions of teacher discriminators and foregoes a student discriminator completely. A gradient aggregator gets gradients from teachers and accumulates them in PATE-style before passing this private gradient to generator for update. With such scheme G-PATE demonstrated impressive ability to generate high dimensional image data with high utility with very low privacy budgets ($\epsilon \le 1)$

DataLens \cite{DBLP:journals/corr/abs-2103-11109} follows the similar idea and, instead of using random projection of gradients before introducing the noise, as in G-PATE \cite{DBLP:journals/corr/abs-1906-09338},  authors employ dimensionality compression that allows to further reduce the noise needed for private aggregation. PATE-TripleGAN takes this solution one step further \cite{jiang2024patetripleganprivacypreservingimagesynthesis} and introduces modifications that allow to use unlabeled private data during GANs training. Namely, they introduce a classifier to pre-classify unlabeled private data.

It is worth mentioning that PATE style algorithms introduced computational and architectural complexity that hinders their adoption in real world and academic research. Additionally, \citet{ganev2025elusivepursuitreproducingpategan} highlight that increased system complexity of many PATE-GAN style implementations leads to implementation bugs that weaken/invalidate privacy guarantees and result in more than expected privacy leakage. In terms of utility, \citet{bie2023private} report significantly better results with DP-Training of GANs compared to PATE-style training of GANs. 

\subsubsection{Methods That Use an Intermediary Representation}\label{sec-image-through-intermediary}
Two concurrent works recently proposed creating DP synthetic images via first creating text captions of private images, then DP-fying these captions either via DP-finetuning (Section \ref{sec-dp-finetuning}) \cite{syntheticblogpost2025} or Private Evolution (Section \ref{private-evolution-text}) \cite{wang2025synthesizeprivacypreservinghighresolutionimages}, and subsequently creating images based on the DP-captions. \citet{syntheticblogpost2025} used standard DP-finetuning, but in a hierarchical manner: they first trained a model to generate privatized album descriptions, and then trained a model to create privatized photos descriptions given the DP-fied album descriptions. The hierarchical generation strategy ensured thematic and character consistency within each album, while circumventing the context window limitations that would be met if large photo albums were modeled in one go. Instead of DP fine-tuning, \citet{wang2025synthesizeprivacypreservinghighresolutionimages} adopted the Private Evolution method. A key challenge with PE is that caption embeddings do not always reflect similarity in the resulting images. To address this, they proposed using embeddings of the generated images during the voting process, albeit at the cost of increased computation.

This promising family of approaches has proven powerful in photo generation settings. Using text as an intermediate representation leverages the main strength of a large-language model. Text is a rich representation that is however much less dimensional than high-quality images \cite{wang2025synthesizeprivacypreservinghighresolutionimages}

The downside of using text as intermediary is increased complexity and compute requirements stemming from reliance on two additional models - one that creates captions and the one that renders images given captions. Additionally, not all image data can be easily captionized - e.g. screenshots of apps on a phone or a monitor and
images with text are notoriously hard to describe and generate based on descriptions.

\subsection{Comparison of the Methods}\label{sec-image-comparison}
We have previously outlined the main methods for creating DP synthetic text data: via DP finetuning of GANs (Section \ref{sec-image-gans}) and Diffusion models (Section \ref{sec-image-diffusion} and DP-training free methods: PATE-style (Section \ref{sec-dp-pate-image}) and private evolution (Section \ref{sec-pe-image}). Each of the aforementioned methods have their strengths and weaknesses, as well as different requirements. Table \ref{tab:compare-image-methods} compares various aspects of each method. 

\begin{specialistbox}{Choosing the method for DP synthetic text data generation}
For large enough volume of sensitive data (>XXK datapoints) with enough compute the method that results in highest fidelity and utility of DP synthetic data is almost always DP-Finetuning of pretrained Diffusion models. Less computationally expensive methods like PE that don't require finetuning can provide reasonable data when sensitive data is somewhat in distribution for the pretraining data or for situations when were strict privacy guarantees (low $\epsilon)$ are needed.
\end{specialistbox}

\begin{compactitem}
    \item \textbf{DP finetuning} of generative models like GANs and Diffusion remains the workhorse method that delivers the best quality  given sufficient amount data, compute and engineering and time investment. 
    However, both GANs and Diffusion models are though hard to DP-finetune, both for computational reasons and due to the nature of the models. For DP-Diffusion, access to relevant large public datasets for pre-training (or access to a pre-trained checkpoint) is often highly beneficial. Diffusion models are more computationally expensive but are likely to provide better utility of synthetic data (e.g current state-of-the-art DP-LDM \cite{ghosh:dpldm}) than GANs based ones. 
    DP-GANs have historically faced more significant utility degradation, although alternative frameworks like DP-Sinkhorn have shown promise in improving quality and stability~\cite{gao:dpsinkhorn}.
    
    \textbf{Since the adoption of DP synthetic data is often \emph{quality-bottlenecked}, the best option given unlimited computational resources and large amount of data is likely DP-finetuning of a diffusion model.}

    \item \textbf{Private evolution} is most applicable when training or fine-tuning large models is computationally infeasible, or when access to powerful foundation models is restricted to inference APIs. Additionally, PE requires an access to a suitable  embedding space (that captures aspects of the final use of the image data) and variation and random API must be available for the image modality. 
    PE is the only method that has a chance of obtaining reasonably quality data on very small private datasets. Additionally, PE can provide more stringent privacy guarantees (low $\epsilon$) that for finetuning methods will result in a very low utility. 
    \item If high resolution/visual quality of generated examples is of paramount importance (as opposed to fidelity of matching the original image distribution), using methods that don't use DP-trained generators will produce the best quality images. For these scenarios, either using PE or image-via-proxy methods like \cite{syntheticblogpost2025,wang2025synthesizeprivacypreservinghighresolutionimages}~(Section \ref{image-alternative-methods}) will like produce the best resolution at the cost of potential lower fidelity (in case of PE) and increased compute/complexity in case of proxy methods.
\end{compactitem}    

The field appears to be converging on the strategy of leveraging large, publicly pre-trained models as the most effective way to achieve high utility in DP image synthesis. Whether this leverage occurs via APIs (as in PE) or through DP fine-tuning (as in DP-Diffusion), the general knowledge captured in these large models seems crucial for offsetting the information loss imposed by DP when learning specifics from private data. Training high-quality, complex generative models like diffusion models purely from limited private data under strong DP guarantees remains largely impractical today.

{\small
\begin{longtable}[htb]{p{3.5cm} p{3.7cm} p{3.7cm} p{3.7cm} }
\caption{High-level comparison of DP finetuning of GANs and Diffusion models, PATE-GANs and private evolution for DP text synthesis. \label{tab:compare-image-methods}} \\
    \toprule

     \multirow{2}{*}{ \textbf{Aspect}} &
      \multicolumn{3}{c}{\textbf{Method}} \\
    
    & \textbf{DP-finetuning (GANs, Diffusion)} &
    \textbf{PATE-GANs} &
    \textbf{Private evolution} \\
    \midrule    
\endfirsthead 

    \toprule
     \multirow{2}{*}{ \textbf{Aspect}} &
      \multicolumn{3}{c}{\textbf{Method}} \\
    & \textbf{DP-finetuning (GANs, Diffusion)} &
    \textbf{PATE-GANs} &
    \textbf{Private evolution} \\
    \midrule 
\endhead 
    
    \multicolumn{4}{l}{\textbf{Amount of input private data}} \\
    \hline

    \textit{Small input quantity (<5K)} &
    Not recommended &
    Not recommended &
    Preferred \\
    \hline

    \textit{Large input quantity ($>$10K)} &
    Preferred &
    Recommended &
    Not recommended \\
    \hline

    \multicolumn{4}{l}{\textbf{ Reliance on Pre-trained Models / Public Data}} \\
    \hline
    \textit{Can benefit from additional public data} &
    Yes both for Diffusion and GANs &
    Distilling PATE teachers into students can utilize public data
    & No \\
    
    \textit{Needs pretrained models} &
    Yes for Diffusion, models should be pretrained to improve quality. &
    No &
    Implicit (via Foundation Model API)  \\
    \hline

     \multicolumn{4}{l}{\textbf{Yield}} \\
    \hline
    &
    Unlimited number of output examples, although with diminishing returns to downstream task performance. &
     Unlimited number of output examples, although with diminishing returns to downstream task performance. &
    In practice, suitable for outputting synthetic dataset of size $\leq$ size of input private dataset. \\
    \hline
    
    \multicolumn{4}{l}{\textbf{Model access required}} \\
    \hline
    
    &
    Weights &
    Weights of Generator and Student network and a number of teachers &
    Generations via API \\
    \hline

    \multicolumn{4}{l}{\textbf{Compute resources and engineering effort}} \\
    \hline

   \textit{ Training of generative models is required} &
    Yes, 1 &
    Yes, training a Generator, A number of teacher discriminators and a distilled student discriminator, in alternating fashion &
    No \\
    \hline

   \textit{ Inference cost multiple per synthetic example} &
    1 (same as regular inference). &
    $\propto$ (number of PATE submodels to aggregate over). &
    $\propto$ (number of iterations) $\times$ (number of variants per sample). \\
    \hline

    \textit{Prompt engineering required} &
    No &
    No &
    Possibly -- if using multimodal (text, image) models, potentially need to craft prompts for initial pure synthetic data and variate templates. \\
    \hline

    \textit{Time to first example} &
    Long (Run finetuning on the entire dataset, then sample) &
    Extra long (train multiple models on disjoint subsets, infer on all of the models to get pseudo labels, distill into a student model, repeat the process for a number of iterations after Generator's gradient updates) &
    Medium (Might require prompt engineering + running variate and embedding on the entire private dataset) \\
    \hline

    \textit{Direct side by side comparison of inputs and outputs} &
    No &
    No &
    No \\
    \hline

    \textit{Resilience to distribution gaps between private data and LLM} &
    High &
    Medium &
    Low \\
    \hline

    \multicolumn{4}{l}{\textbf{Target privacy guarantee}} \\
    \hline

    \textit{High $\epsilon$ (e.g. 10)/large amount of private data} &
    Preferred &
    Recommended &
    Not recommended \\
    \hline

    \textit{Stringent (e.g. $\epsilon < 1$)/small amount of private data} &
    Not recommended (quality will be bad) &
    Not recommended (quality will be extremely bad) &
    Preferred \\
    \hline

    \multicolumn{4}{l}{\textbf{Data persistence requirements}} \\
    \hline

    &
    Entire dataset required at once for training. &
    Entire dataset required at once for training. &
    Single examples can arrive in a streaming fashion, be used to cast votes, and then discarded immediately. \\
    \bottomrule
 
\end{longtable}
}

\subsection{Evaluating Synthetic Image Data Quality}\label{sec-image-quality}

On top of the generic metrics applicable to all data modalities mentioned in Section \ref{sec-fidelity-generic}, below we outlined metrics specific to evaluating image fidelity. 

The fidelity of synthetic image data can be evaluated using several specialized techniques. Originating from the literature on Generative Adversarial Networks (GANs), so-called Inception-based scores are specifically designed to align with human judgment of image quality~\cite{salimans2016improved}. The core principle of these methods involves leveraging a pre-trained Inception model, or more generally Convolutional Neural Network (CNN), trained on public data to extract visually relevant features from both real and synthetic images. By analyzing the distributions of these features, these scores can provide a quantitative measure of how realistic the generated data is.

\paragraph{Inception Score (IS)} \textbf{\cite{NIPS2016_03e7ef47,salimans2016improved}}
The Inception Score (IS) operates on two key principles. First, it assumes that high-quality, realistic images should contain meaningful objects, leading to a confident classification by the Inception model. This translates to a low entropy for the predicted class probability distribution of each individual image. Second, for a diverse set of generated images, the overall distribution of predicted labels should be highly varied, reflecting a wide range of objects. This means the marginal distribution across all images should have high entropy. A significant drawback of the Inception Score is that it only evaluates the generated images in isolation, without directly comparing their statistical properties to those of the real data. This limitation prompted the development of the Fréchet Inception Distance.

\paragraph{Fréchet Inception Distance (FID) \cite{NIPS2017_8a1d6947} }
The Fréchet Inception Distance (FID) improves upon the IS by directly comparing the feature distributions of real and synthetic data. It models the activations from the Inception model's coding layer for both datasets as multivariate Gaussian distributions. The distance between these two distributions is then calculated, providing a measure of their similarity. A lower FID score indicates that the distribution of synthetic data is closer to that of the real data, suggesting higher quality and diversity. The formal definition of the FID is:
 \[||\mu_1 - \mu_2||_2^2 - + \text{Tr}(M_1 + M_2 - 2(M_1 M_2)^{1/2},\]
    where $\mu_1,\mu_2$ are the mean of the two datasets to be compared and $M_1, M_2$ their
    respective covariance matrices of the Inception features for the real and generated datasets, respectively. 

\cite{NIPS2017_8a1d6947} demonstrated that the FID is more robust and correlates better with human judgment of image quality than the Inception Score. Furthermore, by functioning as a true distance metric, it provides a more principled way to measure the dissimilarity between the distributions of real and synthetic data.

\paragraph{MAUVE} is a newer metric that was originally introduced for text domain (Section \ref{sec-text-metrics}, \citet{pillutla2021mauve}) and was later shown to be successful in vision domain as well \cite{JMLR:v24:23-0023}.
Mauve seeks to quantify the tradeoffs between type I (areas of synthetic data that are unlikely under real data) and type II errors (regions missing in synthetic data that are plausible for the real data). 
To calculate Mauve in practice, embeddings of real and synthetic data (obtained via an appropriate embedding model) are clustered into a number of clusters. Cluster assignments for real and synthetic data serve as a low-dimensional representation of real and synthetic data.
A divergence curve that softly measures KL divergence of synthetic vs real and real vs synthetic quantifies Type I and II errors respectively. Area under this curve is the reported MAUVE metric.
Experiments indicate that MAUVE metric induces the same ordering of sampling algorithms as FID, and just like FID, MAUVE ends up accounting for both quality and diversity of synthetic images  \cite{JMLR:v24:23-0023}. 

\paragraph{Discussion}
Recent work by \citet{dpimagebench} points out the lack of fair comparison in the DP image generation literature, due to different papers employing a range of different model architectures, privacy accounting assumptions, public pretraining datasets, hyperparameter ranges, and downstream evaluators. To address these issues, they develop \emph{DPImageBench}, a unified evaluation codebase to compare methods on even footing. They support 12 methods from the literature, 9 evaluation datasets, consistent downstream evaluator training for utility measurement, and 6 fidelty metrics (including FID and IS). Their evaluations report that DP-trained diffusion models employing some form of curriculum (e.g. public pre-training) perform the best overall.

\subsection{Open Questions and Challenges}\label{sec-image-open-questions}
Despite significant progress, several challenges and open questions remain critical for the field of DP synthetic image generation:
\begin{compactenum}
    \item \textbf{Scalability and Fidelity:} Generating very high-resolution images ($>512\times512$) with high diversity and photorealism under strong DP guarantees (e.g., $\eps \le 1$) remains extremely difficult for all methods. DP mechanisms often smooth out fine-grained textures and details.
    \item \textbf{Theoretical Understanding:} A deeper theoretical understanding of why certain techniques (e.g., the effectiveness of pre-training, the benefits of latent space fine-tuning, the dynamics of PE) work well is needed to guide future algorithm design. Understanding better challenging training dynamics of Diffusion or GANs model also has a potential to make DP-finetuning more successful.
    \item  \textbf{Private evolution:} While private evolution has shown promise in domains where foundation models, \cite{lin:pe_images_pdf}, simulators, or public datasets \cite{microsoft:dpsda_repo} are available, effectively applying PE frameworks in low-resource domains remains a challenge.
    \item \textbf{Visual autoregressive models:} given the success of VARs in non-DP settings \cite{sun2024autoregressivemodelbeatsdiffusion}, figuring out effective ways of training them with DP is an important and promising area of research.
\end{compactenum}

\section{DP Synthetic Text data} \label{sec:text}
Text data is information written using a natural language. Text is one of the most prominent data source on the internet and offline. With the rise of power of LLMs, a lot of data collected and trained upon is becoming textual in nature. However while other modalities like tabular and image DP synthetic data had quiet a bit of development prior to generative models expansion (e.g. workload based methods and GANs model respectively), DP synthetic text data really came to a spotlight recently as LLMs became more capable. 
\paragraph{Roadmap}
This section examines methods for generating DP text data. We first focus our discussion on what is a meaningful privacy unit for text data (Section \ref{dp-text-privacy-unit}), then we dive into DP finetuning - the workhorse method for DP synthetic text generation that produces the best quality data given sufficient amount of private data and compute (Section \ref{sec-dp-finetuning}). We also explore developments in methods that don't require expensive finetuning of an LLM - namely DP inference (\ref{sec-dp-inference}) and private evolution (Section \ref{private-evolution-text}). This is followed by a high-level discussion and comparison between the various methods (Section \ref{sec:text-comparison}). Section \ref{sec:text-resampling} discusses resampling from a corpus of DP synthetic text to improve utility. Next, we cover proxy metrics for evaluating synthetic text data (w.r.t. its fidelity and utility) in Section \ref{sec-text-metrics} and conclude with a discussion of open research problems pertinent to text DP synthetic data (Section \ref{sec-text-open-questions}).

\subsection{Privacy Unit for Text Data}\label{dp-text-privacy-unit}

\textit{Example-level privacy}: The majority of papers on DP synthetic text data omit the discussion of the privacy unit, suggesting that example-level privacy level is employed. While for many applications (e.g. image classifications or classical models that work on rows of tabular data \cite{rosenblatt2025differential}) an example is a well-defined semantic concept (e.g. an image or a row), the question of what constitutes a meaningful semantic chunk of text is more ambiguous \cite{brown2022doesmeanlanguagemodel} and example-level DP privacy may not reflect human expectations of what it means for their data to be protected. 

Importantly, the way text data is batched into examples for DP-SGD training to a large extend determines the unit-of-privacy, because per-example clipping is the primary mechanism for controlling sensitivity; hence, changes in example packing and context length could impact the privacy guarantee unless care is taken. For example, given a corpus of Wikipedia articles, an LLM can be trained treating each full article as an example, or by splitting the documents into chunks like paragraphs and treating each of them as a separate example. Employing DP-SGD in these two settings will provide DP guarantees for different privacy units. Smaller privacy units like paragraphs are much easier for DP to protect due to the artificially increased size of the dataset, but provide a weaker guarantee. Thus comparing reported DP synthetic data performance (both utility and fidelity) on different textual privacy units is largely meaningless. 

\textit{User-level privacy}: When each example is associated with a user (and this metadata is available), this is one the most straightforward and perhaps meaningful privacy units to use with DP with text data. To achieve user-level privacy, one can either subsample sensitive data to retain only 1 piece of textual information per user (e.g. keep only 1 text document per user) or allow multiple examples per user and adjust the privacy accounting accordingly. Additional complications can arise when attempting to do such subsampling as capping user contributions can become tricky, especially for multi-owner documents (refer to Section \ref{user-contrib-bounding} for more in-depth discussion). It is worth also mentioning that some algorithms are more amenable to accounting for multiple examples per user (e.g. DP-SGD has a large body of work pertaining to such accounting changes, such as \citet{charles2024finetuninglargelanguagemodels}), while other more recent methods like private evolution (Section \ref{private-evolution-text}) and DP inference (Section \ref{sec-dp-inference}) might be harder to adjust. 

\textit{Sub-unit level of privacy}: \citet{brown2022doesmeanlanguagemodel} highlighted that DP assumes the need to preserve privacy of each user's record. However a record may contain a mix of private and public information - in a sentence like 'My social security number if XXX-XXX-XXXX' all the tokens apart from the actual SSN could be considered public and common, and so revealing them would not be considered a privacy violation. \citet{brown2022doesmeanlanguagemodel} argues that the ideal unit of privacy for DP is actually a ``secret-level'' privacy unit, that in context of DP synthetic data would mean that an addition or removal of any user secret does not change the synthetic data significantly. However realizing this definition is currently unrealistic as it will require an understanding of what constitutes a secret and automatically determining a secret's boundaries. The field of \textit{contextual integrity} \cite{10.5555/1822585} introduces a framework for understanding in which context private information can be shared and offers a direction for answering these questions.

\paragraph{Privacy unit discussion}
While user-level privacy unit is broadly used in industry \cite{xu23gboard} and perhaps the most meaningful privacy unit to use, in several scenarios when modeling text, user-level DP still cannot prevent secret leakage. In free-form text obtained from users interacting in the real world, secrets can be represented across multiple users. Under this view point, protections from example-level and user-level lie on a spectrum depending on how secrets are propogated. User-level is the appropriate fit in the case where secrets are strongly tied to a single user (e.g. personal search history), but not qualitatively stronger than example-level in other scenarios (e.g. from \citet{brown2022doesmeanlanguagemodel}: many people prompting an AI chatbot with same secret document provided as context). 
All of the algorithms that we describe next support example level privacy and can be also adjusted to account for group or user level privacy. 

Finally, caution needs to be exercised when using methods that involve DP training/finetuning of LLM models (Section \ref{sec-dp-finetuning}), as common training optimization techniques with LLMs like \textit{packing} can subtly change the meaning of privacy unit - we cover these peculiarities in Section \ref{sec-gotchas}.

\subsection{DP Training and Finetuning}\label{sec-dp-finetuning}

\textit{DP-Training} (or \textit{DP-finetuning} of already publicly pretrained models) techniques that we explored in images synthetic section~(Section~\ref{sec-image-gans}) are also the workhorse methods for creating DP synthetic text data. 

Many works on DP synthetic text generation explored LLMs ability for open-ended text generation \cite{radford2018improving}. Works like \cite{bommasani2019towards,yue2022synthetic,mattern2022differentially,kurakin2023harnessing,wang2024knowledgesgprivacypreservingsynthetictext,yu2024privacy,carranza-etal-2024-synthetic}~employ text generative models (most often tranformer-based LLMs~\cite{vaswani2017attention}
), DP-Train or DP-finetune them (Section \ref{sec-dpsgd}) on sensitive data and then sample synthetic examples from such models. While this recipe is conceptually straightforward \cite{bommasani2019towards}, practical implementations which achieve high quality synthetic data quality required several modifications to standard open-ended text generation. 

\paragraph{Data preparation for DP finetuning}

Most works format text data for finetuning in the following way. First a prefix $p$ or set of prefixes $\{p_1, \ldots, p_m\}$ is chosen. These prefixes could represent a class label or any other type of conditioning that can be used during generation of synthetic data. For example, in a benchmark task of generation of DP synthetic Yelp reviews, \citet{yue2022synthetic}~used the conditional prefix $\texttt{"Product type: P | Review score: R"}$, while \citet{kurakin2023harnessing} used $\texttt{"[yelp] [label]"}$.  The example for finetuning is then formed by concatenating prefix $p$ with sensitive text $t$ (e.g. Yelp review) that needs to be protected with DP.

\paragraph{Defining training objective} The standard language modeling objective for LLM training (cross-entropy next token prediction loss~\cite{radford2018improving}) can be used without modification for DP-Training \cite{yue2022synthetic,kurakin2023harnessing}. However, some works explored modifications to this objective. 

For example, \citet{kurakin2023harnessing} used the Prefix-LM ~\cite{raffel2020t5} formulation of loss that resulted in improvements in quality of synthetic data.  Assume prefix $p$ and suffix $t$ tokens. Standard language modeling would use causal self-attention (each token depends only on previous tokens) for each of the tokens in $pt$ and cross entropy loss on tokens in $pt$. 
Prefix-LM instead uses bi-directional attention for tokens in the prefix $p$, while causal self-attention is used for tokens in $t$. The generator is trained to minimize cross entropy loss calculated from only tokens in $t$. In the case of finetuning for synthetic data generation, $t$ corresponds to private data. This objective teaches the LLM to produce synthetic data conditioned on the prefix $p$, which is known and always supplied at inference time.

Other attempts at improving conditional generation included works like 
\citet{putta2023differentially}, which introduced an extra discriminator head that attempts to distinguish between the classes used for conditioning (e.g. Yelp ''positive'' and ''negative'' labels). The discriminator's loss is included into overall training objective, resulting in improved separability of classes representations. In a similar spirit, \citet{mattern2022differentially} introduced a loss term to penalize generation of sentences of incorrect class. \citet{kurakin2023harnessing} however argued that proper hyperparameter tuning for DP-training with Prefix-LM loss can achieve better results and is sufficient ~\cite{kurakin2023harnessing}.

\paragraph{DP training setup} To achieve DP, model training could be done with DP-SGD or any of its variants. To reduce the negative effect of DP on utility, proper hyperparameter tuning  \cite{xuechen2021dp_training_llm,yu2021dp_training_llm,Ponomareva_2023}, including batch size, clipping norm, learning rate and number of epochs should be done. Notably, optimal hyperparameters for DP can vary significantly from the optimal hyperparameters in the non-DP setup -- without a sufficient sweep, it is easy to underestimate achievable model utility under DP \cite{xuechen2021dp_training_llm}.

The choice of a checkpoint  (pre-trained or instruction-tuned) has not been explored much in context of DP synthetic data generation, however works like \cite{nasr2025scalable} suggest that pre-training checkpoints are more prone to memorization of the training data (and therefore perhaps are more suited for open-ended text generation). Instruction tuned checkpoints might require different prefixes for finetuning, including finding appropriate natural text prompts: \textit{"Please generate a positive review about a restaurant"}. 

Parameter efficient finetuning (PET) approaches such as LoRA \cite{hu2022lora} has been shown to be beneficial compared to full finetuning with DP, resulting in significant improvements of quality of DP synthetic data \cite{kurakin2023harnessing,yu2021dp_training_llm}.

DP training is typically slower and more memory hungry compared to standard non-DP training. Some of these issues could be alleviated with clever training techniques. Ghost clipping~\cite{xuechen2021dp_training_llm} could be used to improve memory usage and thus potentially increase batch size per step. If desired batch size still does not fit into memory of accelerators, gradient accumulation (also known as virtual batch training) could be used to further increase the effective batch size per training update \cite{Ponomareva_2023}. Gradient accumulation works by splitting desired training batch into smaller chunks which fit into accelerator memory, performing gradient computation for each individual chunk and then aggregating across all chunks to perform update of model parameters.

\paragraph{Sampling} Sampling of DP synthetic data is achieved by choosing one of the training prefixes $p_i$ and then generating continuation of the prefix using DP-finetuned LLM. This process is repeated until desired number of samples in the synthetic dataset is reached. Prefixes frequency can be chosen based on the distribution of labels in the private data (to increase the fidelity of DP synthetic data) or sampled uniformly. 

Given a sampling prefix, sampling can be performed using many common sampling schemes such as temperature sampling, top-k sampling~\cite{fan2018topk} or nucleus sampling~\cite{Holtzman2020topp}.
\citet{kurakin2023harnessing} ablated various sampling approaches and advocated natural sampling, i.e. temperature sampling with $T=1.0$ and no top-k or top-p restrictions, however \citet{pillutla2021mauve} argued that nucleus sampling results in higher quality text than natural (or ancestral) sampling.

\paragraph{Size of the foundational model} 

In general, using larger and more capable models for DP training is expected to improve the fidelity and utility of DP synthetic data at the cost of increased computation, for example   \citet{yu2024privacy} and \citet{kurakin2023harnessing} used 7 and 8 billion parameter LLMs respectively. \citet{tan2025synthesizing} use significantly smaller LLMs (140M) by employing several stage process and publicly pretrained models. In particular, a universal topic model is first pre-trained on large scale public corpora. Given a new private dataset, a topic model produces a topic assignment for each private document, as well as topic distribution histogram (with DP) at the dataset level. Topics are then used as conditional prefixes for DP-finetuning a small LLM, with private documents as targets. During sampling, DP topic histogram is used to generate appropriate proportion of samples from each topic. This process produces DP synthetic text data comparable or slightly better than data from DP-finetuning of a much larger model (1.5B), especially under tight privacy budgets. This work also illustrates that for smaller backbone LLMs, DP finetuning with conditional prefixes, for example derived from clustering private data, makes learning the distribution and subsequent sampling of synthetic data `easier`. 

\subsection{Methods That Avoid DP-Training}

DP finetuning yields high-quality synthetic data but requires finetuning a pretrained LLM. This means that one requires: (1) access to model weights; (2) engineering effort to integrate an efficient DP-SGD implementation into existing (possibly distributed) training pipelines; and (3) sufficient computational resources to run DP-SGD on pretrained LLMs, which is typically significantly more expensive than non-private training.

In this section, we discuss \emph{training-free} methods that can operate with only inference access to a pretrained LLM. The two main classes of methods we discuss are (1)  \emph{DP inference} (Section \ref{sec-dp-inference}), which requires access to model prediction probabilities (\emph{logits access}); and (2) \emph{Private evolution} (Section \ref{private-evolution-text}), which only requires output text (\emph{API access}). 

\subsubsection{DP Inference}\label{sec-dp-inference}

\emph{DP inference} techniques for DP synthetic text generation use LLMs' in-context learning ability and incorporate differential privacy into the LLM's decoding process. These techniques require the ability to query for next-token probabilities from pretrained models prompted with private data, which are then aggregated and released with differential privacy. DP inference belongs to a class of techniques called \textit{Private Prediction}, that we briefly describe first. 

\paragraph{Private prediction.} Recall that producing DP synthetic data via DP finetuning involves learning a generative model of the data and privatizing the release of the \emph{model gradients (and therefore weights) itself}; from such privatized model, arbitrary amounts of synthetic data can be sampled. Hence for generating synthetic data, privatizing model release is \emph{sufficient but not necessary} -- instead one can aim to guarantee privacy of the sampled text by privatizing non-private model outputs directly. This is a well-studied problem in the DP literature referred to as \emph{private prediction} \cite{dwork2018privacy}. 

Since it is a relaxation in the space of admissible algorithms, in theory, private prediction can improve the privacy-utility tradeoff \cite{Ponomareva_2023}. The core problem in introducing DP with private prediction is to ensure that model predictions do not change significantly on neighbouring datasets. While some work obtains bounds for convex losses \cite{dwork2018privacy,van2020trade}, such techniques have not been successfully extended to LLMs. Instead, to achieve private prediction for LLMs, variants of \emph{subsample-and-aggregate} \cite{nissim2007} framework are used.  

The \emph{PATE (Private Aggregation of Teacher Ensembles)} line of work is one example of sample-and-aggregate framework \cite{papernot2017semi,papernot2018scalable}. In its classical form, PATE partitions the private data into several disjoint sets and trains separate models non-privately on each set. At inference time, all models are invoked and their outputs are aggregated and released with DP. Crucial to the analysis is the fact than each example can affect only a single model from the ensemble.

The first studies marrying private prediction and LLMs followed the PATE paradigm. \citet{ginart2022submix} introduced \emph{SubMix}, which, like PATE, non-privately finetunes several GPT-2 models on disjoint partitions of the private dataset and runs them in parallel for inference. SubMix employs a private prediction protocol that is better-suited for language modeling than PATE, and further makes use of public model predictions. \emph{PMixED} \cite{flemings2024differentially} opts to use LoRA for per-partition models for better inference efficiency (LoRA allows all models to fit in HBM simultaneously; further gains are possible via LoRA batching \cite{wen2024batched}). They also introduce a much simpler prediction protocol that offers standard, unconditional DP guarantees.

\begin{algorithm}[t]
\caption{DP Inference}\label{alg:dp-inference}
Denote the vocabulary by $\mathcal X$. Let $\mathcal X^*$ be the set of all strings over $\mathcal X$, and $\Delta(\mathcal X)$ be the set of all probability distributions over $\mathcal X$. For $a,b \in \mathcal X^*$, $ab$ denotes concatenation. DPTokenSelect$(\cdot)$ stands in for any DP mechanism used to select the next token.
\begin{algorithmic}[1]
\Require LLM: $\mathcal{X}^* \to \Delta(\mathcal{X})$, private prompts $\boldsymbol{x}^{(1)}, \dots, \boldsymbol{x}^{(n)} \in \mathcal{X}^*$, response length $T$.
\Ensure Response $\boldsymbol{s} \in \mathcal{X}^*$.
\State $\boldsymbol{s} \gets \emptyset$
\For{$t = 1$ to $T$}
  \For{$i = 1$ to $n$}
    \State $p^{(i)}_t \gets \text{LLM}(\boldsymbol{x}^{(i)}\boldsymbol{s})$
  \EndFor
  \State $\tilde{x}_t \gets \text{DPTokenSelect}(p^{(1)}_t, \dots, p^{(n)}_t)$
  \State $\boldsymbol{s} \gets \boldsymbol{s}\tilde{x}_t$
\EndFor
\State \Return $\boldsymbol{s}$
\end{algorithmic}
\end{algorithm}

\paragraph{DP inference.}
Generating text with PATE-style private prediction requires training a large number of models (>30), and having them all loaded simultaneously for inference. Also, each split should be reasonably large to ensure there is sufficient data to train a model. To address these issues, the next advance in private prediction for LLMs was to employ private data for \textit{in-context learning}. We will refer to this class of techniques as \textit{DP inference}, although the term \emph{differentially private in-context learning} is also employed \cite{wu2024privacy}. 
\citet{wu2024privacy} propose to partition private examples into disjoint contexts to condition inference, rather than training sets for finetuning; PromptPATE \cite{duan2023flocks} uses private predictions to label public examples, which are collected in a ``student prompt'' that can be deployed at test time. \citet{wu2024privacy} employ Gaussian ReportNoisyMax \cite{zhu2022adaptive} for token selection. They demonstrate effectiveness on generative tasks like summarization by generating keywords with DP inference followed by using a public model to post-process results. 

Algorithm \ref{alg:dp-inference} gives pseudo-code for DP inference. In words: DP inference techniques for LLMs directly prompt a pretrained language model with examples from the private dataset, distributing them among many parallel prompts, with each one asking for similar text to the provided example. When generating each token of the response, the next-token prediction produced by each context is aggregated with DP to produce a differentially private output token. As the precise aggregation approach varies among methods, this is abstracted as $\text{DPTokenSelect}(\cdot)$ in Algorithm \ref{alg:dp-inference}, which takes as input next-token probabilities and produces a single token with some privacy cost. That token is then appended to each of the parallel contexts and the process repeats. The final privacy cost is calculated by composing the privacy cost of all calls to $\text{DPTokenSelect}(\cdot)$.

\begin{specialistbox}{Privacy cost of DP Inference}
As opposed to DP finetuning, \emph{the privacy cost of DP inference increases with the number of tokens generated}. This is because the privacy analysis typically proceeds by bounding the privacy cost of generating a single token, and then accumulating these costs via composition.
\end{specialistbox}

\paragraph{Synthetic data generation.}
As shown in Algorithm \ref{alg:dp-inference}, generating synthetic data with DP inference simply requires feeding in sampled tokens in an auto-regressive manner. 
\citet{tang2024privacy} first demonstrate generating \emph{few-shot synthetic data} with DP inference for classification and extraction tasks. They use the Gaussian mechanism with subsampling for token selection, and demonstrate generating $\approx12$ examples, which suffices for effective few-shot learning. \citet{amin2024private} introduce algorithmic improvements (clipped logits exponential mechanism for token sampling, as well as parallel composition) to scale generation to $\approx$ thousands of examples. Increased quantity enables one to finetune downstream BERT models on generated synthetic data; quantity also improves downstream in-context learning accuracy. \citet{gao2025data} turn their attention to adaptively selecting the clipping radius for noise addition, noting that for certain tokens, predictions are highly similar and therefore less noise can be added; they demonstrate utility improvements for few-shot generation. \citet{flemings2025differentially} introduce a PMixED-inspired prediction protocol and show improvements for in-context few-shot generation when targeting more difficult generative tasks. \citet{hong2024dp} skips synthetic data points as an intermediatary for in-context learning and directly learns a classification prompt (which may include some synthetic data); they sample tokens with the LimitedDomain mechanism \cite{durfee2019practical}.

Synthetic data should capture facets of your source data beyond task accuracy. \citet{cohen2023hot} introduce \emph{HotPATE}, a drop-in replacement for PATE which aims to preserve the diversity of the source data during generation. PATE targeted classification tasks and hence did not consider this a design requirement. The key technical tool of \citet{cohen2023hot} is \emph{coordinated sampling}: rather than having all teachers sample tokens independently before aggregation, HotPATE uses shared public randomness when sampling to maximize agreement between teachers; this lets it better preserve the tail of the data distribution.~\citet{amin2025clustering} compute representativeness metrics (i.e. MAUVE \cite{pillutla2021mauve}) on data generated from DP inference, and find it is poor. Indeed, most prior work generated a small number of examples for in-context learning and could not compute dataset-level metrics for representativeness. \citet{amin2025clustering} show that clustering data and forming same-cluster batches improves representativeness.

\paragraph{Leveraging public information.}
A theme explored in several works is to use predictions made by a public model to improve the privacy/utility tradeoff of DP inference. The conceptual appeal is the following: in text, many tokens are predictable without private information, and therefore we should not ''pay'' privacy cost for them. The privacy cost of text generation should not scale with the \emph{number of tokens produced}, but rather the \emph{surplus information content} introduced by the private data. 

\citet{flemings2024differentially}, \citet{ginart2022submix}, and \citet{flemings2024adaptively} mix predictions made from the public model with the private model. For the latter two, which employ data-dependent analyses, public predictions aligning with private predictions result in charging lower $\epsilon$. \citet{tang2024privacy} filters token selection candidates to the top-$k$ choices of a public model; this reduces the frequency of noise-induced error when selecting tokens from a large vocabulary, but introduces bias. \citet{amin2024private} and \citet{koga2024privacy} use AboveThreshold \citet{DworkRothBook:2014} to query whether the public prediction aligns with the private prediction, and \emph{only consumes privacy budget to use the private prediction if the check fails}; hence the privacy cost scales with the number of tokens where public and private predictions differ. \emph{InvisibleInk} \cite{vinod2025invisibleink} employs the exponential mechanism for token sampling and uses the public prediction to determine where to center the clipping range. This modification allows for a significant reduction of privacy budget, unlocking generation with a much smaller batch size. \citet{amin2025clustering} employs clusters derived from public data to batch together similar private data when running inference, which improves representativeness.

\paragraph{Data-dependent analysis.} \citet{papernot2017semi} introduced a \emph{data-dependent} privacy analysis in which the privacy guarantee is a function of the input dataset, instead of a worst-case upper bound over all datasets. Accepting this relaxation can lead to significant gains in data volume. Roughly speaking, these analyses allow one to take advantage of \emph{consensus} among predictions. In particular, on tokens where many teachers agree, one can argue the counterfactual impact of any user's data is small; however this scenario cannot be assumed to hold in worst-case datasets considered in unconditional DP analyses.  \citet{duan2023flocks} inherits PATE's analysis and provide such a  guarantee; they report $\epsilon=0.147$. SubMix \cite{ginart2022submix} presents a data-dependent analysis. Coordinated sampling in HotPATE \cite{cohen2023hot} improves teacher agreement which results in tighter data-dependent guarantees. \emph{AdaPMixED} \cite{flemings2024adaptively} computes data-dependent $\epsilon$ and introduces a screening step to defer expensive queries to public model. They demonstrate $16\times$ smaller $\epsilon$ while maintaining similar utility as the unconditional DP variant \cite{flemings2024differentially}. \citet{amin2025clustering} introduce a coordinate-wise median-based token selection algorithm designed to take advantage of prediction consensus, and prove a data-dependent and \emph{ex-post DP} guarantee  \cite{ligett2017accuracy}.

\paragraph{User-level guarantees.} To support user-level guarantees, the input to the base DP token selection mechanism (before applying subsampling and composition) must satisfy that \emph{a user's data is restricted to appearing in a single model context}. A user's data appearing in multiple contexts necessitates the application of group privacy bounds.

\paragraph{Comparison of methods.}
Table \ref{tab:dp-inference-methods} presents a comparison Private prediction, including DP inference methods, along several main axes: partioning strategy, token selection mechanism, task, and the privacy guarantee.
\begin{itemize}
\item A simplified version of the method described in \citet{amin2024private}  offers the best privacy-utility trade-offs, as it is capable of generating thousands of examples at reasonable privacy budget. The simplified version notably employs parallel composition, KV cache reuse, and exponential mechanism sampling; but does not implement the use of AboveThreshold with an additional public query.  This method is recommended because it is relatively simple, is capable of generating a large quantity of data, and offers the standard unconditional DP guarantee.  Results for this approach are reported as \emph{Baseline++} in \citet{amin2025clustering}.
\item Extensions of this approach to adopt the public query logit clipping proposed by \citet{vinod2025invisibleink} results in further improvements in privacy budget and compute efficiency. 
\item For the above methods, the most straghforward choice of the model for synthetic data generation is a pretrained model (instead of instruction-tuned) with a generic prompt that puts sensitive examples directly into context with no additional description. This minimal approach performs well (see \emph{Baseline++} results in \citet{amin2025clustering}) and requires no prompt engineering. Including at least 2 serial examples per context improves results \cite{tang2024privacy,amin2025clustering}. 
\end{itemize}

{\footnotesize
\begin{longtable}[htbp]{p{3.2cm} p{1.4cm} p{1.8cm} p{1.7cm} p{1.3cm} p{3cm}}
\caption{A comparison of DP inference methods in the literature. We recommend a simplified version of the method of \citet{amin2024private} (omitting the use of public predictions) as a simple, strong baseline method for generating large quantities of synthetic data via DP inference. \label{tab:dp-inference-methods}
} \\
\toprule
\textbf{Method Paper} & \textbf{Partition} & \textbf{Mechanism} &\textbf{Task} & \textbf{Privacy} & \textbf{Approach} \\
\toprule
\citet{ginart2022submix} & Train sets & SubMix & Next-token prediction & Data-dependent & Privacy by sampling \\
\cmidrule{1-6}
\citet{wu2024privacy} & Contexts & Gaussian Noisy Max & Query answering, generative response & Standard &  Paritions examples into contexts\\
\cmidrule{1-6}
\citet{duan2023flocks} & Contexts & PATE &  Query answering & Data-dependent & Label public prompts \& deploy  \\
\cmidrule{1-6}
\citet{cohen2023hot} & Contexts & HotPATE & Query answering & Data-dependent & Coordinated sampling to improve PATE diversity \\
\cmidrule{1-6}
\citet{flemings2024differentially} & Train set & PMixED & Next-token prediction & Standard & Mix public distribution with private, then sample\\
\cmidrule{1-6}
\citet{flemings2024adaptively}  & Train set & AdaPMixED & Next-token prediction & Data-dependent & Data-dependent extension of PMixED \\
\bottomrule
\toprule
\citet{tang2024privacy}  & Contexts & Gaussian Noisy Max & Few shot synthetic data & Standard & Generate a few synthetic examples to be used for ICL \\
\midrule
\citet{hong2024dp}  & Contexts & Limited-Domain & Prompt generation & Standard & Generate a prompt directly instead of few-shot examples \\
\midrule
\citet{amin2024private}  & Contexts & Exponential & Synthetic data & Standard & Privacy by sampling and parallel composition to generate much more data \\
\midrule
\citet{gao2025data} & Contexts & GoodRadius + Gaussian & Few shot synthetic data & Standard & Pick per-token adaptive clipping radius \\
\midrule
\citet{flemings2025differentially} & Contexts & PMixED & Few shot synthetic data & Standard & Mix public distribution with private, then sample \\
\midrule
\citet{amin2025clustering}  & Contexts & Median & Synthetic data & Ex-post data-dependent & Form batches from pre-clustering data to improve representativeness \\
\midrule
\citet{vinod2025invisibleink} & Contexts & Exponential & Synthetic data & Standard & Recenter clipping range with public prediction for better privacy \\
\bottomrule
\end{longtable}
}

\subsubsection{Private Evolution for Text}\label{private-evolution-text}

\begin{algorithm}[H]
    \caption{Private Evolution for Text (Aug-PE, \citet{xie2024differentially})} 
        \label{alg:aug_pe}

    \begin{algorithmic}[1] 
        \Require Private data $D$, text embedding model  $\Phi$, target number of synthetic samples $N$, evolution rounds $T$, population size multiplier $L$ (number of rewrites per chosen synthetic sample).
        \Ensure Synthetic dataset $\hat{D}$.
        \State Initialize $\hat{D}_{0}$ of size $(L+1)\cdot N$ with Random API.
        \For{$t = 0, \dots, T-1$}
            \State $E_{t}=\Phi(\hat{D}_{t})$ \textsl{\scriptsize //Embedding calculation for  synthetic samples.}
            \State Let $\Phi(p), p\in D$ vote for the nearest embedding in $E_{t}$.
            \State Privatize the voting results with $(\epsilon, \delta)$-DP to get a DP histogram $H_{t}$.
            \State $\hat{H}_{t}$ = $H_{t}/\text{sum}(H_{t})$ \textsl{\scriptsize //Histogram normalization.}  
            \State Get $\hat{D}_{t}^{'}$: top-$N$ samples from $\hat{D}_{t}$ based on $\hat{H}_{t}$.
            \If{$t<T-1$}
                \State $\hat{D}_{t+1}$ of size $(L+1)\cdot N$:  $L$ variants for each $z\in \hat{D}_{t}^{'}$ via Variation API and $\hat{D}_{t}^{'}$.
            \Else
                \State \Return $\hat{D}_{t}^{'}$.
            \EndIf
        \EndFor
    \end{algorithmic}
\end{algorithm}

Private Evolution (PE) is an emerging approach for generating DP synthetic data using API-only (e.g. inference) access to foundation models. Its key advantage lies in leveraging the powerful capabilities of these models without requiring the design of DP-specific training or sampling algorithms, an often technically challenging or even infeasible task, particularly when the models are closed-source.

Originally introduced for DP image synthesis \cite{lin:pe_images_pdf}, PE was later extended to text by \citet{xie2024differentially} who proposed Augmented Private Evolution (Aug-PE). Aug-PE enhances PE with techniques specifically tailored to text generation, improving the quality and relevance of synthetic outputs. Despite domain differences, both PE and Aug-PE follow the same high-level pipeline: they start with an initial population of synthetic samples generated via prompting (this pool can also be public data of similar to sensitive data distribution), then iteratively refines these samples using a DP histogram-based evolutionary process that gradually aligns the distribution with the private dataset. We provide a  pseudocode of the Aug-PE algorithm from \citet{xie2024differentially} in Algorithm \ref{alg:aug_pe}.

Before detailing Aug-PE's workflow, we highlight two key APIs required to be built using the target foundation model: (1) \textbf{Random API}: Generating an initial population of synthetic samples without access to private data. Alternatively, access to some public data of similar distribution to the sensitive data and (2) \textbf{Variation API}:  Modifies/rewrites existing synthetic samples to introduce diversity. With these two functionalities in place, the Aug-PE process proceeds as follows.

\paragraph{Initialization.} The process begins by generating an initial pool of synthetic samples, typically through prompting the foundation model. Although Aug-PE can technically start with arbitrary synthetic samples, empirical results show that using well-designed prompts informed by public priors leads to more relevant and effective outputs. These priors help guide the model toward producing samples that better reflect the target data domain. For example, in generating differentially private synthetic Yelp reviews \citet{xie2024differentially} used prompts with contextually appropriate phrases such as “reviews for steakhouse restaurants”—drawn exclusively from public sources. Importantly, these prompts must not incorporate any private data to ensure privacy is preserved. Given a target of $N$ synthetic samples, Aug-PE generates an initial population of size $L \times N$, where $L$ is a small constant (e.g. single or low double digit). This differs from the original PE for images, where the initial and target population sizes are identical.

\paragraph{Iterative Refinement.} Once the initial population is constructed, Aug-PE refines it through repeated variation and voting steps. During variation, each sample is modified using one of two strategies: prompting the model to rephrase the text or masking random tokens and asking the model to fill in the blanks. In the voting step, an embedding is computed for each synthetic candidate sample using a text embedding model—either directly from the synthetic datapoint itself or by first generating a few variants via the Variation API and aggregating their embeddings. Each private data point then votes for its nearest candidate based on embedding space distance, producing a histogram of votes. 

This histogram is privatized with differential privacy, and the samples with the top $N$ vote counts are selected to form the next generation (an alternative strategy is to sample the synthetic datapoints with replacement according to probabilities induced by the histogram of votes, similar to what was done for images  in Section \ref{sec-pe-image}, which can however result in the same synthetic samples being chosen multiple times; this isn't desirable in this version of the algorithms, because each selected example is kept in the pool as-is). If $L > 1$, these $N$ samples are expanded back to $(L+1) \cdot N$ using the Variation API, rewriting each chosen synthetic sample $L$ times. Note that this is different from PE for Images (Section \ref{sec-pe-image}), where only variants (but not the $N$ synthetic examples themselves) were added to the pool for the next round. This is done to increase the chance of retaining high quality synthetic examples \citet{xie2024differentially}.

This refinement process is repeated for a fixed number of iterations. At the final iteration, the top $N$ samples are directly returned as the final synthetic dataset.

\paragraph{User-level privacy unit adjustments}
In the original algorithm, each private example uses only 1 vote (that can be either fully allocated to a closest synthetic example or distributed between a number of closest synthetic examples, depending on hyperparameter \textit{number of nearest neighbours}). This means that l2 sensitivity is $1$ and gaussian noise introduced to the histogram depends on this sensitivity. 
When user-level privacy is required and each user contributes multiple samples, the vote counts from each user can be normalized to sum to one (in l2 norm). This ensures that the sensitivity per user remains bounded by one. 

\paragraph{Discussion}
The befits of PE lie in its simplicity and practicality: it is an intuitive algorithm that is easy to explain and that relies solely on inference using off-the-shelf LLMs. This eliminates the need for computationally expensive DP fine-tuning of LLM checkpoints, does not require access to logits and allows to potentially use a more powerful model that can be effectively DP-finetuned.

However, the effectiveness of PE hinges on two key components: the initial pool of synthetic data and the rewrite prompt, both of which must be manually crafted by the user. In the absence of publicly available data to seed the initial pool, the authors suggest manually creating a prompt to generate data that mimics the private dataset. Yet constructing either the initial or rewrite prompts requires some public prior knowledge about the private data, which may be difficult to obtain in some settings. For instance, the prompts proposed in \cite{xie2024differentially} reflect a deep understanding of the private data’s distribution, such as Yelp review categories, rating scales, or the typical format of research papers.

This raises a critical question: how can high-quality prompts be designed for sensitive datasets when no human-readable samples are available? It remains unclear whether structural insights, like categories or data schemas, can be derived in a privacy-preserving way without human inspection. Furthermore, since PE does not involve fine-tuning the LLM’s weights, it assumes that the private data’s general domain is already covered in the LLM’s training corpus. As a result, PE is unlikely to perform well in domains absent from pretraining. For instance, generating coherent legal text if legal language was not part of the original training data will be a challenging task for PE.

Recent approach for PE introduced by \citet{hou2025private} introduces updates to the foundational model via policy optimization tuning. The key idea is that voting information can be used to guide the model to prefer synthetic examples that obtain better scores than synthetic examples generated with the same prompts but receiving worse scores. With this approach the foundational model however is no longer fixed and access to model weights and tuning of the model and to compute to perform the udpates is required. 

\subsection{Summary and Comparison of Methods}\label{sec:text-comparison}
We have outlined 3 main methods for creating DP synthetic text data: via DP finetuning (Section \ref{sec-dp-finetuning}); DP inference (Section \ref{sec-dp-inference}); or private evolution algorithms (Section \ref{private-evolution-text}. We provide a detailed comparison of the methods in this section, highlighting similarities and differences.

\subsubsection{Unified View of DP Text Synthesis Methods}

First, we present a unified view of the discussed methods. Existing methods can be viewed under the following recipe:
\begin{itemize}
    \item Start from an existing non-private approach to generate synthetic text.
    \item Identify the \emph{iterative primitive} in the approach used to \emph{extract information from real data}.
    \item DP-fy that primitive.
\end{itemize}

Specific examples are presented in Table \ref{tab:summary-text-methods}. The primitive used to extract information from real data must do so via aggregates (e.g. batch gradient) which allows it to be amenable to DP. The iterative nature of the primitive means that proving privacy guarantees of the primitive suffices to guarantee privacy of the entire algorithm.

\begin{longtable}[htbp]{p{3.5cm} p{3.5cm} p{3.5cm} p{3.5cm}}
\caption{For each DP text synthesis method, we can identify a base non-private method, as well as the core iterative primitive being DP-ified. \label{tab:summary-text-methods}
}  \\
\toprule
& \textbf{DP finetuning} & \textbf{DP inference} & \textbf{Private evolution} \\
\midrule
\textbf{Comparison} & Finetuning & Inference & Evolutionary prompting  \\
\midrule
\textbf{Primitive} & Gradient  & Token sampling & Selection \\
\bottomrule
\end{longtable}

\begin{specialistbox}{Unified view of DP text synthesis methods}
Existing approaches to DP text synthesis start from a non-private text generation procedure and identify an iterative primitive used to extract information from real data to DP-fy. The effectiveness of the method is predicted by its \emph{original effectiveness} plus the degree of \emph{distortion introduced by the DP substitute of the primitive}.
\end{specialistbox}

\subsubsection{Comparison}

Each of the aforementioned methods have their strengths and weaknesses and different requirements. Table \ref{tab:compare-text-methods} provides a detailed comparison of these methods along various axis like amount data, computer resources needed, yield, type of access needed etc. 

\begin{specialistbox}{Choosing the method for DP synthetic text data generation}
For large enough volume of sensitive data (>XX K datapoints) with enough compute the method that results in highest fidelity and utility of DP synthetic data is almost always DP-Finetuning. Less computationally expensive methods that don't require LLM's finetuning can provide reasonable data when sensitive data is somewhat in distribution for the pretraining data.
\end{specialistbox}

\begin{compactitem}
    \item \textbf{DP finetuning} is the workhorse method that delivers the best quality  given sufficient amount data, compute and engineering and time investment. \textbf{Since the adoption of DP synthetic data is often \emph{quality-bottlenecked}, this is likely the option that best fits your needs}. It does require significant engineering expertise to implement and run DP-finetuning and edit access to model checkpoint weights, in turn it delivers the best quality when given access to a significant amount of data. DP-finetuning results in high fidelity synthetic data even if private data to mimic is significantly out of the distribution for LLM's pretraining data.
    \item \textbf{DP inference} is suitable for fast prototyping but is limited to generating short pieces of text only, due to its privacy cost accumulating with each output token.
    \item \textbf{Private evolution} is a strong candidate when one wants very stringent privacy guarantees ($\epsilon <1$ and/or has access to very limited data. In fact, that it the only method that can work with very few datapoints (e.g. <1K). It is also extremely easy to explain and does not egress sensitive data to an LLM for inference (only using sensitive data's embeddings). It does require access to high quality public data similar to the sensitive data and significant work for prompt engineering.
\end{compactitem}

{\small
\begin{longtable}[htb]{p{3.5cm} p{3.7cm} p{3.7cm} p{3.7cm}}
\caption{{Comparison of DP finetuning, DP inference, and private evolution for DP text synthesis.} \label{tab:compare-text-methods}
} \\
    \toprule

     \multirow{2}{*}{ \textbf{Aspect}} &
      \multicolumn{3}{c}{\textbf{Method}} \\
    
    & \textbf{DP finetuning} &
    \textbf{DP inference} &
    \textbf{Private evolution} \\
\midrule    
\endfirsthead 

\toprule

     \multirow{2}{*}{ \textbf{Aspect}} &
      \multicolumn{3}{c}{\textbf{Method}} \\
    
    & \textbf{DP finetuning} &
    \textbf{DP inference} &
    \textbf{Private evolution} \\
\midrule   
\endhead 

    \multicolumn{4}{l}{\textbf{Amount of input private data}} \\
    \hline

    \textit{Small input quantity (<5K)} &
    Not recommended &
    Not recommended &
    Preferred \\
    \hline

    \textit{Large input quantity ($>$10K)} &
    Preferred &
    Not recommended &
    Not recommended \\
    \hline

    \multicolumn{4}{l}{\textbf{Yield}} \\
    \hline
    &
    Unlimited number of output examples, although with diminishing returns to downstream task performance. &
    $\approx$ 1K input private examples $\rightarrow$ $\approx$ 25 synthetic examples out (200 tokens per example at $\epsilon=10$). &
    In practice, suitable for outputting synthetic dataset of size $\leq$ size of input private dataset. \\
    \hline
    
    \multicolumn{4}{l}{\textbf{Model access required}} \\
    \hline
    
    &
    Weights &
    Per-token prediction logits &
    Generations via API \\
    \hline

    \multicolumn{4}{l}{\textbf{Compute resources and engineering effort}} \\
    \hline

   \textit{ Training of LLM is required} &
    Yes &
    No &
    No \\
    \hline

   \textit{ Inference cost multiple per synthetic example} &
    1 (same as regular inference). &
    $\propto$ (number of contexts to aggregate over). &
    $\propto$ (number of iterations) $\times$ (number of rewrites). \\
    \hline

    \textit{Prompt engineering required} &
    No &
    No &
    Yes -- craft prompts for initial pure synthetic data and rewrite templates. \\
    \hline

    \textit{Time to first example} &
    Long (Run finetuning on the entire dataset, then sample) &
    Short (Batch-by-batch inference) &
    Medium (Requires prompt engineering + running rewriting and embedding on the entire dataset) \\
    \hline

    \textit{Direct side by side comparison of inputs and outputs} &
    No &
    Yes (input batch vs. generated output) &
    No \\
    \hline

    \textit{Resilience to distribution gaps between private data and LLM} &
    High &
    Medium &
    Low \\
    \hline

    \multicolumn{4}{l}{\textbf{Target privacy guarantee}} \\
    \hline

    \textit{High $\epsilon$ (e.g. 10)/large amount of private data} &
    Preferred &
    Not recommended &
    Not recommended \\
    \hline

    \textit{Stringent (e.g. $\epsilon < 1$)/small amount of private data} &
    Not recommended (quality will be bad) &
    Not recommended (short synthetic outputs and little data) &
    Preferred \\
    \hline

    \multicolumn{4}{l}{\textbf{Data persistence requirements}} \\
    \hline

    &
    Entire dataset required at once for training. &
    A batch of data required to generate synthetic examples for that batch. &
    Single examples can arrive in a streaming fashion, be used to cast votes, and then discarded immediately. \\
    \bottomrule
\end{longtable}
}

\subsection{Resampling DP Synthetic Text}\label{sec:text-resampling}

There are situations where it is beneficial to resample a subset from an initial differentially private (DP) synthetic dataset. This is often due to: (1) a distributional gap between the DP data and the original source data, caused by factors such as noise perturbation, clipping bias in DP algorithms, or suboptimal DP training hyperparameters; or (2) the need to focus on specific subsets of the data, such as synthetic text with a particular sentiment or length, even when the overall DP data distribution aligns well with the original training data. Additionally, one might wish to filter out noisy signals originating from the training corpus. These challenges have driven the development of resampling techniques for DP synthetic data, especially in the tabular data domain \cite{neunhoeffer2021private,liu2021iterative}. For example, \citet{wang2023post} proposed a method that privatizes the correlation matrix of an initial DP dataset and then resamples synthetic data to better match the underlying distribution of the source data.

\citet{xie2024differentially,yu2024privacy} extend the idea of resampling to the text domain using private voting idea similar to PE: first it constructs a histogram to represent the target data distribution, then resamples the initial DP data according to this histogram. \citet{xie2024differentially} construct the histogram by assigning each bin to an initial synthetic sample, with each private sample voting for the nearest synthetic sample in the embedding space. The histogram is released with DP guarantees and subsequently used to guide the resampling process. It is noteworthy that when the number of private samples is small, or when the size of the dataset targeted for filtering is large, releasing the histogram under DP can result in a poor signal-to-noise ratio.

To address this, \citet{yu2024privacy} propose clustering the initial DP synthetic samples in the embedding space, so that each bin represents a cluster of synthetic samples. This aggregation unites votes from private samples, thereby improving the signal-to-noise ratio in the DP histogram. Synthetic examples are then sampled from synthetic clusters that received highest number of private votes, and the number of synthetic samples is proportional to the number of votes a cluster received.

It is important to remember that such postprocessing uses the  sensitive data and requires privacy budget allocation. When combining with core technique like DP finetuning using the same sensitive data for both to improve fidelity, it is recommended to allocate the majority budget to DP finetuning and use the rest for DP postprocessing, e.g. for a target $\epsilon=10, \delta$ DP finetuning can be used for create DP synthetic pool of data with $\epsilon=9, \delta/2$ and postprocessing can use remaining $\epsilon=1, \delta/2$ budget to chose the synthetic samples that most closely resemble the sensitive data.

\subsection{Evaluating Synthetic Text Data Quality}\label{sec-text-metrics}
There are several fidelity metrics that can be used for comparing real and synthetic text data. 

\paragraph{Statistics-based metrics.} Tried-and-tested \textit{statistics-based metrics}, which compare statistics on real and synthetic data, can be useful for detecting problems in data preprocessing or training. Relevant statistics are often selected in an ad-hoc fashion, depending on the task at hand. For example, identifying significant difference of distributions of sequence lengths on real and synthetic data can signal context length or sampling problems \cite{kurakin2023harnessing}. \citet{kurakin2023harnessing} use n-grams statistics like unigram/bigram histograms and compare the area under their divergence frontiers \cite{sajjadi2018assessinggenerativemodelsprecision}, which is conceptually similar to MAUVE \cite{pillutla2021mauve}. Additionally, \citet{kurakin2023harnessing} find that the ordering of model candidates induced by test-set perplexity is highly correlated with downstream model performance and is even comparable with correlation of leading metrics like MAUVE. 

\paragraph{MAUVE.} \textit{MAUVE} \cite{pillutla2021mauve} is a recent and widely adopted metric that compares two text distributions. The idea behind this metric is to quantify the tradeoffs between Type I errors (areas of probability distribution in synthetic data that is unlikely under real data) and Type II errors (lack of synthetic data in the regions that are plausible for the real data). 

A practical algorithm for calculating MAUVE score first embeds real and synthetic data and clusters into $k$ clusters (using k-means) jointly. An optional step before clustering can involve doing PCA on the embeddings. Cluster assignments for real and synthetic data are counted, resulting in two histograms of support size $k$, which form the representation of real and synthetic distributions. KL divergence between two representations can quantify type I and type II errors, however it is $\infty$ when supports of histograms differ. Therefore a divergence curve that softly (by varying mixing weight) measures KL divergence of synthetic vs real and real vs synthetic data using the aforementioned representation is built. Each point on the curve quantifies tradeoffs between Type I and II errors, and the area under the curve is the reported MAUVE metric \citet{pillutla2021mauve}.

MAUVE exhibits behavior that is in line with expected human judgment. For example, larger models in general produce synthetic text that is higher quality and more similar to human created text, as judged by the human raters, and MAUVE values increase with model size increases. Similarly, MAUVE values are correlated with human judgments when comparing various decoding algorithms. MAUVE also correctly captures the fact that generating longer synthetic text is harder, as text becomes incoherent and inconsistent the longer the model decodes for. 

Practically speaking, the higher the MAUVE value, the more similar two measured text distributions. While 1.0 is the upper limit of MAUVE, measuring MAUVE from samples from the same distribution will not necessarily result in the value of 1.0, as MAUVE scores depend on the sample size. MAUVE should be first measured on two disjoint subsets of the real data (\emph{real vs. real}) to calculate a practical upper bound value that synthetic data will be measured against. Calculating \emph{synthetic vs. real} MAUVE should be done with the same sample sizes. In general, papers reporting MAUVE should indicate sample sizes and MAUVE algorithm hyperparameters used (e.g. the number of clusters $k$, whether PCA was used on the embeddings, etc.) to allow for fair comparison. While values of MAUVE on different sample sizes are incomparable, the ranking induced by MAUVE scores is consistent. Because of randomness induces by k-means and sampling of data to calculate MAUVE, we suggest to run several trials of MAUVE calculation with different random samples of real and synthetic data (with a fixed sample size) and report mean and standard error. 

Additionally, MAUVE strongly depends on the quality of the embeddings, with 'stronger' embeddings increasing the gap between \emph{real vs. real} and \emph{synthetic vs. real} values. Ideally, the embedding used captures the information that is essential for the downstream use of the synthetic data. This also somewhat blurs the line between MAUVE being a fidelity metric or utility metric. In practice it is rare to have access to an embedding from the downstream model at the stage of synthetic data generation, so encoders like RoBERTa \cite{DBLP:journals/corr/abs-1907-11692}, T5 \cite{raffel2020t5} or task-specific Gecko embeddings \cite{lee2024geckoversatiletextembeddings} can be used.

Finally, it is worth pointing out that MAUVE is a generic metric that can work with different modalities given an appropriate embedding model, and is now also commonly used for image data fidelity evaluation. 

\subsection{Open Questions and Challenges}\label{sec-text-open-questions}

Despite recent advances in DP synthetic text generation, several challenges remain in the field, including but not limited to the following.

\paragraph{Generation of long DP synthetic text.} Most existing work focuses on generating synthetic text up to a few hundred tokens \cite{yue2022synthetic,kurakin2023harnessing,yu2024privacy,xie2024differentially}, which is significantly shorter than the context window length of modern LLMs. \citet{yue2022synthetic} find that the length distribution of DP-generated text tends to be shorter than that of real data, likely due to the disparate impact of differential privacy \citet{bagdasaryan2019differential}. Additionally, it is known that in open-ended text generation, data becomes less coherent and consistent the longer the output is. This suggests that generating long DP synthetic text may pose additional challenges and might require different modelling techniques, for example breaking up large piece of text into overlapping chunks and finetuning (and subsequently sampling) such smaller chunks. 

\paragraph{DP synthetic text without pre-trained LLMs.}
Recent advances in DP text generation heavily rely on pre-trained foundational LLMs, which are trained on broad and often uncurated corpora. As a result, the DP guarantees typically apply only to the target dataset used for generating synthetic data, not to the pre-training data of the foundational model itself. Although using foundation models has become standard practice, there are concerns about potential privacy risks stemming from the pre-training data \cite{bommasani2021opportunities,tramer2022position}. Developing methods for DP text generation that do not depend on powerful pre-trained LLMs remains an open research challenge.

\paragraph{Challenges specific to main methods}
For DP-finetuning, the current state-of-the-art utilizes standard pretrained or post-trained checkpoint. However pretraining checkpoints are trained with teacher forcing next-token loss, encouraging the model to learn to generate the exact sequence of training data. This objective is arguabably more suited for post-training tasks like summarizaton or translation but less so for open-ended text generation that does not require exact matching of the target text, rather a generation of a text that is similar but not exactly the same. Additionally, essentially the same loss is used for DP-finetuning. Research info best ways of pretraining (and DP-finetuning) LLMs to maximize the utility and fidelity of open ended text generation is clearly needed. Perhaps training objectives similar to RL can be used to choose better generated examples, instead of enforcing the generation of the exact training example, as it is done currently. 

DP finetuning of models like Diffusion have been used for text generation \cite{li2022diffusionlm} and have demonstrated impressive results in non DP setting. Attempting such models with DP for synthetic data generation is a promissing direction to explore.

For private evolution, finding a way to "variate" text that does not require extensive prompt engineering can greatly expand the adoption and improve algorithm's utility.

DP inference methods have still a long way to go before effective generation of long text is achieved, which will require significant advancing in accounting such that more tokens can be generated for the same privacy budget.

\paragraph{DP Synthetic data generation in online manner:}
Online synthetic data genereation (e.g. in a streaming fashion when private data is available on instance-by-instance basis) is an underexplored area of DP synthetic text generation. Both PE and private inference methods are potentially suitable for this application, however to the best of our knowledge, such setups has not been yet been explored in the literature.

\section{DP Synthetic Data in Federated Learning}\label{sec-fl}
So far, when discussing DP synthetic data, we focused on \textit{centralized setting} where a trusted central server accesses a privacy-sensitive source dataset and produces DP synthetic version of this collection. In this section we discuss privacy-preserving synthetic data generation from \textit{sources decentralized} across a distributed system. User data frequently originates on edge devices (e.g. user phones or internal company storage), where local storage offers users greater control. Additionally, in sensitive domains like finance and healthcare, privacy and business considerations may hinder data sharing among organizations. While this decentralization introduces challenges compared to centralized data storage with a trusted provider, it also creates opportunities for distributing computational costs. In this section,  we specifically focus on the challenges and algorithms for generating synthetic data within the federated learning (FL) framework~\cite{mcmahan2017communication}, and discuss methods and challenges across various data modalities including tabular, text, and image, similar to centralized setting (Sections \ref{sec:tabular}-\ref{sec:text}). 

\paragraph{Roadmap}
This section is organized as follows. In Section \ref{sec-fl-primer} we give a brief overview of FL in context of ML. Next, we discuss the interplay between FL and DP in Section \ref{sec:dpfl}, and DP training algorithms in \Cref{sec:dpftrl-fl}. Section \ref{sec:app_fl} describes what DP synthetic data means in context of FL and outlines unique challenges one encounters when attempting to generate DP synthetic data. We briefly survey methods for FL DP synthetic data generation in Section \ref{sec:method_fl} and further compare and discuss them in Section \ref{sec:fl-discussion}. Section \ref{sec:fl-eval} outlines synthetic data evaluation peculiarities in context of FL.

\begin{figure}[htbp]
\centering
    \includegraphics[width=0.8\textwidth]{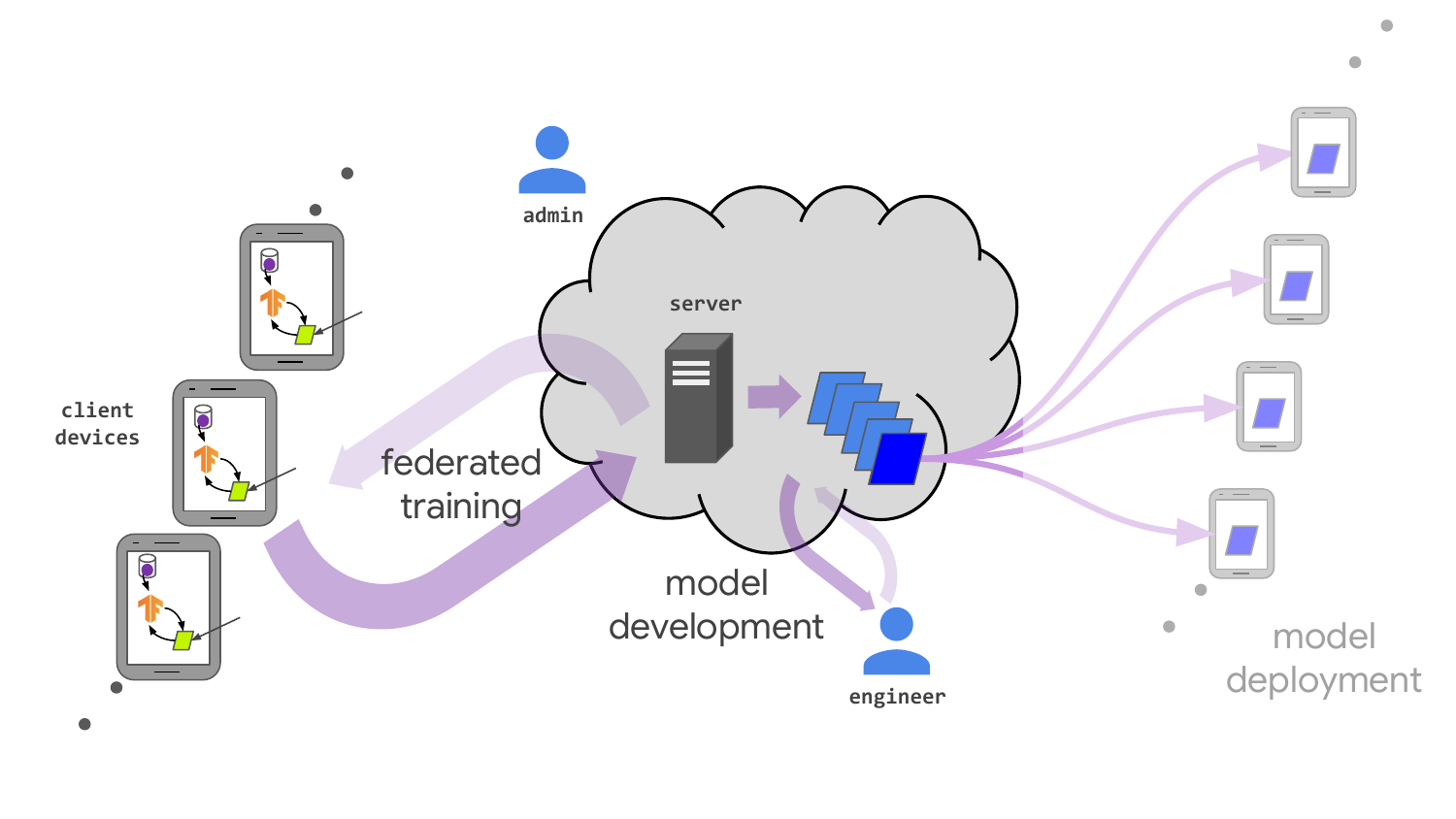}
    \caption{A typical cross-device federated learning system for mobile applications. The figure is adapted from \cite{kairouz2021advances,zhang23flgboard} with permission. Private user data are decentralized on client devices, and focused updates are ephemerally aggregated in line with the data minimization principle \cite{bonawitz2022federated}, an approach that can be further enhanced with techniques like encryption and differential privacy to achieve a multi-faceted model of privacy.
    }\label{fig:fl}. 
\end{figure}

\subsection{Federated Learning Primer}\label{sec-fl-primer}
Federated learning is a machine learning paradigm for decentralized data motivated by privacy protection \cite{mcmahan2017communication,kairouz2021advances,bonawitz2022federated}. \citet{daly2024federated} proposed this updated definition: \emph{\textbf{Federated learning} (FL) is a machine learning setting where multiple entities (clients) collaborate in
solving a machine learning problem, under the coordination of a service provider. A complete FL system should enable clients to maintain full control over their data, the set of workloads allowed to access their data, and the anonymization properties of those workloads. FL systems should provide appropriate transparency and control to the users whose data is managed by FL clients.}
FL is built on the privacy principle of data minimization, as the immediate aggregation of only focused updates that improve a specific model aims to minimize the exposure of data. Clients, who own and control their data, can range from mobile devices in cross-device FL to organizations with richer resources in cross-silo FL. The server, in turn, must orchestrate the machine learning process in a trusted manner. A typical machine learning process in FL is to train a task-specific model with decentralized data, and the final trained model is deployed to the clients for application use. \Cref{fig:fl} illustrates a typical federated learning system where data are decentralized on mobile devices. 

\begin{algorithm}[htbp]
\caption{A FL algorithm and \colorbox{lightgray}{its DP version}(when modifications in grey are applied) used to train production language models in a cross-device FL system, as presented in \cite{mcmahan2024hassle}. The DP version adapts FedAvg~\cite{mcmahan2017communication} with DP-FTRL~\cite{kairouz21practical} to ensure  user-level DP. Client-side optimization defaults to SGD to conserve edge device resources, while the server optimizer (ServerOpt) can use standard methods like momentum SGD. Standard DP parameters ($\zeta$, $\sigma$) are used; details on the correlated noise function are available in \cite{mcmahan2024hassle}}.
\label{alg:dpfl}
\begin{algorithmic}[1] 
\Require number of clients per round $m$, learning rate $\eta_{c}$ for clients and $\eta_{s}$ for server side updates, momentum $\beta=0.9$, total number of rounds $T$, \colorbox{lightgray}{clipping norm $\zeta$, noise multiplier $\sigma$, min-separation b}
\Ensure trained DP model (e.g., synthetic data generator) $W^{n}$
\State Initialize server side model $W^{0}$ (e.g. randomly or starting from pretrained weights)
\State Initialize server optimizer state $\optstate$
\State \HiLi{Initialize correlated noise state $\noisestate$ with $\sigma \zeta$}
\For{each round $t=1, 2, \ldots, n$}
    \State $\mathcal{Q}^t \leftarrow$ (at least $m$ users that did not participate in the previous $b$ rounds)
    \For{each user $i \in \mathcal{Q}^t$ \textbf{in parallel}}
        \State $\Delta^{t}_i \leftarrow$ ClientUpdate$(i, W^{t-1})$ 
        \Comment{Client receives a local copy of global model and returns update} 
    \EndFor
    \State \HiLi{$\tilde{\Delta}^t, \noisestate \leftarrow$AddCorrNoise$(\noisestate, \sum_{i \in \mathcal{Q}^t} \Delta^{t}_i)$} \Comment{DP-fy aggregated update with correlated noise} 
    \State $W^{t}, \optstate \leftarrow$ServerOpt$(W^{t-1}, \frac{1}{m}\tilde{\Delta}^t, \eta_{s}, \beta, \optstate)$
\EndFor

\Function{ClientUpdate}{$i$, $W_i$}
    \State $W_i^{(0)} \leftarrow W_i$, \,\, $\mathcal{G} \leftarrow $ (batches of user $i$'s local data)
    \For{batch $g \in \mathcal{G}$}
        \State $W_i \leftarrow W_i - \eta_c \nabla \ell(W_i;g)$  \Comment{Stochastic gradient descent} 
    \EndFor
    \State $\Delta \leftarrow W_i - W_i^{(0)}$
    \State \HiLiLong{$\Delta' \leftarrow \Delta \cdot \min{\left(1, \frac{\zeta}{||\Delta||}\right)}$} \Comment{$L_2$ norm clipping to control sensitivity for DP} 
    \State \Return $\Delta'$
\EndFunction
\end{algorithmic}
\end{algorithm}

While powerful, the FL paradigm presents several well-studied challenges. One of them is achieving the communication and computational efficiency in the distributed system with decentralized data. Additionally, the heterogeneous data on clients makes optimization of an ML objective on the server hard to converge, making training hard. Finally, privacy and security considerations are much more complex and nuanced in a complex system like FL.  

Among the many algorithms for training ML models in FL, Federated Averaging (FedAvg)~\cite{mcmahan2017communication} is one of the most popular ones. Algorithm \ref{alg:dpfl} presents both FedAvg algorithm (without highlighted lines) and its DP version (via DP-FTRL, with highlighted lines taken into account). 
FedAvg algorithm has two training loops: the inner loop of locally updating local models (copy of a global model) using clients data, and the outer loop of applying the updates aggregated from clients to the global model $W$. 
In each communication round of the outer loop, stochastic gradient descent is used on local clients to compute updates in cross-device FL~\Cref{fig:fl}, and momentum optimizer is used on the server when applying the (DP-fied) update aggregated from $m$ clients. 
FedAvg promotes data minimization and communication efficiency by distributing local updates on clients and infrequently aggregating client updates for the global model. While the aggregation step can incorporate DP and encryption methods to boost privacy in FL, the inherent data heterogeneity across clients can still pose a challenge to the FedAvg training paradigm. 

See \cite{kairouz2021advances} for a more detailed overview of federated learning. 
The clear delineation of responsibilities between clients and server in FL, where users maintain full control over their data, is a critical considerations for our subsequent discussion on differential privacy (DP) and synthetic data generation. 

\subsection{Differential Privacy in Federated Learning} \label{sec:dpfl}
Combining FL with DP provides robust privacy protection by leveraging both data minimization and formal anonymization guarantees~\cite{bonawitz2022federated}.
In the distributed paradigm of FL, a key consideration of DP is whether the server is trusted. \textbf{Central DP} is defined on the global learning process and the entire collection of decentralized data. For example, training an ML model or creating a DP synthetic data using the information about all the data from the clients without explicitly storing and pooling all clients data on the server. The server is trusted and can access intermediate results (e.g., aggregated but not noised model updates) in addition to the released results with DP guarantees (e.g., all model checkpoints $\{y^t\}$). \textbf{Local DP} is defined on each client's local data and learning process (e.g. each client trains and shares their own ML model or creates and shares a DP synthetic version of their own data). The server does not need to be trusted as the client updates have effective DP guarantees before aggregation. 
While local DP has the advantage of not requiring a trusted server, it struggles to maintain usable model (or data) utility \cite{kairouz2021advances}. Enhancing the FL system to reduce necessary trust on server with strong privacy-utility trade-off is an active area of research, exploring among other ideas aggregation in trusted execution environments (TEEs) \citep{daly2024federated}. In practice, a central DP guarantee is reported, and Figure \ref{fig:dpfl} illustrates how a central DP FL algorithm is deployed in cross-device FL. As we will discuss next in Section \ref{sec:distributed-trust}, achieving these central guarantees without fully trusting the server is an active area of innovation, relying on secure aggregation and TEEs~\citep{daly2024federated}.

\subsubsection{Distributing Trust: Secure Aggregation and TEEs}\label{sec:distributed-trust}
To recover the utility of central DP without relying on a fully trusted central server, a suite of approaches known as distributed DP is often employed~\citep{bittau17prochlo, kairouz2021distributed, agarwal2021skellam}. In a standard distributed DP setup, clients compute minimal application-specific reports, perturb them with a small amount of random noise, and execute secure aggregation protocol, a cryptographic multiparty computation protocol that allows the server to see the sum of high dimensional vectors without revealing the summands \cite{bonawitz2017practical, bell2020secure}. While the noise added by any individual client is typically insufficient for a meaningful local DP guarantee, the server only ever sees the aggregated output. This aggregate benefits from the total sum of noise added across all clients, yielding a strong central DP guarantee even against an honest-but-curious server.

However, physically distributing DP mechanisms in this way introduces significant practical hurdles. Multi-round cryptographic protocols are communication and compute intensive, and they are susceptible to client dropouts, a ubiquitous issue in cross-device FL that can silently compromise the formal privacy guarantee if client noise is lost. Furthermore, standard secure aggregation protocols struggle to efficiently distribute the correlated Gaussian noise required across training rounds for algorithms like DP-FTRL, making strong formal guarantees difficult to achieve at scale in production. Some novel research tries to develop stateful secure aggregation protocols to handle this setting but more work is needed to ensure these techniques are efficient and scalable in practice \citep{ball2024secure}. 

To overcome the fragility and overhead of cryptographic distributed DP, recent advancements leverage hardware-based Trusted Execution Environments (TEEs) via frameworks like Confidential Federated Computations (CFC)~\citep{eichner2024confidential}. TEEs create a secure, isolated, and publicly auditable enclave on the server where data is processed in a tamper-proof manner, inaccessible even to the server operator. This architecture relies on a cryptographic ``chain of trust:'' user devices encrypt data such that it can only be decrypted by a TEE that successfully attests to running a verified, open-source DP algorithm. By shifting the trust from the server operator to the hardware and the auditable code, CFC allows for the deployment of high-utility algorithms, including those using correlated noise—without sacrificing verifiability. This approach has already been successfully deployed for Confidential Federated Analytics on Google Keyboard (Gboard) to discover out-of-vocabulary words with tight privacy bounds~\citep{confidentialfa}. More recently, it was applied to Android's Recorder application where a open weight Gemma model was loaded in a TEE an used to label recording transcripts and only release their histogram with strong DP protections \citep{cheu2025toward}. While CFC's primary production use case has been analytics, applying TEE-backed confidential computations to federated model training and synthetic data generation represents a highly promising frontier for closing the privacy-utility gap in federated learning settings.

\subsubsection{Privacy Unit for FL}\label{fl-privacy-unit}
Apart from the choice of central, local, or distributed DP, a second crucial aspect is the choice of the privacy unit, similar to the discussion in the centralized setting of ML (Section \ref{sec-privacy-unit-general}). \textbf{User-level privacy\footnote{User-level privacy or more technically, device-level DP, as we use the device as a proxy user identifier} unit} is often a natural fit and practically achievable for cross-device FL in the \textbf{central DP setting} as data is already siloed by user devices, making each user an intuitive unit for privacy protection. For example \citet{gboard_dp_blogpost} launched production on-device language models for mobile virtual keyboard trained with FL.

Achieving \textbf{user-level DP for local DP is particularly challenging and is currently not practical} in cross-device FL because the corresponding noise is added on every client for each user. \footnote{Local DP setting is instead can be used with example-level privacy guarantees to preserve usable utility.}

Unless otherwise specified, from now on we focus our discussion on \textbf{user-level central DP} when describing methods of DP synthetic data generation in FL.

\begin{figure}[htbp]
\centering
    \includegraphics[width=\textwidth]{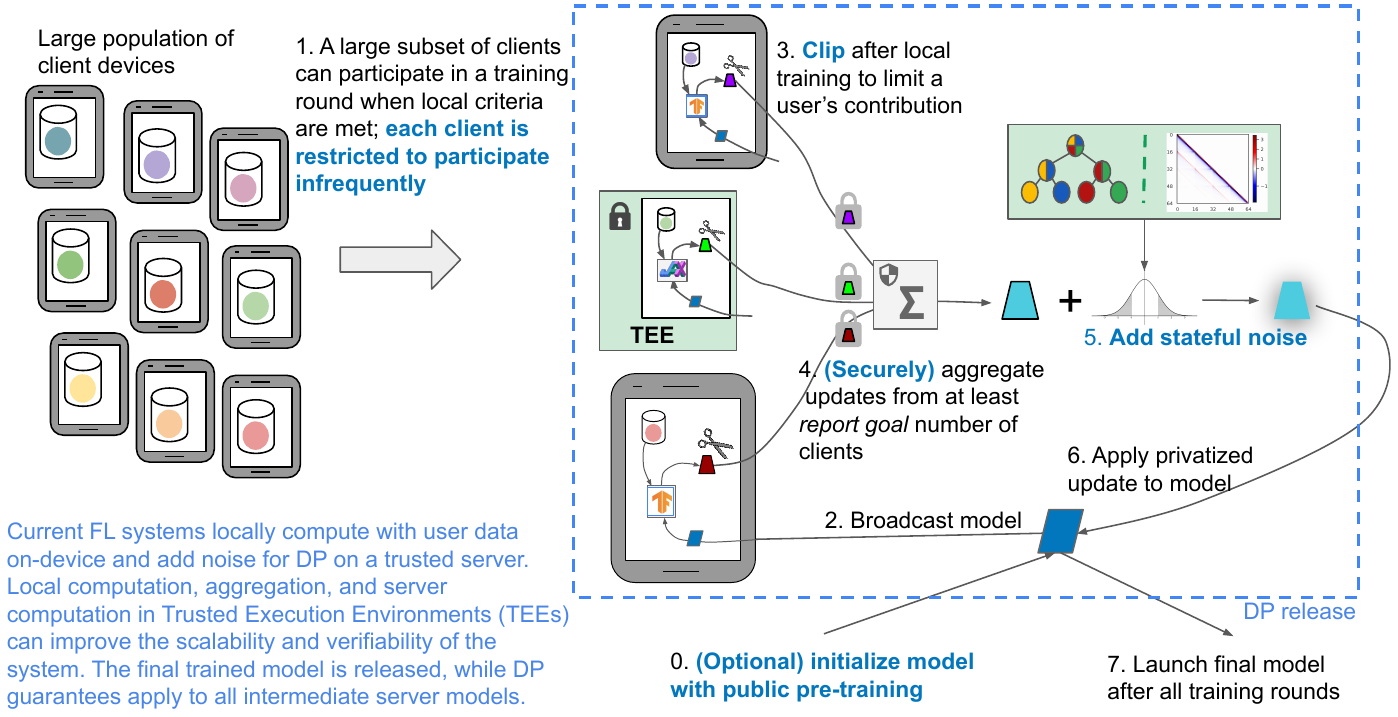}
    \caption{Federated learning with differential privacy deploying Algorithm \ref{alg:dpfl}. The architecture is based on current production systems with on-device model training~\cite{xu23gboard}, and the integration of Trusted Execution Environments for enhanced scalability, verifiability, and privacy in future systems~\cite{daly2024federated}. } \label{fig:dpfl} 
\end{figure}

\subsection{DP-Training in FL} \label{sec:dpftrl-fl}

Model training is an important technique in FL and it is also a fundamental component for DP synthetic data generation, echoing the discussion in the centralized setting (Sections \ref{sec:image} and \ref{sec:text}). Next we provide a deeper dive of the DP FL algorithm in Algorithm \ref{alg:dpfl} for model training as an example on adapting a DP algorithm to the FL setting, as well as preparing background for more discussion on synthetic data algorithms in FL in \cref{sec:method_fl}. 

Algorithm \ref{alg:dpfl} (with highligted lines) presents a DP-training version in FL setting. It is based on the generalized FedAvg (Section \ref{sec-fl-primer}). Instead of using the DP-SGD (Algorithm \ref{algo:dpsgd}) for global model training in the central DP setting, Algorithm \ref{alg:dpfl} adapts the DP-FTRL algorithm~\cite{kairouz21practical,mcmahan2024hassle} to FL. DP-FTRL uses correlated noise to achieve strong DP guarantee with high model utility without relying on data sampling for privacy amplification. In contrast, DP-SGD adds independent noise in each round, but relies on Poisson or specific form of sampling of clients to achieve strong DP guarantees. DP-FTRL overcomes a significant practical challenge in current cross-device FL: the inability to sample clients because clients are only available when restrictive criteria are met (e.g., charging, on an unmetered network, and idle)~\cite{kairouz2021advances,wang2021field}. Each client update $\Delta_i^t$ is constructed from model training on clients' local data, and then being clipped to control $\ell_2$ sensitivity. The client update process before aggregation not only improves communication efficiency in FL, but also naturally fits for user-level DP. 
While we mostly focus on central DP setting, we want to mention that Generalized FedAvg algorithm can also be adapted to  local DP setting and/or example-level privacy unit.  Instead of using DP-FTRL for server-side optimization as in Algorithm \ref{alg:dpfl}, if the noise is added in the CliendUpdate function for each $\Delta_i^t$, local DP can be achieved.
\paragraph{Local DP with user-level privacy unit}
 As the noise is added to each $\Delta_i^t$ for user-level local DP, significantly more clients ($n^2$ for local DP vs $n$ in central DP)  are needed to get similar signal-to-noise ratio comparing to adding noise on aggregated $\sum_i \Delta_i^t$ in central DP.
 As we mentioned before, local DP on each user device with user privacy unit struggles to achieve usable utility and \textbf{is currently not practical}. Local DP with user-level DP in cross-silo FL, where each organization has a large number of users, is achievable.
\paragraph{Local DP with example-level privacy unit}
To get example-level local DP, a straightforward approach is to use DP-SGD (Algorithm \ref{algo:dpsgd}) to replace SGD in the ClientUpdate function of the FedAvg algorithm. 

Though not directly comparable, user-level DP is fundamentally more challenging than example-level DP (Section \ref{sec:priv_unit}), and local DP is more challenging than central DP. 

\subsection{DP Synthetic Data in FL} \label{sec:app_fl}

The value of (DP) synthetic data is amplified in FL, as it enables privacy-preserving modeling on decentralized data. 
Synthetic data can be generated to mirror either the global distribution of the entire decentralized data (\textbf{global synthetic data}) or the local distributions of individual, heterogeneous clients (\textbf{client synthetic data}). In the literature, it is more common to use central DP for global synthetic data, and local DP for client synthetic data, as discussed in Section \ref{sec:method_fl}.

The applications of synthetic data in FL are numerous. It can serve as a proxy for decentralized data, which can then be safely released outside of the conventional FL cycle (Figure \ref{fig:dpfl}) thanks to the post-processing property of DP. For instance, early work used global synthetic data for debugging applications on private domains with unseen data~\cite{augenstein2020generative}. More recent work highlights the advantages of using global synthetic data as a proxy to flexibly combine public and private knowledge, potentially simplifying the system stack and reducing maintenance costs~\cite{houpre,zhang2025synthesizing}. Furthermore, \citet{zhang2025synthesizing} highlight this proxy data can be inspected  and then used in standard datacenter training pipelines to train much larger production models than are feasible in current FL systems.

Synthetic data can also be used to directly improve the FL training process itself. In cross-silo FL, client synthetic data with local DP guarantees has been used to improve federated optimization in image applications on academic benchmark datasets~\cite{xiong2023feddm,abacha2024synthetic}. In cross-device FL, global synthetic data has been used in pre-training to enhance the DP-FL training of on-device production language models~\cite{wu2024prompt}.

Generating DP synthetic data in FL is more challenging than in datacenter settings due to the inherent principle of data minimization.  We highlight challenges that are mostly relevant for synthetic data generation, mirroring the more general FL challenges in \cite{wang2021field}:
\begin{itemize}
    \item Communication and computation efficiency. Generating high-quality synthetic data is a non-trivial task and often requires extensive computation resources for both fidelity and privacy. While methods like FedAvg designed for communication and computation efficiency are particularly effective and widely used \cite{wang2022unreasonable}, synthetic data generation algorithms are often far more resource-intensive. This is particularly acute in cross-device FL, where users' mobile devices have limited computational power and network bandwidth~\cite{kairouz2021advances}. Current production cross-device FL systems support reliable training models of less than 100M parameters; although next-generation systems with TEEs may offer more resources, efficiency remains a primary consideration for supporting real-world applications~\cite{daly2024federated}.
    \item Data heterogeneity and system complexity: Heterogeneous client data that affects the model training speed and quality is extensively studied in FL~\cite{kairouz2021advances,wang2021field}. The system complexity of the distributed training paradigm on decentralized data often poses challenges, for example, the trust on server. The characteristics of the system can be challenging for achieving strong privacy-utility trade-off, and require system algorithm co-design such as the DP-FTRL algorithm. These factors, already central to FL as briefly discussed in the previous sections, add further layers of difficulty to the task of generating high-fidelity synthetic data that accurately reflects diverse client distributions while upholding strong privacy guarantees.
\end{itemize}

Overcoming these challenges presents a significant opportunity. Harnessing the vast amount of data born and siloed on user devices in a privacy-preserving manner could unblock substantial improvements in both the model capacity and user experience of next-generation LLMs, especially for mobile applications. 

\subsection{Methods for Federated DP Synthetic Data} \label{sec:method_fl}

{\footnotesize
\begin{longtable}[htbp]{p{3.5cm} p{1.5cm} p{1.5cm} p{3.5cm} p{3.5cm}}
\caption{DP synthetic data methods in FL. This list does not include methods that are potentially feasible in FL but have not been applied in FL yet, or methods in FL do not use DP.  DP-in-MPC (multi party computation) is central DP with reduced trust on the server. Dataset distillation is a learning process that trains the input instead of the model weights, using fixed model and backpropagating to the input. This input will become synthetic data. Dataset distillation can be viewed as a compression of real data into smaller amount of synthetic data. \label{tab:fl_methods}
} \\
\toprule
\textbf{Method Paper} & \textbf{Modality} & \textbf{Data Distribution} & \textbf{DP Setting} & \textbf{Key DP Method} \\
\midrule
\endfirsthead 

\toprule
\textbf{Method Paper} & \textbf{Modality} & \textbf{Data Distribution} & \textbf{DP Setting} & \textbf{Key DP Method} \\
\midrule
\endhead 

\citet{pentyala2024caps} & Tabular & Global & DP-in-MPC & select-measure-estimate\\
\hline
\citet{maddock2024flaim} & Tabular & Global & Central, example-level &  select-measure-estimate \\
\hline
\citet{xiong2023feddm} & Image & Client & Local, example-level & DP dataset distillation \\
\hline
\citet{abacha2024synthetic}  & Image & Client & Local, example-level & private evolution \\
\hline
\citet{chen:gswgan} & Image & Global & Local, example-level & DP model training \\
\hline
\citet{augenstein2020generative} & Image \& Text & Global & Central, user-level & DP model training \\
\hline 
\citet{houpre,hou2025private} & Text & Global & Central, user-level & private evolution \\
\hline
\citet{wu2024prompt,zhang2025synthesizing} & Text & Global & Central, user-level & post-processing with small DP LMs \\
\bottomrule

\end{longtable}
}

Many methods for generating DP synthetic data can be adapted for FL, provided that the constraints on communication, computation, and the inherent privacy-utility tradeoff are carefully considered. We discuss the key algorithms of DP synthetic data for tabular, image, and text in FL and assume readers have gained familiarity with the methods from the corresponding modalities (Sections \ref{sec:tabular}, \ref{sec:image} and \ref{sec:text}). Table \ref{tab:fl_methods} provides a summary of DP synthetic data papers in FL we discuss next.

\subsubsection{Methods for Global and Client Synthetic Data}

We discuss methods for generating global synthetic data that can capture the combined distribution from decentralized data, as well as client synthetic data that can capture the heterogeneous client data at a finer grain. More specifically, using $\mathcal{P}$ to denote the distribution on the population of clients, using $\mathcal{D}_i$ to denote the data distribution of client $i \sim \mathcal{P}$, global synthetic data are sampled from the global distribution, $X \approx\{x | x \sim \mathcal{D}_i, i \sim \mathcal{P} \}$; client synthetic data are sampled separately, $\mathcal{X} = \{ X_i | X_i \approx \{ x \sim \mathcal{D}_i\ | i \sim \mathcal{P}\}\}$. The interpretation of $\mathcal{P}$ is essential: if we view $\mathcal{P}$ as a discrete distribution over the specific users participating in FL, then user-level DP will not be possible, as discussed below. On the other hand, if we view $\mathcal{P}$ as a distribution over say user profiles, then we may be able to first sample DP synthetic user profiles, and then sample a client dataset for each such profile.

\paragraph{Methods for client synthetic data}
Obtaining client synthetic data can be done in local DP setting by using client's data to finetune a local model (e.g. LLM) with DP-SGD and then subsequently sampling from it. Client synthetic data techniques are therefore often direct analogs to methods in datacenter settings that we explored in previous sections. These techniques apply DP synthetic data algorithms to client data, optionally incorporating a global model that has already learned the commonalities across clients. Only few works discuss client synthetic data, with the focus of using such data for improving FL model training for images~\cite{xiong2023feddm,abacha2024synthetic}.

Client synthetic data in these studies provide only example-level privacy. Beyond the general challenges with local DP discussed in  \cref{sec:dpftrl-fl},  there is a more fundamental limitation when generating synthetic data for each (real) user under user-level DP: the client synthetic data can only learn commonalities across users, and cannot reveal information of a specific user. That is, the user-level DP guarantee says that the synthetic client data for one user must be indistinguishable from the synthetic client data of any other user, defeating the point of modeling heterogeneous client data distributions.
In cross-silo FL, user-level DP for a client can be possibly achieved as each client may contain data from many users. Alternatively, in cross-device FL, the recent hierarchical synthetic data generation approach can be adopted \cite{syntheticblogpost2025,hu2025actgarl}: client synthetic data can be generated from synthetic user profiles. Such client synthetic data from synthetic users are compatible with user-level DP, and can satisfy both central DP and local DP. These client synthetic data are useful for improving privacy-preserving model training in federated learning.

Additionally, global DP models can be used on clients as teacher models for knowledge distillation to improve FL model training \cite{yao2021local,zhu2021data}. In this setting clients use teacher model to synthesize labels or augment features on their own data, and uses the augmented data to improve FL training without explicitly collecting synthetic data.  

\paragraph{Methods for global synthetic data}
Obtaining global synthetic data can be done in many ways. One potential approach is to combine local DP client synthetic data from all clients. A better alternative is to use the central DP model trained with Algorithm \ref{alg:dpfl} and use it to sample global synthetic data.  The latter approach is more commonly used and provides better utility and fidelity of the DP synthetic data. 

Our primary focus from now on, unless mentioned otherwise, will be on methods for generating \textit{global synthetic data} that reflects the overall distribution of data across the users. We will discuss tabular, image, and text data, respectively.

\subsubsection{DP Synthetic Tabular Data in FL}
 Synthesizing tabular data is often less resource-intensive and many methods can be straightforwardly federated. For 
 example, many mechanisms from the select-measure-estimate paradigm (Section \ref{select-measure-estimate}) can be distributed with federated analytics. The key insight is that these mechanisms only depend on the data through a specific set of low-order marginals, which can readily be computed by federated analytics systems. After these sufficient statistics have been collected, no more server to client communication is necessary to run the remainder of the mechanism.  Additional work in this space includes \citet{maddock2024flaim}, who demonstrated how to address data heterogeneity in FL setting. Additionally, \citet{pentyala2024caps} have proposed secure multi-party computation protocols to replace the trusted aggregator in workload-based tabular synthesis methods.

\subsubsection{DP Synthetic Image and Text Data in FL}
The key algorithms for image and text, just as in central setting (Sections \ref{sec:image} and \ref{sec:text}) can be categorized similarly into DP 
training/finetuning, DP inference, and Private Evolution.

\paragraph{DP Model Training.}
Early work on synthesizing image data in FL focused on Generative Adversarial Networks (GANs), carefully selecting components (e.g., discriminators) that could be federated~\cite{augenstein2020generative,chen:gswgan,cai2021generative}.  

More recently, synthetic text data came into a spotlight. The popular approach of DP fine-tuning Large Language Models (LLMs)~\cite{yue2022synthetic,kurakin2023harnessing} is difficult to apply directly as is in \cref{alg:dpfl} because of the computation requirement of updating local LLMs on clients. While communication costs can be addressed with compression, the primary bottleneck is local computation. Parameter-efficient fine-tuning (PEFT) methods like LoRA or prompt tuning can help~\cite{hu2022lora,lester2021power,cho2024heterogeneous,kurakin2023harnessing}, but they often still require backpropagation through billion-parameter models, which is computationally expensive, if not impossible, in many FL environments.

\paragraph{DP Inference methods.}
DP Inference-based methods are unexplored for DP synthetic data in FL. One naive approach is to aggregate in DP manner the outputs from client models when training a central model for generating global synthetic data (Algorithm \ref{alg:dpfl}). However, PATE-style aggregation  (\cref{sec-dp-pate-image}) would need public data, and/or a large amount of clients reliably connected at the same time. Private inference for text similar to ~\cite{amin2024private,amin2025clustering} that utilize in-context learning and generates a sample token by token (Section \ref{sec:text}) are challenging in FL because generating a single sample in central DP will require multiple communication rounds for sequential decoding and sufficient number of clients connected at the same time. 

\paragraph{Methods that use LLM inference only}
Prompt-based methods that use foundation models without retraining or finetuning \cite{wu2024prompt,zhang2025synthesizing}, including private evolution algorithms~\cite{xie2024differentially,houpre,hou2025private}, are more lightweight as they only require exchanging DP statistics in each round. 
\citet{wu2024prompt,zhang2025synthesizing} train a small DP LM (less than 10 million parameters) to reweight and filter synthetic data generated by prompting public LLMs.
\citet{houpre,hou2025private} extends private evolution algorithms (Algorithms \ref{alg:image_pe} and \ref{alg:aug_pe}) by sending candidate synthetic data to clients, aggregating the voting histogram from clients,  and selecting seed synthetic data for the next round. 
While these methods show promising early results, they often fail to capture detailed domain information due to inability to update the model that generates the synthetic data, leading to a utility gap on downstream tasks compared to DP fine-tuning on real data \cite{wu2024prompt} or to suboptimal privacy-utility tradeoffs~\cite{xie2024differentially,tan2025synthesizing}. \citet{hou2025private} incorporates model updates into PE: authors update PE with policy optimization approach, where pairs of synthetic examples generated by the same prompt but having different voting scores used to guiding the LLM towards generating better synthetic data via a policy optimization algorithm with LoRA finetuning. 

 Recent research demonstrates the potential of combining DP fine-tuning with lightweight DP statistics calculated on private data (e.g. topics of the sensitive data) ~\cite{yu2024privacy,tan2025synthesizing}. Notably, \citet{tan2025synthesizing} generated high-quality DP synthetic data by fine-tuning a relatively small model (140 million parameters) that was conditioned on DP statistics of data aspects like topics. This hybrid approach represents a promising direction for balancing privacy, utility, and computational feasibility in federated text synthesis.

\subsection{Evaluating Synthetic Data in FL}\label{sec:fl-eval}
The evaluation of DP synthetic data in FL shares many considerations with the centralized setting. While metrics specific to data types like tabular, image and text data are covered in their dedicated sections (Sections \ref{tabular-metrics}, \ref{sec-image-quality}, \ref{sec-text-metrics}), this section discusses evaluation methods that are commonly employed due to the decentralized nature of FL.
\begin{itemize}
    \item \textit{Human inspection}: The value of direct human inspection should not be underestimated. As discussed in Section \ref{sec:app_fl}, a common application of synthetic data in FL is for debugging and inspection, as the underlying decentralized data cannot be directly observed. For all applications, developer "eyeballing" of canonical synthetic samples and performing extensive data analysis are invaluable evaluation steps.
    \item \textit{Downstream Model Performance}: For production applications, the most definitive evaluation often comes from deploying a downstream model trained on synthetic data and measuring its performance through live A/B testing~\cite{wu2024prompt,zhang2025synthesizing}. An insightful intermediate step is federated evaluation, where the model trained on synthetic data is sent to clients to be evaluated on their real, locally stored data, providing a high-fidelity measure of performance without centralizing the test data~\cite{wu2024prompt}.
    \item \textit{Distributional Fidelity metrics}: Another approach is to measure the difference between the distribution of the synthetic data and the real user data, using scores like MAUVE~\cite{pillutla2021mauve} or FID~\cite{NIPS2017_8a1d6947}. While straightforward for academic benchmarks where a real evaluation dataset is available, this can be challenging in a practical FL system. It requires either sending synthetic data statistics to clients for local computation and aggregation, or first collecting differentially private statistics from clients to compute the score on the server, both of which add complexity.
\end{itemize}

\subsection{Discussion}\label{sec:fl-discussion}
Due to computational constraints of on-device learning in FL, methods that avoid DP training of large models remain the most practical as of now~\cite{synthfed_blogpost}. 

As a starting point, prompting LLMs with domain prior knowledge about client's  data \cite{wu2024prompt,zhang2025synthesizing} is a practical and effective method, although it results in simple synthetic data, not DP synthetic data that reflects the real data distribution. In this setting, synthetic data is entirely artificial (obtained from public models).  
For higher-quality data suitable for debugging or developer inspection, private evolution algorithms that select and evolve instances of such public data ~\cite{xie2024differentially,houpre} offer promising improvements in FL settings. For applications where maximizing the performance of downstream models is the primary objective, DP fine-tuning methods are most effective, particularly when combined with DP statistics~\cite{yu2024privacy,tan2025synthesizing}.
However, realizing the full potential of DP fine-tuning in practice will require not only algorithmic development but also critical system-level improvements to address the communication and computation constraints inherent in FL~\cite{daly2024federated}.

Algorithmic development for both current and future FL systems remains an important and active research direction. For research, it is an open question whether the recommendations from existing studies that considered only smaller LLMs (up to 13 billion parameters~\cite{yu2024privacy}) will scale effectively to much larger models. For production, a significant engineering challenge lies in advancing beyond current FL systems, which largely support only prompt-based or lightweight methods, to build future infrastructure that offers more resources while maintaining strong privacy protections. Finally, the extent to which DP synthetic data can fully represent the utility of decentralized data across all possible use cases, and how much resources are needed to generate such synthetic data remains a fundamental open question.

\section{Practical Components of DP Synthetic Data Generation System}\label{sec-practical}
This section outlines how one can go about building a DP synthetic data generation system end-to-end. We will touch upon important questions that should be answered, including which privacy guarantees to target (Section \ref{privacy-guarantees-decisions}). We will explore data preparation  including user level contribution bounding (Section \ref{user-contrib-bounding}), and safe sensitive data handling practices (Section \ref{sec-safe-data-handling}). We will describe a crucial component that should be in place before any DP synthetic data is used in downstream tasks or shared -- namely \textit{empirical privacy testing} (Section \ref{sec:empirical-privacy-auditing}). Finally, we will conclude with a discussion of lineage tracking (Section \ref{lineage-tracking}). 

\subsection{Privacy-Related Decisions}\label{privacy-guarantees-decisions}
There are several important decisions that need to be made by the designer of a system that will produce and employ DP synthetic data. Namely, one must decide what privacy guarantees to target and use that to select an appropriate privacy unit.

\paragraph{Choice of privacy unit} is one of the most important choices to make. We outlined 
possible privacy units in Section \ref{sec-privacy-unit-general} and provided discussions on privacy units for each data modality (Sections \ref{sec-tabular-unit}, \ref{dp-image-privacy-unit}, \ref{dp-text-privacy-unit} and \ref{fl-privacy-unit}). While the appropriate choice is always application specific, we emphasize that in most cases \textit{example-level privacy unit carries limited meaning in real world applications}, and \textbf{user-level} or even larger privacy units like \textbf{group/organization privacy units} are preferred. It is also important to remember that ideal privacy unit might be not achievable for various reasons, including a 
technical inability to identify the same user across multiple data records in order to ensure user-level privacy. We advocate for all practical compromises to be explicitly documented and released with the DP synthetic data. 

\paragraph{Target DP parameters ($\epsilon, \delta$)}
The $\delta$ parameter is usually set to be less than inverse size of the private dataset being protected (e.g. $1/n^{1.1}$, where $n$ is the number of privacy units within the dataset). For choosing the $\epsilon$ parameter, \citet{Ponomareva_2023} outlined 3 tiers of privacy guarantees in context of ML models trained with DP, which are also applicable to DP Synthetic data. 

\begin{specialistbox}{Target $\epsilon$ values for DP synthetic data \cite{Ponomareva_2023}}
We assume \textbf{user privacy unit (or example-level where a single user or other appropriate group contributes at most one
example)} with the add-or-remove or zero-out adjacency.
\begin{compactenum}
    
\item 
\textbf{Tier 1: Strong formal privacy guarantees.} Strong privacy protection achieved under $\epsilon \le 1$ directly from the DP definition. Such low values however often result in very low quality (utility and fidelity) of synthetic data, sometimes to a point of being too low to be useful. 
\item \textbf{Tier 2: Reasonable privacy guarantees.} $\epsilon \le 10$. The values from this range are currently widely used for DP synthetic data generation in academic papers and production applications. Empirical privacy auditing (Section \ref{sec:empirical-privacy-auditing}) and training data preprocessing (Section \ref{data-prep}) are highly encouraged in this regime. 
\item \textbf{Tier 3: Weak to no formal privacy guarantees.}  $\epsilon >10$. While any level of formal DP is an improvement over synthetic data with no formal DP guarantees, 
DP guarantee on its own for values of $\epsilon$ from this tier is vacuous. Additional measures to ensure sufficient randomization, including empirical privacy auditing and training data preprocessing are paramount. 
\end{compactenum}

\textbf{Practitioners are encouraged to choose the lowest achievable tier from the aforementioned tiers.}

\end{specialistbox}

Finally, it is important to remember that both the method that is chosen for DP synthetic data creation and the modality of data at hand will determine which tiers may be feasible to achieve at reasonable privacy-utility-computation tradeoffs. For example, private evolution methods for images (Section \ref{sec-pe-image}) can (with a high-quality set of seed data) potentially achieve low $\epsilon$ values including Tier 2 and even Tier 1 level guarantees, as can workload-based methods (Section \ref{sec-workfload-based}) for tabular data and DP-training smaller language models~\cite{gboard_dp_blogpost}. For small language models, DP training can achieve good utility for small $\epsilon$ values from Tier 1 and 2 \cite{gboard_dp_blogpost}. 
However, for DP-training based methods using large models like diffusion models and LLMs on complex modalities like text or image, such low $\epsilon$ values are currently not practically obtainable. 

\subsection{User Contribution Bounding}\label{user-contrib-bounding}
As we have seen in previous sections, user-level privacy often provides the most straightforward and semantically meaningful privacy guarantees. However, many algorithms, for example DP-SGD, in their vanilla forms assume example-level privacy. Further, most of the tried and tested libraries for DP training like Opacus \cite{DBLP:journals/corr/abs-2109-12298}, Tensorflow Privacy \cite{tensorflow2015-whitepaper}, Jax privacy \cite{jax-privacy2022github} provide implementations for example-level DP only.

In order to support a user-level privacy unit, a practitioner has several choices. One involves reducing the problem to example-level privacy and accounting for user-level unit post factum. Straightforward application of group-level privacy \cite{650758} (as covered in discussion in Section \ref{sec-privacy-unit-general}) 
requires a bound on group size $b$, which represents the maximum number of examples each user might contribute to the dataset. To obtain this bound, data can be subsampled to ensure \textit{data level user contribution bounding}. Once the data is subsampled, any example-level DP algorithm like DP-SGD can be used with $\epsilon/b, \delta/b$ values to account for groups of size $b$ \cite{Vadhan2017}. Further, the accounting in some cases like for DP-SGD can be significantly improved further, as we discuss next.

An alternative to reframing the problem to example-level privacy is to employ \textit{algorithms that directly support user (or larger) privacy units} and ensure contribution bounding as part of the algorithms.
 
While next we cover these two types of solutions in the context of user level privacy, similar recipe can be applied to larger privacy units like groups of users, organizations etc.

\paragraph{User contribution bounding during preprocessing}
User contribution bounding during preprocessing involves selecting a maximum-size collection of examples from the input such that each user is associated with at most $b$ of the selected examples. When $b=1$, problems with a user-level privacy unit are essentially reframed as problems with an example-level privacy unit and all the algorithms we discussed previously can be used without additional modification. However, selecting more than 1 datapoint per user is often beneficial --- this results in a larger and possibly more diverse training dataset (which in turn allows to use less noise to ensure the same level of privacy, providing better utility). Additionally, users with more data points might have higher quality data, and removing most of their datapoints via subsampling to 1 example per user can be detrimental to data quality \cite{aminContributions19}.

For \textit{single-owner data}, such data selection can be achieved independently across users as they do not share any examples \cite{charles2024finetuninglargelanguagemodels}. Each user can pick $b$ of its own examples without interfering with the selection process of other users. 

For \textit{multi-attribution scenarios}, where one example can be associated with multiple users, bounding user contributions is more involved. 
\citet{ganesh2025itsdatatooprivate} propose a sequential greedy algorithm for the task of maximally selecting instances from a multi-attribution dataset subject to a user contribution bound. \citet{ganesh2025itsdatatooprivate} show that the result of their greedy algorithm is not too far from the optimum upper bound on the number of selected items achieved through a linear programming formulation. 
Naturally, sequential algorithms can be limited in their scalability as they must process all examples in one machine one by one.
Recent work by \citet{cohenaddad2025scalablecontributionboundingachieve} addressed this issue introducing a  scalable, distributed algorithm to address data level user-level differential privacy contribution bounding problem. This algorithm can be implemented in large-scale parallel and distributed architectures including in MapReduce-like frameworks \cite{dean2008mapreduce,zaharia2016apache}. The authors modeled the data ownership as a hypergraph, where users are vertices and examples are hyperedges. This modeling allows to cast the example contribution bounding problem as maximizing the number of records while ensuring that each user's participation in the selected records does not exceed a predefined budget $b$. This algorithm operates in rounds. In each round, unsaturated users propose their most preferred examples up to their remaining capacity (using an arbitrary but consistent ranking function). An example is only added to the final set if all its participating users unanimously propose it. Given that the algorithm can process each round in parallel it can scale to massive datasets beyond the capacity of a single machine. 
Once the data is selected, any example-level algorithm like DP-SGD, Private inference or Private Evolution can be used, however some modifications to accounting will still be required to account for at most $b$ samples per user. Improvements over group privacy exist for DP-SGD algorithm. 
\citet{charles2024finetuninglargelanguagemodels} propose DP-SGD-ELS which employs modified accounting to standard DP SGD algorithm. Instead of relying on group privacy, they leverage the Mixture-of-Gaussians mechanism to derive optimal accounting for DP-SGD.  This can decrease compute costs or improve accuracy at a fixed epsilon (alternatively, when holding the number of training iterations and batch size constant, increasing number of samples per user does not significantly affect the epsilon, so larger $b$ can be used at no cost in the level of noise introduced by DP-SGD and the same $\epsilon$ guarantees). The suggested number of samples per user $b$ is chosen as a median number of samples across the users, although it can be treated as a hyperparameter and further tuned. 

The beauty of data sampling contribution bounding lies in the fact that it does not require changes in training algorithms or data ingestion, allowing the use of established frameworks like  Opacus \cite{DBLP:journals/corr/abs-2109-12298}, Tensorflow Privacy \cite{tensorflow2015-whitepaper}, Jax privacy \cite{jax-privacy2022github} and others.  

\paragraph{Algorithms that directly support user-level contribution bounding} Along with previously mentioned DP-SGD-ELS,  \citet{charles2024finetuninglargelanguagemodels} introduces a variant DP-SGD-ULS (user-level sampling and per user gradient clipping) for handling user privacy unit directly in the single-attribution case. While this algorithm foregoes the need for contribution bounding during pre-processing, the  data needs to be partitioned and sampled per user during DP-Training. Additionally, clipping of gradients happens on a per-user level as well. ULS often outperforms ELS (data subsampled) variant in fixed compute setting with the magnitude of the improvement is the largest when the data between users is highly variable, and in small $\epsilon$ and large compute budget. The downsides are significant modifications needed to the training data ingestion pipeline (propagating user ids), batch preparation (sampling per user data) and actual training algorithm (e.g changing the clipping of DP-SGD from per example to per user etc). ULS is not readily available in any of the well known DP-Training libraries so far. That being said, ULS is closely related to differential privacy in federated learning discussed in \cref{sec:dpftrl-fl}. For example, ULS can be constructed from \cref{alg:dpfl} by using one step of gradient descent in $ClientUpdate$, and carefully adjusting the client sampling strategy to construct $\mathcal{Q}^t$ and noise mechanism in aggregating client updates. Previous work~\cite{mcmahan2018learningdifferentiallyprivaterecurrent,xu2023learning} have implemented such algorithms by combining TensorFlow Privacy~\cite{tensorflow2015-whitepaper} and TensorFlow Federated~\cite{tfflib} libraries.

\subsection{Pitfalls to Avoid When Implementing DP-training}\label{sec-gotchas}
As we have seen previously, for modalities like image and text, DP-training based methods remain the workhorse methods for creating DP synthetic data. These methods rely on obtaining DP models via DP-Training algorithms like DP-SGD. However, care should be taken when swapping the optimizer to a DP-SGD-based one. 
We will outline several known angles that need to be considered when choosing a model and its parameters, however we want to stress that in order to obtain a proper DP model, the synthetic data creator needs to have a deep understanding of the model they are using. While ideally frameworks that implement privacy should validate any privacy critical assumptions, currently this is not sufficient. 
Please refer to \cite{Ponomareva_2023} (Section 5.5.1) for additional discussion of model components that can invalidate privacy guarantees.

\paragraph{Packing}
For LLM models, packing is one of the most commonly used way of improving training efficiency. The training time of LLMs depends on the sequence length of the longest examples in the batch. When a batch contains examples of various lengths, short examples can be padded up to a maximum example length. However such padding wastes compute, so \textit{packing} is often employed instead. Packing appends several shorter training example, forming one example of length up to the maximum context sequence length \cite{raffel2020t5}. 

In Section \ref{dp-text-privacy-unit} we previously outlined that the choice of how to split the text into instances affects the meaning of privacy unit and obtained guarantees. Packing may similarly subtly change the interpretation of the unit of privacy if off-the-shelf algorithm is not adjusted. In the simplest case, when multiple examples are packed together, masking is commonly done to prevent tokens from different examples attending to each other \cite{chung2022scalinginstructionfinetunedlanguagemodels}. For implementation that don't do such masking and examples can attend to each other, an analogy of DP with microbatching can be drawn (\cite{Ponomareva_2023} Section 5.6). This means that sensitivity calculation needs to be adjusted to ensure DP guarantees apply to the original unit of privacy. Such packing also violates Poisson sampling assumptions, which can be handled with adjustment of the noise injected during the training. A microbatching analogy can be also drawn for packing algorithms that create any data dependency and choose which examples to pack together based on other examples. For example, \cite{staniszewski2025structured} introduces packing algorithms that packs together documents that are similar in the embedding space.

Some commonly used in pretraining packing algorithms like \textit{concat-then-split} \cite{grain2023github} can additionally end up splitting one example into two, which can be handled with the help of \textit{group privacy} \cite{DworkRothBook:2014,charles2024finetuninglargelanguagemodels} or sensitivity adjustments.

We recommend to keep the packing off when using off-the-shelf DP finetuning methods for text synthetic data generation. If packing is used, careful sensitivity and accounting adjustements as outlined above will be needed 

\paragraph{Inconsistencies between accounting assumptions and practice} Accounting for DP-SGD-like algorithms often relies on privacy amplification via subsampling.  The most commonly used amplifications theorems assume Poisson sampling of examples (where each example is selected independently with constant probability, leading to variable-sized batches). However most actual implementations shuffle the data (often incompletely) and then cycle over the fixed-size batches \cite{Ponomareva_2023}. To the best of our knowledge, so far only Opacus \cite{DBLP:journals/corr/abs-2109-12298} implements proper Poisson sampling. This gap between infrastructure and theory has led to the common practice in academic papers of training with shuffled data, but doing privacy accounting assuming Poisson sampling. While this practice can lead to a fair comparison of algorithms in a research setting, it should be avoid in production settings including the generation of DP synthetic data from real user data.

Recent theoretical \cite{chua2024privatedpsgdimplementations,chua2024scalabledpsgdshufflingvs} and empirical \cite{annamalai2025shuffleshuffleauditingdpsgd}
works show that the gap between the actual $\epsilon$ due to shuffling and one computed assuming Poisson sampling can be significant (up to 4x), especially in low noise regimes (few finetuning steps/epochs). If proper Poisson sampling can't be used, DP introduced noise should be increased to account for shuffling. 

There are other potential sources of inconsistencies that can worsen actual privacy guarantees --- for example using microbatching and clipping at microbatch level, but not accounting for doubled sensitivity (\cite{Ponomareva_2023} Section 5.6).

\paragraph{Components that create inter-example dependencies}
Per-example gradient clipping (more technically, clipping the gradient associated with each row in a minibatch) is the main mechanism in DP-SGD algorithms to limit the contribution of each training example to the overall gradients and the resulting model. However, any components in modern models that create cross-example dependencies within the batch may break the ability of per-example clipping to bound the sensitivity; or alternatively, a robust implementation of per-example clipping will behave as if the gradients are all computed on batches of size one, and then these are clipped and summed, but this will essentially disable the cross-example components of the loss, likely breaking model training.

We highlight three well-known approaches that could trigger such issues:

\begin{compactenum}
    
\item \textit{Load balancing losses} are often used in Mixture of Experts (MOE) training. \textit{Load balancing losses} are commonly used in an attempt to nudge the gates towards distributing the tokens evenly among experts, to ensure inference time efficiency. The routing of each token when load balancing loss is enabled depends on other examples from the batches and assignment of their tokens to the experts. Such losses should be used during pre-training of MOE but not during DP-finetuning.

\item Similar to load balancing losses, \textit{capped routing} \cite{DBLP:journals/corr/abs-2006-16668} which defines a maximum number of tokens tokens that each expert can be assigned for the batch. If more tokens are routed to the expert than its maximum capacity, these extra tokens are not routed to an expert for computation and are instead handled by residual connections after MOE layers. In this setting the prediction of a token from an example depends on whether other tokens from other examples already filled experts' capacity, again introducing an inter example dependency in a batch. Megablocks \cite{gale2022megablocksefficientsparsetraining} is an efficient implementation of uncapped routing that voids token dropping, eliminating inter-examples dependencies and therefore should be preferred over capped routing. 

\item \textit{BatchNormalization layers}, though less common in modern LLM architectures, are still widely used~\cite{DBLP:journals/corr/IoffeS15}.  During training, BatchNormalization layers calculate the mean and standard deviation of activations in the batch and uses these statistics to renormalize layer's input \cite{Ponomareva_2023}.

\end{compactenum}

\paragraph{Tuning the training recipe}
Addressing the pitfalls above requires architecture changes or special care to avoid inadvertently violating the assumptions of the DP analysis, potentially leading to an invalid DP claim. On the other hand, the common pitfull of not carefully re-tuning training hyperparameters is likely to result in a model with very poor utility. DP Training requires additional hyperparameter tuning, including typically a substantial batch size increase, finding an appropriate clipping norm and retuning learning rate and number of epochs \cite{Ponomareva_2023}. On top of this, peculiarities like adversarial training (e.g. in GANs) can make DP trained models more unstable, requiring additional tuning or early stopping.  

\subsection{Best-Effort PII Removal}\label{data-prep}
While applying DP during synthetic data generation is a well established and theoretically grounded technique for protecting sensitive information, DP is not a silver bullet. Known limitations of DP include the fact that shared secrets that are repeated between privacy units are not protected with the same strength of privacy guarantees. Additionally, using DP with high values of $\epsilon$ leads to theoretical guarantees that, on their own, indicate very little privacy. Moreover, explaining the nuances of DP to non-specialists can be challenging. Finally, a production system should guard against incorrect implementations of DP algorithms (see Section \ref{sec-gotchas}) that could invalidate the protections.  

Therefore, DP should be coupled with simple and straightforward techniques that promote data minimization. Techniques like best-effort PII removal, which we discuss next, are \textbf{not sufficient on their own} without DP to ensure privacy of the sensitive data. They instead provide an additional level of reassurance and for modalities like text or tabular data are easy to implement and computationally cheap to apply. 

\paragraph{PII removal}
One preprocessing step that can be used to reduce the chance that DP synthetic data contains sensitive data is \textit{Personally Identifiable Information (PII)} removal from the original sensitive data used for training. Various commerical services (e.g. Google Cloud Sensitive Data Protection \cite{cloudpii} and Azure PII detection service \cite{azurepii}) offer text PII detection and removal capabilities. While the methods behind those services are not nessesarily public, in general PII removal techniques can be categorized into rule-based and ML assisted. 

Rule-based techniques work by matching the data against a large number of regex-based rules that capture PII that conforms to the expected format (e.g. SSNs, email addresses, phone numbers, employee IDs etc). Such techniques are cheap, simple and effective, however the coverage is defined by the breadth of predefined set of rules used.  ML based solutions can be broadly categorized into classifier based that perform Named Entity Recognition (e.g. Name, Address, Phone number, etc) \cite{vakili-etal-2022-downstream} and LLM assisted models that use LLM capabilities. Classifier based models are 
usually expected to have some explainability component, so simpler models like Trees or Naive Bayes can be used. LLM-based solutions include finetuning LLMs for PII removal or using prompt engineering \cite{llmpii}.

It is also possible to combine the various methods for PII removal to maximize the true positive and minimize false negatives. It is worth noticing that after PII detection, one can undertake either PII masking or replacement. 
Traditional masking of PII (e.g. replace the names with placeholders like <NAME>) might hinder the resulting DP synthetic data performance. Ideally PII detection is also accompanied with replacement using random but semantically meaningful values (e.g. replacing names and addresses with randomly generated/selected names and addresses). 

Finally, it is worth pointing out that some modalities like text enjoy established body of work related to PII removal, whereas for example detecting and removal PII in image data 
might be non-trivial. While embedded metadata like geo location, creator of the image etc. can be removed easily, identifying and removing PII from the actual visual representation of an image \cite{imagescans} appears challenging. 

\subsection{Empirical Privacy Auditing} \label{sec:empirical-privacy-auditing}
Theoretical privacy guarantees, while powerful, are not always a complete picture of a system's real-world privacy posture. Empirical Privacy Auditing (EPA) is a direct, practical assessment of a system's privacy properties, conducted by actively mounting privacy attacks against a model or its outputs (e.g., data synthesized from the model). EPA techniques probe a trained ML model to measure how much it has ``memorized'' about its sensitive training data. These methods, which include membership inference and reconstruction attacks \cite{shokri2017membership,carlini2021extracting}, essentially act as a stress test by attempting to determine if a specific person's data was used for training, or by trying to recreate the original data points that are unique to a single person from the model's outputs. If needed, EPA can produce a (high probability) lower bound on the DP's $\epsilon$,\footnote{See \ref{sec:audit-anatomy} for the rationale behind lower bounding $\epsilon$.}, quantifying the empirical leakage under a class of realistic attacks. EPA is essential for several reasons that bridge the gap between theory and practice. 

First, EPA serves as a critical end-to-end system validation. Modern machine learning pipelines, especially those implementing differentially private training, are extraordinarily complex. A correct implementation requires careful handling of per-example gradient clipping, noise generation and addition, and privacy accounting across many iterations and potentially distributed machines. A subtle bug in any of these components can silently and completely invalidate the theoretical privacy guarantee~\cite{namatevs2025privacy}. An empirical audit, by directly attacking the final artifact (the model or its data), functions as a holistic test that can detect such implementation flaws, providing confidence that the system as a whole is behaving as expected~\cite{nasr2023tight}. 

Second, EPA helps to quantify the conservatism of theoretical DP. For example, the provable $(\epsilon, \delta)$ guarantees in the context of DP training in Section \ref{sec-dp-training-image} and Section \ref{sec-dp-finetuning}: (a) hold with respect to worst-case neighboring datasets, (b) assume an adversary who has open box (white-box) access to all intermediate model checkpoints, including the parameters of the final trained or fine-tuned model, and (c) ignore many sources of randomness (e.g. random model initialization or random shuffling and batching). Relaxing some (or all) of these assumptions leads to a more realistic threat model that better captures real-life risks. However, this leads to an intractable analysis of the provable privacy parameters.  This is particularly relevant in common scenarios where models are trained to a high theoretical $\epsilon$ (e.g., $\epsilon=10$). Such a level of privacy protection is formally almost vacuous, but may still offer substantial practical privacy, given the formidable challenges of achieving lower $\epsilon$ values in large generative models trained on vast datasets without significantly compromising model quality or incurring prohibitive computational costs. EPA provides a methodology to measure this practical level of privacy, which is invaluable for stakeholders making trade-offs between privacy and model utility~\cite{hafner2024empirical}.

A deeper layer to this issue is the phenomenon of ``empirical privacy variance''~\cite{hu2025empirical}. Recent work has shown that while the same theoretical $(\epsilon, \delta)$-DP guarantee can be achieved through multiple distinct hyperparameter configurations (e.g., different combinations of batch size, learning rate, and number of training iterations), these different configurations can lead to significantly different levels of empirical privacy leakage. Thus, the theoretical $(\epsilon, \delta)$ pair alone may be insufficient to fully characterize the practical privacy risk of a model, making direct empirical measurement necessary for a complete understanding.

Finally, EPA is a powerful tool for communicating privacy to stakeholders, key opinion formers, and the end users. The formal definitions of DP are notoriously difficult for non-experts, including product managers, legal teams, and end-users, to interpret. A statement such as ``the model is $(\epsilon=8, \delta=10^{-5})$-differentially private'' offers little intuition about the concrete risks. In contrast, an empirical finding like ``we performed a suite of tests demonstrating that the model is incapable of approximately reconstructing any realistic sensitive training data'' is far more accessible and provides a more tangible assurance of safety. EPA can provide the evidentiary basis for these more intuitive and convincing privacy claims, while also allowing privacy experts who are familiar with DP to obtain an empirical estimate of its parameters. 

\subsubsection{The Anatomy of a Standard Audit} \label{sec:audit-anatomy}
While specific techniques vary, a typical empirical privacy audit follows a well-defined, three-step workflow.

\paragraph{Canary insertion.} The process begins by identifying or creating a set of target examples, known as ``canaries,'' that will be the subject of the attack. These can be a reserved subset of the actual training data (e.g., for training data regurgitation tests) or, more commonly, artificially generated records that are inserted into the training set. To simulate a worst-case scenario for a user-level privacy guarantee with a contribution bound of $b$, each canary is often replicated at least $b$ times in the training data. This represents a ``secret'' that is present in every example contributed by the most heavily-represented user.\footnote{One may also elect to use a number of canary repetitions that is a multiple of the contribution bound (say $4b$), to demonstrate protection of secrets that are shared among a small group of users. Secrets about a user that may be shared with large groups or appear in training examples that are not owned by the user are harder to protect using DP, however \textit{group privacy} (Section \ref{sec-privacy-unit-general}) still offers way to quantify protection for such situations.
} Depending on the attack metric, a corresponding set of ``held-out'' canaries, drawn from the same distribution but not included in training, may also be required for comparison. The design of effective canaries is a critical and nuanced task, which is explored in detail in Section \ref{sec:canary_design} below.

\paragraph{Attack execution.} With the canaries inserted into the training data, the system is trained, producing a private model or synthetic dataset. A hypothetical adversary then mounts an attack to detect the canaries' influence. The two most common classes of attack are \textit{reconstruction attacks} and \textit{membership inference attacks} (MIAs). In a reconstruction attack, the adversary attempts to use the model to regenerate each canary, for example, by providing a prefix of a canary and prompting the model to complete it. If the adversary does not have access to the model and is only given data synthesized from it, the adversary can search for (partial) matches between the canary and  data synthesized from the model. In an MIA, the adversary is given a canary and must guess whether it was part of the training set (a ``member'') or the held-out set (a ``non-member''). Of the two, MIA is generally considered the stronger and more fundamental. Intuitively, an adversary performing MIA only needs to guess a single bit of information (whether the example was in the training data) to ``win'', whereas reconstruction usually requires determining more than one bit. The existence of a successful reconstruction attack directly entails a successful MIA: guess that the canary is part of the training set if it is reconstructed from its prefix to some degree of accuracy~\cite{shokri2017membership}.

\paragraph{Risk quantification.} The final step is to measure the success of the attack. For reconstruction attacks, this could be the rate of exact or partial success in reconstructing the canaries. A held-out set of i.i.d.\ canaries is often employed here to test if the difference between reconstruction rates on held-in vs.\ held-out canaries are statistically significant. For MIAs, a held-out set of i.i.d.\ canaries is always necessary, and success is quantified using standard binary classification metrics. While simple accuracy and even Area Under the Receiver Operating Characteristic Curve (AUROC) can be misleading,\footnote{Metrics like accuracy and AUROC are problematic because it is possible to have low average-case leakage even when the attacker could reconstruct the data of a small set of users with high confidence, something which would rightly be considered a serious privacy violation.} a more robust metric is the True Positive Rate (TPR) at a fixed, low False Positive Rate (FPR), such as TPR@1\%FPR~\cite{carlini2022membership}. A high TPR at a low FPR indicates that the attacker can confidently identify members without making many false accusations, representing a significant privacy breach. These empirical attack results can also be used to compute a statistical lower bound on the effective privacy loss parameter $\epsilon$, providing a direct empirical counterpart to the theoretical guarantee~\cite{namatevs2025privacy}.

It may be counterintuitive that a \emph{lower bound} on the privacy parameter is typically what is reported in a privacy auditing scenario, and the reason for this deserves some discussion. A concrete membership inference attack is a specification of neighboring datasets $\mathbb{D}$ and $\mathbb{D}'$, and a classifier $F: \text{Range}(\mathcal{M}) \rightarrow \{0, 1\}$. Often $F$ takes the form of a score-threshold rule: $F(o)=\mathbb{I}\{s(o)\geq \tau\}$ for some scoring function $s$ and threshold $\tau$, where higher values of $s$ reflect greater confidence that the underlying dataset was $\mathbb{D}'$ and not $\mathbb{D}$. The TPR of the attack is then $\Pr[s(\mathcal{M}(\mathbb{D}')) \geq \tau]$ and the FPR is $\Pr[s(\mathcal{M}(\mathbb{D})) \geq \tau]$, where the randomness is taken over the mechanism $\mathcal{M}$. Without loss of generality, the ``true'' $\varepsilon$ of the mechanism is:
\[ \sup_{\mathbb{D}, \mathbb{D}'} \sup_s \sup_\tau \log \frac{\Pr[s(\mathcal{M}(\mathbb{D'})) \geq \tau]}{\Pr[s(\mathcal{M}(\mathbb{D})) \geq \tau]}.\]
The challenge comes from estimating these probabilities from samples. In particular, for any finite set of samples, there will be some setting of $\tau$ for which $\Pr[s(\mathcal{M}(\mathbb{D})) \geq \tau]=0$, so the estimated $\varepsilon$ will be infinite. More generally, when the empirical FPR is very low, we will have extreme variance in our empirical $\varepsilon$ estimate. The accepted solution is to use a statistical upper bound on the $FPR$ so that the estimate remains finite and well-behaved. This can be thought of a smoothing technique so as not to falsely declare a mechanism as non-private due only to estimation error. Thus the challenge of the auditor is to choose $\mathbb{D}, \mathbb{D}'$, and $s$, and then optimize $\tau$ to obtain the highest achievable lower bound on $\varepsilon$. To the extent that the bound remains low despite the auditor's best efforts, we have greater confidence that the mechanism is truly private.

\subsubsection{Auditing the Synthetic Data vs.\ Auditing the Model}
Ultimately our goal is to demonstrate that the synthetic data is anonymous, that is, no user-specific information contained in the training data is present in the synthetic data (explicitly or implicitly). However, if the data comes from a DP finetuned model (e.g. Sections \ref{sec-dp-training-image} and \ref{sec-dp-finetuning}), it is also possible to verify the model's privacy, which thereby guarantees the privacy of the data. This choice reflects a deeper decision about the assumed power of the adversary and has significant implications for the scope and strength of the audit's conclusions.

The primary advantage of a data audit is that it assumes a smaller and more realistic attack surface. The end product that is released is a static dataset, not a queryable model. This aligns with a threat model where the synthetic data itself is released, but any artifacts constructed along the way are ephemeral or private. In this setting, a real attacker who wishes to reconstruct the suffix of a certain prefix they know is in the training data (perhaps ``Brendan McMahan's credit card number is ...'') cannot simply query the model with that prefix as the prompt. Of course, if that prefix happens to occur in the synthetic data, then they could still learn the private information, but careful choices of prompts or post-processing of the synthetic data might defeat such an attack even if it were possible given access to the synthetic data generating model.  To reflect these limitations on a real attacker, we change the class of hypothetical attacks we might conduct as part of EPA; for example, we can no longer directly compute the loss of the model on a canary, instead we might look for features of canaries that are also present in the synthetic data. 

Data auditing is also a more generally applicable methodology, since it does not depend on the mode of synthetic data generation. The same data audit methods apply equally well to datasets generated by private evolution or DP inference, whereas model-based audits only work by interacting with the DP finetuned model.

Model auditing (for synthetic data created via DP-training or finetuning methods) comes with a different set of pros and cons. First, a model audit tests against a more powerful adversary who can interact with the model directly. If a model can withstand direct interrogation---for instance, through carefully crafted prompts or inspection of its internal state---it provides a strong guarantee that no private information has been memorized in a way that could be exploited. This choice to test against a stronger threat model is a core principle of robust security and privacy engineering. The flip side is that the adversary implied by a model audit may be \emph{too} strong: a model vulnerability may not necessarily translate to a practical data vulnerability. As a general recommendation, model auditing may be preferred when it is possible to pass a stringent model audit while still meeting synthetic data quality requirements.

Second, model auditing offers superior scalability and reusability. Once a generative model has been audited and certified as ``anonymous'' to a certain standard, an arbitrary amount of synthetic data can be generated from it without the need for additional audits: the privacy guarantee is a property of the generator itself. In contrast, a data audit is a property of a specific data artifact. If a new batch of synthetic data is generated, it must be re-audited from scratch, and one must account for the cumulative empirical privacy loss of multiple releases.

The field of privacy auditing and memorization attacks is rapidly evolving. Model auditing results may be impacted by additional training such as alignment or SFT, even on non-private data~\cite{nasr2025scalable,borkar2025privacy}. Attacks should be chosen to be as strong as possible given the full training recipe and the way the model is intended to be used. According to current best practices, synthetic data should be generated directly from a model that has passed a privacy audit, without further processing of the model.

\subsubsection{A Taxonomy of Privacy Attacks for Empirical Auditing}
The landscape of privacy attacks is diverse and rapidly evolving. Here we will discuss some of the more influential model and data attacks.

\paragraph{Model-based auditing.}
The foundational paradigm of model-based MIAs is built on the tendency of machine learning models to overfit to their training data. Consequently, members of the training set tend to exhibit lower loss values or produce prediction vectors with lower entropy compared to non-members. A simple, though often effective MIA is the \emph{loss attack}, which classifies a example as a member if its loss on the target model is below a certain threshold. In the case of DP finetuning, this can be significantly improved by considering the \emph{likelihood ratio}, which looks at the \emph{difference} in example log-likelihood between the finetuned and pretrained model~\cite{kandpal2023user}.

The use of likelihood ratio partially mitigates the failure of the loss attack to account for the \emph{intrinsic complexity} of an example. A more sophisticated mechanism to achieve this is the LiRA attack proposed by \citet{carlini2022membership}. LiRA operates by calibrating the confidence score of a target point $x$ on the target model. It does this by comparing the target model's output to the outputs from a large ensemble of ``reference'' or ``shadow'' models. Crucially, these reference models are trained on datasets that are similar to the target model's training data, but some are trained with the target point $x$, and some are trained without it. By observing the distribution of confidence scores from both sets of models, the attacker can compute a likelihood ratio score for the target point. A high score suggests it is much more likely that the target model's output was generated by a model that had seen $x$ during training.

Several model-based reconstruction attacks have been proposed. For generative models trained without DP, very simple methods can be used to successfully extract significant amounts of training data, for example prompting an LLM with an example prefix and using greedy or beam-search decoding~\cite{carlini2021extracting}. Even when the model is not generative, for example a text classification model, a loss-based MIA can be used to prune and rank candidate inputs for non-trivial reconstruction success~\cite{elmahdy2024deconstructing}. Attribute inference attacks aim to deduce sensitive attributes of a data record in the training set, even if the attacker does not have the full record, for example~\cite{fredrikson2015model}.

\paragraph{Data-based auditing.}
The most direct and intuitive form of data-based auditing is to search for verbatim or near-verbatim copies of sensitive data. For text, this is most commonly done using $n$-gram similarity attacks. \citet{meeus2025canarys} propose a method in which both the canary text and the generated synthetic text are decomposed into $n$-grams. Then the collection of $n$-grams from the canary is compared to the collection from the synthetic text using a set-based
similarity metric such as the Jaccard similarity or the Cosine similarity. A high similarity score indicates that a significant portion of the canary's content has been copied into the synthetic output. This method is widely used for tasks like plagiarism detection and identifying near-duplicates in databases, making it a natural fit for detecting blatant data copying in an EPA context. However, its primary limitation is its focus on syntactic overlap. It is excellent at detecting verbatim leakage of a fixed number of tokens but may fail to detect more subtle semantic leakage, where a model learns the meaning of a secret and rephrases it using different words. Data-based auditing is still very much nascent but important area of research.

\subsubsection{Canary Design} \label{sec:canary_design}
The success of an empirical privacy audit is not solely dependent on the attack algorithm; it is also critically reliant on the design of the ``canaries'' used as bait. The properties of these inserted (or identified) records determine the strength of the signal the attack attempts to detect. A weak, poorly designed canary will lead to a weak attack, which in turn produces a loose and uninformative empirical bound on privacy leakage. Therefore, the strategic design of canaries is an active and crucial area of research~\cite{panda2025privacy}. A core challenge in canary design is navigating the trade-off between the \textbf{strength} of the attack signal and the \textbf{naturalness} of the canary (and hence the realism of the risk).

\paragraph{Canary strength.} The key criterion is to choose canaries that will maximize the success of the attacker, giving the highest confidence that any privacy violations that could occur in practice will be detected by the audit. Often the most powerful attack signal is generated by canaries that are highly out-of-distribution relative to the rest of the training data. These examples are ``surprising'' to the model, forcing it to dedicate significant capacity to learn them. This results in large, distinct gradients during training that resist being ``unlearned'' by other data. However, canary strength also depends on the form of model finetuning, and the choice of model vs.\ data auditing. For example, if parameter efficient finetuning (e.g., LoRA) is used, the model may not have sufficient capacity to memorize maximally out-of-distribution canaries. Also, for a data-based audit, a canary consisting of extremely unlikely tokens may have a very low likelihood of being generated. \citet{meeus2025canarys} proposed to use an in-distribution prefix (to maximize the chance of being generated) with an out-of-distribution suffix (to maximize the memorization signal once generation begins).

\paragraph{Naturalness.} There are also arguments for using ``natural'' canaries that are closer to the real data distribution. First, they serve an important communicative purpose. Demonstrating that a model does not leak a canary that looks like realistic user data (e.g., a plausible name and address) is more convincing to users, regulators, and other non-expert stakeholders than demonstrating protection for a string of random tokens. Second, natural canaries are less likely to degrade model utility. Injecting highly out-of-distribution examples into the training data can disrupt the learning process and harm the performance of the final model on its primary task, an undesirable side effect of the auditing process itself.\footnote{If strong out-of-distribution canaries are taking a toll on model quality, an alternative solution is to train two models with identical training setups but for the fact that the auditing model $\mathcal{M}_A$ is trained with canaries added, while the generating model $\mathcal{M}_G$ is trained on only real data. If we make the reasonable assumption that the mere presence of canaries does not significantly affect privacy properties, and $\mathcal{M}_A$ passes the privacy audit, we can confidently generate from $\mathcal{M}_G$. However, it is preferred to use a single model if possible.}

Unfortunately, these criteria are expected often to be in conflict. In case it is not possible to satisfy both goals, it is recommended to audit with several sets of canaries to demonstrate robustness to a range of worst-case and realistic threats.

\subsubsection{Concrete Instantiations of Empirical Privacy Auditing}
While the three step auditing framework of canary insertion, attack execution, and risk quantification is broadly applicable, its concrete instantiation varies depending on the specific synthetic data generation mechanism and the data modality. These variations are crucial for designing effective and informative privacy audits. While a deep-dive into the specific auditing literature for various generation mechanisms and data modalities is out of scope for this manuscript, we provide an overview for the sake of completeness. The reader is encouraged to look into the references below for more details.

\paragraph{The influence of the data generation method.}
The nature of an audit is most directly influenced by the synthetic data generation process. For the widely researched area of DP-trained or DP-finetuned models, the canary-based model and data audits described previously are the standard approach. The attacks typically target the model's memorization of specific examples introduced during the training or finetuning phase. Auditing becomes more nuanced for private prediction based methods. These approaches, such as PATE or DP inference with large language models, introduce privacy at the prediction stage rather than during training on the sensitive dataset. For PATE style algorithms, which involve an ensemble of models trained on disjoint data subsets, canary based attacks remain applicable to the teacher models. \citet{chadha2024auditing} introduce a systematic framework for auditing private prediction protocols, providing a precise privacy analysis for specific instances and demonstrating that the empirical privacy leakage can be close to the analytical bounds. For DP inference methods that leverage the in-context learning capabilities of LLMs, audits must contend with the risk of leaking information from the sensitive examples provided in the prompt. Prior work has demonstrated successful membership and reconstruction attacks against in-context learning in non-private settings, highlighting the inherent risks~\cite{duan2023flocks,morris2024language}.

Finally, methods that use foundational models for inference only like Private Evolution, use private data solely for a voting step to select or refine publicly generated candidates. These methods present a different challenge. Since the sensitive data is never used to update model parameters, the privacy leakage is more constrained and controlled by the DP guarantees of the voting mechanism. These methods are likely the most secure against direct data reconstruction. However, they can still leak private attributes if a publicly generated candidate happens to be semantically close to a private record and is selected by the private vote. More research is needed to fully quantify the empirical privacy risks for this class of methods.

\paragraph{The influence of the data modality.}
The data modality also shapes the specifics of the attack execution and canary design. 

For synthetic \textbf{tabular data}, a rich body of work exists. Membership inference attacks often employ a shadow modeling technique, which is conceptually similar to canary insertion. This involves training many ``shadow'' models on datasets similar to the private one, some containing the target record and some not. A meta-classifier is then trained on the synthetic data from these shadow models to predict membership in the original dataset~\cite{DBLP:journals/corr/abs-2011-07018,sidorenko2025privacypreservingtabularsyntheticdata}. Other attacks are tailored to the generation algorithm itself; for instance, \citet{golob2025privacyvulnerabilitiesmarginalsbasedsynthetic} presented a successful membership attack targeting methods that rely on publishing private marginals.

For synthetic \textbf{image data}, auditing research has revealed significant vulnerabilities, particularly in modern generative models. Diffusion models, for example, have been shown to memorize and reproduce training images, including photographs and trademarked logos, at a much higher rate than older models like GANs~\cite{carlini2023extracting}. More recent work has conducted a comprehensive privacy analysis of (non DP) Image AutoRegressive (IAR) models, a newer architecture that can achieve higher image quality and generation speed than diffusion models. \citet{kowalczuk2025privacy} developed a novel membership inference attack for IARs that achieves a remarkably high success rate (86.38\% TPR@1\%FPR), demonstrating that IARs are significantly more vulnerable to privacy attacks than diffusion models. Their work also presents a data extraction attack capable of reconstructing up to 698 verbatim training images from a large IAR model, underscoring a critical trade off between model performance and empirical privacy.

Finally, for synthetic \textbf{text data}, auditing has evolved from early demonstrations of memorization to sophisticated methods targeting modern LLMs. The reconstruction attacks we discussed previously, where an adversary prompts a model with a prefix of a sensitive sequence and observes if the model completes it verbatim, are often employed with text. This approach was famously used to extract hundreds of unique training sequences from (non-DP) GPT-2, including personally identifiable information, even when the sequences appeared in only a single training document~\cite{carlini2021extracting}. The success of these attacks is heavily influenced by data duplication within the training corpus; models are significantly more likely to memorize and reproduce sequences that appear multiple times, even within a single document~\cite{kandpal2023user}. More recent work has demonstrated that even state-of-the-art, aligned models like ChatGPT can be induced to divulge gigabytes of their training data through a novel ``divergence attack,'' which causes the model to bypass its safety alignment and revert to its base language modeling objective~\cite{nasr2025scalable}.

Beyond direct reconstruction, membership inference attacks for text aim to determine if a specific token sequence was used in training. While classic loss-based MIAs often fail on large language models~\cite{duan2023flocks}, more advanced techniques have shown success. The Likelihood Ratio Attack (LiRA), for instance, improves upon simple loss thresholding by calibrating an example's loss against the distribution of losses from shadow models, making it a more robust measure for membership~\cite{carlini2022membership}. When auditing synthetic data specifically, where an adversary only has access to the generated text and not the model, attacks must adapt. \citet{meeus2025canarys} demonstrate that MIAs based on $n$-gram frequencies in the synthetic corpus can successfully infer membership, suggesting that memorized secrets subtly alter the statistical properties of the generated text. They also find that standard out-of-distribution canaries are less effective in this data-auditing setting, as they are unlikely to be generated. To address this, they propose a hybrid canary design with an in-distribution prefix to increase the likelihood of generation and a high-perplexity suffix to ensure a strong memorization signal. The ongoing development of such attacks highlights a clear need for robust, specialized (or automated) auditing methodologies for LLM-generated text.

\subsection{Security and Data Minimization During DP Synthetic Data Generation}\label{sec-safe-data-handling}

We have so far emphasized the need for anonymization and auditing algorithms to ensure that the derived synthetic data does not leak any private information. Equally important is the need for rigorous security protocols, strict access control lists, and policies to make sure that user data is protected to the highest standards, as in, if the security of the user data is compromised, even by engineers responsible for generating the synthetic data, the anonymization of the DP-fied synthetic data would not hold. This section discusses some of the security and data minimization best practices. 

A cornerstone of secure data handling is strict access control and environmental isolation. All sensitive user data and associated compute workloads, including dataset preprocessing, user contribution bounding, algorithms like differentially private model training or fine-tuning or private evolution, and subsequent auditing, must operate within highly sandboxed and isolated environments. Direct human access to raw sensitive data should be eliminated wherever possible, with instead only explicitly-reviewed and allow-listed code to run on the data via an auditable process. Furthermore, intermediate sensitive data should be subject to strict Time-To-Live (TTL) policies, ensuring its automatic deletion immediately after its designated purpose is fulfilled. This minimizes the window of exposure and reduces the potential impact of a security breach.

Complementing these controls is an unwavering commitment to data minimization \cite{bonawitz2022federated}. Any preprocessing or post-processing required for model training or synthetic data generation should be designed to remove unnecessary information as early as possible in the pipeline, with all unneeded artifacts being promptly discarded. Adopting these best practices creates a multi-layered defense, significantly strengthening the overall privacy posture of generative AI system

\subsection{Lineage Tracking of Synthetic Data Usage\color{red}{*}} \label{lineage-tracking}

So far when talked about DP guarantees of DP synthetic data $S$, these guarantees ($\epsilon, \delta$) covered a single run of an algorithm (whether it is DP-finetuning, Private Evolution or Private inference) on a private dataset $D$  to create DP synthetic dataset. 
While resulting synthetic data $S$ can be used freely (to train any other model, to perform analyses, to combine it with any datasources) due to postprocessing,
all other uses of the private dataset $D$ (or datasets that come from the same source as dataset $D$), including possible training runs used to tune the hyperparameters of the algorithm, or running different algorithms or training other models are not covered by this guarantee. However such isolated guarantees might be insufficient for real world applications. Suppose a company wants to make a DP guaranteed that holds across \textit{all} GenAI/ML/Synthethic Data uses of
a particular user data source.
For example, a company might want to periodically create (e.g. monthly) DP synthetic datasets as private data of users interacting with a banking assistance comes in. Since such datasets can be potentially combined, providing an isolated guarantee for each dataset will not be sufficient to quantify the risks of such much broader use of user data source. Providing guarantees using \textit{composition rules} (e.g. in the simplest form, essentially summing up $\epsilon$s and $\delta$s) might be too pessimistic. 

Applications of DP synthetic data are still nascent and at the current stage it might be not practical yet to aim for such stronger global DP claims. However such real world applications should be considered as DP synthetic data use becomes truly widespread and we gather more knowledge of whether empirical risks compose in similar manner to formal privacy guarantees across multiple uses of the same source of data. When this happens, data lineage will become essential for supporting such usecases. 

The term \textit{data lineage} refers to information about where the data comes from, what it is based on and how it evolves over time \cite{Ikeda2009DataLA}.
In this section we will explore what information should be tracked for DP synthetic datasets to unlock aspirational use of private data sources. While there has been no work specifically on DP synthetic data lineage, we draw inspiration from generic lineage research surveys like \cite{Ikeda2009DataLA} and works like \cite{10.1145/1376616.1376716}.

\citet{Ikeda2009DataLA} suggests categorizing the needs of lineage tracking along several axes. First is \textit{where-lineage}, which can be used to understand what inputs the output depends on, and \textit{how} lineage, which can be used for tracking how inputs were transformed to produce the output. Each type of lineage can be tracked at a coarse (dataset) or fine-grained (example) granularitiy. For DP synthetic data, we recommend tracking  \textit{dataset-level where} lineage and \textit{dataset-level how lineage}. Most of the implementation work of lineage tracking would fall into the former. Indeed, the how lineage will simply describe what DP method was used for producing such data (whether it is DP-finetuning, DP inference, Private evolution or a mix of the above), what privacy unit was used and other details outlined in 5.3.3 Section of \cite{Ponomareva_2023}.

Next, we will provide some motivational examples that demonstrate that careful lineage tracking is important for use of private data sources. 

\subsubsection{Motivating Examples}
The problems that can arise from untracked DP synthetic data can be broadly categorized as problems pertinent to any synthetic data (DP or not, whether it is created using real data or purely synthetic etc.) and problems \textbf{specific to DP} synthetic data. While we briefly discuss the first case, our focus will be on DP-specific needs for lineage tracking. 

\paragraph{Potential problems with any synthetic data usage}
Train-eval cross-contamination is a significant risk (where the same or very similar data ends up in both datasets), potentially leading to over-fiting and over-estimation of model performance. An example of such problematic use would be when synthetic data derived from some sensitive data being included in model training dataset, while a portion of the original sensitive data is retained for evaluation of the aforementioned model. In this case DP synthetic data was influenced by the original sensitive data but can't be deduplicated against it (most likely it won't contain exact matches). Ideally the DP synthetic data should be different enough from the source data that the evaluation results would be similar whether the eval data was exactly the same data used to produce the DP synthetic data, or a separate held-out set; but empirical study is needed to test this hypothesis. 

A related issue could be synthetic data and the real data from which it was derived both ending up in a the training data, leading to over-weighting this data in the training mixture and potential quality degradation.

\paragraph{DP-specific problems}
As the creation and use of DP synthetic data becomes more common-place, as we forsee, a lack of rigorous lineage tracking could lead to a gradual weakening privacy guarantees. Next we provide a non-exhastive list of potential scenarios where this could occur. 

\textbf{E0}: The most basic example is, given a private dataset $D$, running two synthetic data algorithms (e.g. DP finetuning and PE) to obtain datasets $S_1$ and $S_2$. If both algorithms provide ($\epsilon, \delta$) guarantees, combination of $S_1$ and $S_2$ will have formal guarantees of $(2\epsilon, 2\delta)$

This example demonstrates that there is a close connection between the need for lineage tracking and the importance of carefully disclosing exactly what ``private data touches'' are covered by a specific DP guarantee.  Quoting \citet{Ponomareva_2023}, \emph{Private data can be accessed for many reasons during the process of building and deploying ML models.  DP guarantees should include a description of which of these data uses are covered and which are not. E.g., does the DP guarantee apply (only) to a single training run or does it also cover hyperparameter tuning?}. Expanding this to the cases considered here, as long as such precise statements are provided, the above scenario E0 would not violate either of the individual DP guarantees, but it should also be clear that if one wanted a DP guarantee that covers both synthetic datasets $S_1$ and $S_2$, one would have to compose across the two guarantees.  While this scenario might be relatively easy to detect, careful lineage tracking is necessary to surface potentially more subtle versions of the same basic issue, that we explore with the help of the next few examples.

\textbf{E1}: For problems that can arise with DP synthetic data built using \textit{example-level privacy unit}, consider the following scenario. DP synthetic data $S_{D_1}$ was produced using a model $A$ and data $D_1$ with privacy guarantees $\epsilon$, and data $S_{D_1}$ later included into a model $B$ (e.g. during pre-training). We now seek to produce a new version of such synthetic dataset $S_{D_2}$, using a more recent dataset $D_2$, where intersection $D_1$ and $D_2$ is not empty. We now take a model $B$ and DP-finetune it with data ${D_2}$ with $\epsilon$ and generate synthetic dataset $S_{D_2}$. 
This is problematic from DP point of view, as the intersection of $D_1$ and $D_2$ is not protected with the same $\epsilon$, as we are now essentially doing twice the number of DP-SGD steps on this portion of the data. This problem can be resolved by de-duplicating $D_2$ against $D_1$ before producing the DP synthetic data, but this requires (example-level) lineage tracking.

\textbf{E2}: For a user-level privacy unit, the most straightforward example of weakening of privacy guarantees is as follows. Imagine a user \textit{U} contributed their data, for example their reviews of restaurants, to a sensitive dataset $D_1$ ($x$ data points were sampled for each user for inclusion into the dataset). $D_1$ was then used to create DP synthetic version of this dataset $S_{D_1}$ with user-level guarantees of $\epsilon$. Such synthetic data, among other data sources, was then used to train a model $A$ for classifying any reviews as positive or negative. This model was then subsequently finetuned on a restaurant-only reviews dataset $D_2$ with differential privacy ($\epsilon$) to obtain a model $B$ specifically for restaurants. If the user \textit{U} contributed their data to both datasets $D_1$ and $D_2$, even though no single review from the user \textit{U} appears in both of the datasets (further, intersection of $D_1$ and $D_2$ is empty) the user-level privacy guarantees $\epsilon$ no longer hold since essentially a user contributed 2$x$ datapoints to model $B$ for training. This case can't be detected by duplicates checks of $D_1$ against $D_2$. A more modern twist on this scenario would be a foundational generative model $A$ trained on synthetic data $S_{D_1}$ and then subsequently finetuned by a team responsible for restaurant reviews classification with DP using ${D_2}$.

\textbf{E3}: Finally, consider the case of periodically updated synthetic data dumps. 
DP synthetic data was built using a month worth of logs of users interaction with some system (for example, all search queries of users in January) with privacy guarantees $\epsilon$. The training dataset for January DP synthetic logs was prepared to ensure user-level privacy: at most $x$ datapoints were sampled per user. Then the new DP synthetic dataset was prepared using February data by following the same procedure (by sampling $x$ datapoints for each user who interacted with the system in February). This process continued for the whole year, resulting in 12 DP synthetic datasets.
Clearly, each monthly DP synthetic dataset is $\epsilon$-DP w.r.t to \textit{user-month privacy unit}. However the combination of datasets is not $\epsilon$ DP w.r.t \textit{user-level privacy unit}. To see it, imagine first that there was no users whose data appeared in more than 1 month dump. In this case, using parallel composition the combination of all DP synthetic datasets would have been $\epsilon-DP$ w.r.t user level privacy unit. However if there are users that had samples in more than 1 month's dump of data, then the resulting combined synthetic dataset is in worst case $12\epsilon$ w.r.t user-level privacy unit.

\paragraph{Discussion}
An important open question is the degree to which the degradation in the formal DP guarantee in these lineage scenarios actually translates to significantly increased real-world privacy risks. The strongest possible form of DP guarantee would jointly protect \emph{every data access of the totality of a user's privacy data} (across possibly multiple services, and even across multiple companies and organizations). This is an impossibly high bar. Thus, while conceptually the need for lineage tracking is clear, as with the discussion of the unit of privacy, we also encourage pragmatism in calibrating the need for DP guarantees that span multiple ``touches'' of data to the real-world risks, and we hope that further work in empirical privacy auditing will help guide such decisions.

\subsubsection{Potential Solutions}\label{potential-solutions}
In this section we will outline \textit{what} we needs to be tracked to be able to perform lineage checks before using sensitive user data with Differential Privacy. 

We won't prescribe how to technically perform such lineage tracking. There are numerous decisions to be made here: what system to use for such tracking (build an in-house system for automatic tracking or manual entry tracking or using a commercial software), how the data should be stored (whether it is an entry in a database, a graph representation etc.) and retrieved efficiently (e.g. via a recursive query to a database or alternative methods like in \cite{10.1145/1376616.1376716} that use an optimized structure).

\paragraph{Information to track}
For each DP synthetic datasource we create, we  need to keep track of 
\begin{compactenum}
\item \textit{Reference to the original sensitive dataset} (e.g. internal dataset name or its id)
\item \textit{Privacy unit} - whether it is example/user level/sentence level etc.
\item \textit{DP-related information} (method used for creating DP synthetic data, privacy guarantees details outlined in Section 5.3.3 of \cite{Ponomareva_2023})
\item \textit{Information about any LLM checkpoints} that were used in creating DP synthetic data (and implicitly, its lineage information like which original and DP synthetic datasets it was trained on).
\item \textit{Information necessary to run check that would ensure there is no additional loss in privacy from additional training on the source data}:
    \begin{compactenum}
    \item \textbf{For user level privacy unit}
    \begin{compactenum}
        \item \textit{User ids} of users who contributed the data to the dataset that was used for building DP synthetic data. These user ids will be used for de-duplication against each new related data-source. Even if the DP synthetic data is fully anonymous, this set of user ids is potentially highly privacy sensitive (consider a DP synthetic dataset derived from a database of medical histories, for example). Thus, while use of the DP synthetic data might be relatively unencumbered, the strong security and access controls discussed in Section \ref{sec-safe-data-handling} might be necessary to track the lineage information, raising additional systems challenges.
        \item Maximum number of examples per user.
    \end{compactenum}
    \item \textbf{For example-level privacy unit}, if the original dataset is persisted indefinitely, saving the reference against the dataset would suffice. If the dataset is a subject to being removed (for example, due to time-to-live requirements), an alternative to running the date duplicate check is to demand that each DP synthetic dataset is generated using unique time frame (so lineage tracking will save the time-frame used for DP synthetic dataset creation).
\end{compactenum}
\end{compactenum}

\paragraph{Verifications prior to DP synthetic dataset creation}
Tracking the lineage should go hand-in-hand with the checks that need to be implemented before creating any new DP synthetic dataset. Namely
\begin{compactenum}
\item \textbf{For user-level privacy unit}

\begin{compactenum}
\item Verify that the LLM checkpoint has not been built using related DP synthetic dataset (datasets which source is related to the current sensitive dataset that will be dp-fied) with the intersecting user ids. This check is recursive - for any DP synthetic dataset that was used for the checkpoint, the check should run using all the checkpoints in the lineage graph.
\item Verify for each original sensitive dataset, if there was a related sensitive dataset that was used to create DP synthetic data. If yes, verify that user ids used are non-intersecting with the current' sensitive dataset user ids (consider the \textbf{E3} case).
\end{compactenum}

\item \textbf{For example-level privacy unit} the checks similar to the above should verify that some portion of the data was not already used in any DP synthetic data and any of the checkpoints in the lineage graph.

\end{compactenum}

\section{Conclusion}

Recent advances in DP synthetic data have made the technology a viable option for deriving value from sensitive datasets without compromising privacy. When the private source data will only be used for one specific purpose (extracting one statistic, or training a single model), then DP technology specialized to that purposes may still be preferable in terms of utility, privacy, and computation. However, this is rarely the case in the real world. DP synthetic data shines because it can offer a drop-in replacement for the original data, which can subsequently be used for many downstream tasks: plugged into state-of-the-art ML pipelines (without any need for complex DP modifications), inspected and categorized by humans, shared broadly to researchers to further societal goals (for example, environmental or medical data), and more. We hope this work makes clear the range of sophisticated techniques already available for developing DP synthetic data.

Nevertheless, significant challenges and open problems remain. While scattered proofs-of-concept exist, generating high-quality and privacy-safe DP synthetic data requires expert knowledge and complex infrastructure, and as long as this is the case, usage will be limited. There is a clear need to more fully automate and systematize the creation of DP synthetic data in order to make the technology more widely available. A related challenge is that for many domains (e.g., large-scale text or image generation), the source dataset needs to be large (in terms of the number of privacy units it contains) in order to achieve good utility of synthetic data, and substantial computational resources are required for techniques like DP fine-tuning. Bringing down these costs is an obvious and important research direction.

Beyond the purely technical challenges, DP synthetic data deserves a broader dialogue among researchers, privacy and legal experts, regulators, and even end users. As hopefully some of our discussion around the choice of privacy unit and the role of lineage tracking show, a particular DP guarantee can easily be both much stronger than necessary (providing protections against attacks that are far more powerful than are realistically expected), but also fail to address other privacy risks that could be a concern (for example, if a particular secret is contained in many different privacy units of data). 

Stepping back, we cannot escape that privacy is, at the end of the day, a humanistic value, not a mathematical construct. While DP provides a rigorous foundation for reasoning about privacy, as with any mathematical definition, it cannot fully capture the nuance, richness, and even subjectiveness of Privacy in the real world. Questions like ``What is a reasonable privacy unit and choice of $\epsilon$ for this application?'', ``How big are the privacy risks, and what costs are associated with them? How do we balance those risks against the value this data can offer?'', and the like require rich discussion. Our hope is that this work can provide a technical foundation for such conversations, with a goal of establishing best-practices that can provide assurances to synthetic data producers that they are meeting an appropriate standard-of-care for privacy-sensitive data.

\begin{acks}
The authors would like to thank Alessandro Epasto for user contribution bounding discussions, Umar Syed for reviewing the paper, Silvio Lattanzi, Andrew Tomkins, Daniel Ramage, Erik Vee, Raluca Ada Popa for helping navigate internal processes, Badih Ghazi for participating in initial discussions and brainstorming. 
\end{acks}

\printbibliography

%






\end{document}